\documentclass[sigplan,10pt]{acmart}
\acmSubmissionID{133 {(\it Revision from Eurosys 2024 Fall Paper \#580)}}

\AtBeginDocument{%
  }

\setcopyright{acmcopyright}

\acmYear{2025}\copyrightyear{2025}
\setcopyright{rightsretained}
\acmConference[EuroSys '25]{Twentieth European Conference on Computer Systems}{March 30--April 3, 2025}{Rotterdam, Netherlands}
\acmBooktitle{Twentieth European Conference on Computer Systems (EuroSys '25), March 30--April 3, 2025, Rotterdam, Netherlands}
\acmDOI{10.1145/3689031.3696071}
\acmISBN{979-8-4007-1196-1/25/03}

\usepackage{algorithm}
\usepackage{algpseudocode}
\usepackage{adjustbox}
\usepackage{subcaption}
\usepackage{textcomp}
\usepackage{xcolor}
\usepackage{tabularx} 
\usepackage{multirow}
\usepackage{hyperref}
\usepackage{textcomp}

\usepackage{verbatim}
\usepackage{comment}
\usepackage{makecell}

\usepackage{balance}

\usepackage{xspace}
\newcommand{\sysname}{Faro\xspace}
\newcommand{\markpolicy}{Mark/Cocktail/Barista\xspace}
\newcommand{\evalenvfullname}{IBM Cloud Virtual Private Cloud (VPC)~\cite{ibmcloud_vpc}\xspace}

\newcommand{\pureresprop}{\makecell{Linearly \\ Proportional}}
\newcommand{\analyticmodel}{\makecell{Analytical \\ Model}}

\usepackage[normalem]{ulem}

\usepackage{etoolbox}

\newcommand{\ignore}[1]{\xspace}

\newbool{IsPrintComment}
\booltrue{IsPrintComment}

\newcommand{\igc}[1]
{
    \ifbool{IsPrintComment}
    {%
        {\bf \color{cyan} IG comment: #1}
    }{}
}

\newcommand{\bj}[1]
{
    \ifbool{IsPrintComment}
    {%
        {\bf \color{violet} BJ: #1}
    }{}
}

\newcommand{\igt}[1]
{
    \ifbool{IsPrintComment}
    {%
        {\color{red} #1}
    }{}
}

\newcommand{\ay}[1]
{
    \ifbool{IsPrintComment}
    {%
        {\bf \color{teal} AY: #1}
    }{}
}

\newcommand{\cw}[1]
{
    \ifbool{IsPrintComment}
    {%
        {\bf \color{brown} CW: #1}
    }{}
}

\newcommand{\da}[1]
{
    \ifbool{IsPrintComment}
    {%
        {\bf \color{orange} DA: #1}
    }{}
}

\newcommand{\change}[1]
{%
    #1%
}

\newcommand{\bjt}[1]
{%
    #1%
}

\newcommand{\changenew}[2]
{%
    #2%
}

\newcommand{\todo}[1]
{
    \ifbool{IsPrintComment}
    {%
        {\bf \color{red} TODO: #1}
    }{}
}

\newcommand{\squishlist} 
{
    \begin{list}{$\bullet$}
    {
        \setlength{\itemsep}{0pt}      \setlength{\parsep}{3pt}
        \setlength{\topsep}{3pt}       \setlength{\partopsep}{0pt}
        \setlength{\leftmargin}{1.5em} \setlength{\labelwidth}{1em}
        \setlength{\labelsep}{0.5em}
    }
}

\newcommand{\squishend}
{
    \end{list}
}

\usepackage{enumitem}
\newlist{todolist}{itemize}{2}
\setlist[todolist]{label=$\square$}
\usepackage{pifont}
\newcommand{\cmark}{\ding{51}}%
\newcommand{\xmark}{\ding{55}}%

\let\oldhref\href
\renewcommand{\href}[2]{\oldhref{#1}{\color{blue}\underline{#2}}}

\usepackage{soul}

\DeclareMathOperator*{\argmax}{argmax}
\DeclareMathOperator*{\minimize}{minimize}
\DeclareMathOperator*{\maximize}{maximize}

\newcommand{\sloimprovement}{$2.3\times$--$23\times$}
\newcommand{\sloimprovementho}{$1.1\times$--$1.5\times$}
\newcommand{\utilimprovement}{$1.7\times$--$13.1\times$}

\newcommand{\oversubscribed}{oversubscribed}
\newcommand{\Oversubscribed}{Oversubscribed}
\newcommand{\undersubscribed}{undersubscribed}
\newcommand{\Undersubscribed}{Undersubscribed}
\newcommand{\HO}{HO}
\newcommand{\SO}{SO}
\newcommand{\UnderToOver}{Under-to Over-subscribed}

\newcommand{\Section}{Sec.}
\newcommand{\Figure}{Figure}
\newcommand{\Equation}{Eq.}

\begin{document}


\title[\sysname: SLO-Awareness for  On-Premises   Containerized ML Inference Clusters]{
A House United Within Itself:  SLO-Awareness for  On-Premises  Containerized ML Inference Clusters via  \sysname{}}


\author{Beomyeol Jeon}
\email{bj2@illinois.edu}
\affiliation{%
  \institution{University of Illinois Urbana-Champaign}
  \city{Urbana}
  \state{Illinois}
  \country{USA}
  \postcode{61801}
}


\author{Chen Wang}
\email{Chen.Wang1@ibm.com}
\affiliation{%
  \institution{IBM Research}
  \city{Yorktown Heights}
  \state{New York}
  \country{USA}
  \postcode{10598}
}

\author{Diana Arroyo}
\email{darroyo@ibm.com}
\affiliation{%
  \institution{IBM Research}
  \city{Yorktown Heights}
  \state{New York}
  \country{USA}
  \postcode{10598}
}

\author{Alaa Youssef}
\email{asyousse@ibm.com}
\affiliation{%
  \institution{IBM Research}
  \city{Yorktown Heights}
  \state{New York}
  \country{USA}
  \postcode{10598}
}

\author{Indranil Gupta}
\email{indy@illinois.edu}
\affiliation{%
  \institution{University of Illinois Urbana-Champaign}
  \city{Urbana}
  \state{Illinois}
  \country{USA}
  \postcode{61801}
}


\begin{abstract}
This paper tackles the challenge of running multiple ML  inference jobs (models) under time-varying workloads, on a {\it constrained on-premises} production cluster. Our system \sysname takes in latency Service Level Objectives (SLOs) for each job, auto-distills them into utility functions, ``sloppifies'' these utility functions to make them amenable to mathematical optimization, automatically predicts workload via probabilistic prediction, and dynamically makes implicit cross-job resource allocations, in order to satisfy cluster-wide objectives, e.g., total utility, fairness, and other hybrid variants. 
A major challenge  \sysname{} tackles is that using precise utilities and high-fidelity predictors, can be too slow (and in a sense too precise!) for the fast adaptation we require. \sysname's solution is to  ``sloppify'' (relax) its multiple design components to achieve fast adaptation without overly degrading solution quality. \sysname is implemented in a stack consisting of Ray Serve running atop a Kubernetes 
cluster. 
Trace-driven cluster deployments show that \sysname  achieves {\sloimprovement{}}
lower SLO violations compared to state-of-the-art systems.

\end{abstract}



%

\maketitle

\section{Introduction}

\begin{table*}[t]
    \centering
    \caption{\it Multi-tenancy with SLOs for ML Inference Clusters: Prior Work vs. \sysname{}.}
    \vspace*{-0.2cm}
    \adjustbox{max width=\linewidth}{
    \begin{tabular}{c c c c c c c c c c}
        \toprule
         & Ray Serve~\cite{ray_serve_autoscaler} & K8s HPA~\cite{k8s_hpa} & INFaaS~\cite{infaas} & Swayam~\cite{swayam} & Barista~\cite{barista} & MArk~\cite{mark} & Cocktail~\cite{cocktail} & Cilantro~\cite{cilantro} & \textbf{Faro} \\
        \midrule
        Fast Adaptation 
                    &\cmark  &\cmark &\cmark  &\cmark  &\cmark  &\cmark  &\cmark &\xmark &\cmark \\
        SLO-aware              &$\sim$ &$\sim$  &\cmark  &\cmark  &\cmark  &\cmark  &\cmark &\cmark &\cmark \\
        Predictive autoscaling &\xmark  &\xmark &\xmark  &\cmark  &\cmark  &\cmark  &\cmark &\cmark &\cmark \\
        Multitenancy           &\xmark  &\xmark &\xmark  &$\sim$  &\xmark  &\xmark  &\xmark &\cmark &\cmark \\
        Fairness               &\xmark  &\xmark &\xmark  &\cmark  &\xmark  &\xmark  &\xmark &\cmark &\cmark \\
        Limited resource       &\xmark  &\xmark &\xmark  &\xmark  &\xmark  &\xmark  &\xmark &\cmark &\cmark \\
        \makecell{Autoscaling \\ Mechanism} &\pureresprop  &\pureresprop & Additive &\analyticmodel &\pureresprop &\pureresprop &\pureresprop &\makecell{Learning Model-\\based Optimization} &\makecell{Analytical Model-\\based Optimization} \\
        \bottomrule
    \end{tabular}
    }
    \label{tab:comparison}
\end{table*}

Enterprises are increasingly deploying {\it on-premises} (or on-premise or on-prem) clusters wherein ML (Machine Learning) inference engines (e.g., Ray Serve~\cite{ray}, TensorFlow Serving~\cite{tf_serving}, Clipper~\cite{clipper}, etc.) run directly over a container
orchestration framework (e.g., Kubernetes (K8s)~\cite{kubernetes}, OpenShift~\cite{openshift}, etc.). 
This layering offers key advantages arising from containerization such as streamlining of application development workflows, portability of applications, and isolation. 
In the ML inference context we use the terms {\it application} or {\it job}  to refer to a single pre-trained model. Each job receives a dynamically-varying workload stream of queries.
Today it is common to overprovision ML clusters---often each team inside a company (or even each job) gets its own dedicated cluster.  This overprovisioning practice has led to developers requesting more resources than their job needs, and consequently underutilization of overprovisioned resources in production ML clusters~\cite{mlaas_in_the_workload} and other clusters~\cite{quasar, mao2016resource, borg, tetris}.

In order to reduce capital expenses and operating expenses ({\it CapEx} and {\it OpEx} respectively in industry terms) incurred by such siloed infrastructure, {\it consolidated clusters} are becoming popular. 
Multiple teams in an organization share a common cluster, and jobs can {\it autoscale} resource allocation to meet latency requirements as query rates evolve.   
Table~\ref{tab:comparison} shows autoscaling features of state-of-the-art systems. 
Native frameworks like Ray Serve and K8s support adaptation only when developers specify low-level metrics like CPU utilization, queue lengths, etc.---but such metrics are hard for any developer to estimate. What is needed is a way for the system to adaptively assure {\it human (developer)-facing  SLOs} (Service Level Objectives) containing metrics like latency, request drop penalties, etc. 

In ML inference clusters, existing techniques for SLO satisfaction  either minimize resource usage (e.g., Swayam~\cite{swayam}) or minimize the cost of cloud resources (e.g.,  Barista~\cite{barista},  Mark~\cite{mark}, 
INFaaS\cite{infaas}). Cocktail~\cite{cocktail} upscales but cannot downscale. 
Together, these systems either: (i) tackle only a {\it single} job at a time (\cite{barista,mark,cocktail}, K8s Horizontal Pod Autoscaler (HPA)~\cite{k8s_hpa}, and Ray Serve~\cite{ray_serve}), 
or (ii) support {\it multi-tenancy} (i.e., support multiple jobs) but only with unlimited resources available, as in a cloud, e.g., Swayam~\cite{swayam}. 
None of these techniques are extendable to {\it resource-constrained on-premises clusters}.
Merely generalizing single-job scaling techniques to multi-tenancy~\cite{swayam} misses opportunities to exploit  {\it cross-job} resource allocation decisions to maximize cluster-level objectives. 
Cilantro~\cite{cilantro} does multi-tenant scheduling in constrained clusters, but (as we show soon) is unable to adapt quickly for SLOs in ML inference workloads.

We present \sysname{}\footnote{Italian word for {\it Lighthouse}.}, the first SLO-aware autoscaling framework dedicated to fixed-size on-premises ML inference clusters where multiple ML inference jobs with SLOs face dynamic workloads and compete for limited resources.
\sysname{} allows each job to specify its (developer-facing) latency SLO (instead of the developer being forced to guess lower-level metrics). \sysname{} then autoscales each job to maximize cluster-wide utility.
\sysname{} only does autoscaling (i.e., changing the number of replicas for inference jobs), but not scheduling. We integrated \sysname{} with Ray Serve. Together they sit over the K8s  scheduler~\cite{kubernetes}, which schedules replicas to physical/virtual machines.
Production trace-driven cluster deployment experiments and simulations show \sysname{} achieves \sloimprovement{}
lower SLO violations over state-of-the-art.

\section{Motivation: Challenges, Contributions}
\label{sec:motivation}

\begin{figure}[t]
    \centering
    \includegraphics[width=\linewidth]
    {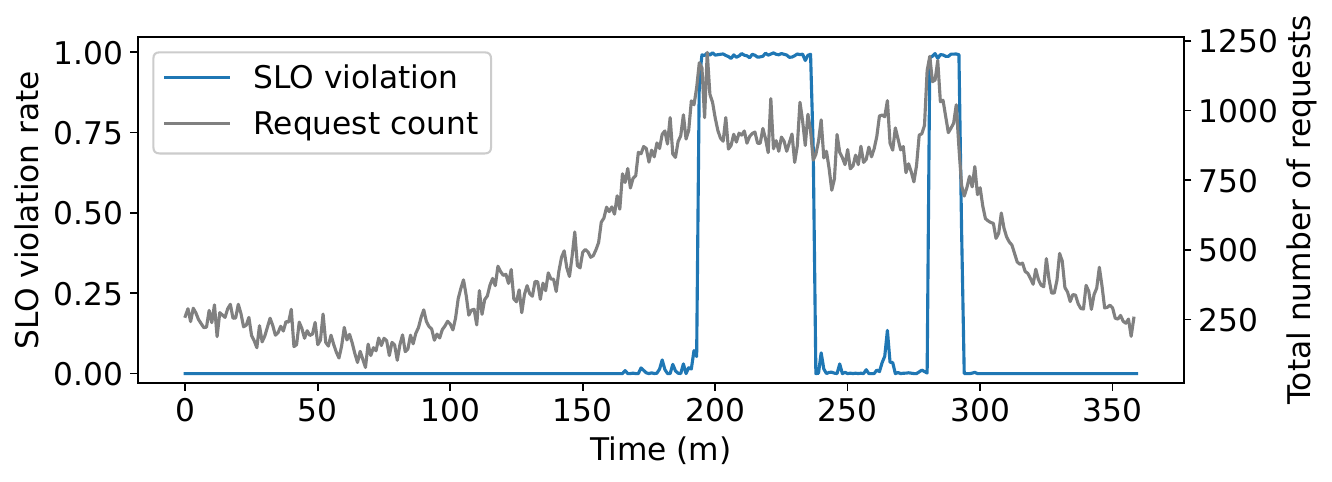}
    \vspace*{-0.4cm}
    \vspace*{-0.1cm}
    \vspace*{-0.1cm}
    \vspace*{-0.1cm}
    \caption{\it Single ML Inference Job (i.e., Model): SLO violation with a fixed-size job and time-varying workload (Azure workload explained in Experimental section). }
    \vspace*{-0.2cm}
    \label{fig:motivation_without_autoscaling}
\end{figure}

\noindent {\bf Challenges in Autoscaling ML: }
While there has been much work on autoscaling in non-ML domains~\cite{autopilot, serverless_autoscaling, k8s_hpa, tiny_autoscaler, Xue_2022, cloudscale, henge, autoscale, cilantro, jockey, morpheus}, the ML domain has some advantages and some differences that have motivated the design of myriad ML-specific autoscaling  solutions~\cite{ray_serve_autoscaler, swayam, mark, cocktail, barista, infaas, ml_inference_for_clouds} (Table~\ref{tab:comparison}). These exploit the advantage in the ML domain that ML inference times, for a given model, are remarkably stable and predictable~\cite{swayam}. 
ML autoscaling is not ``easier'' than generic autoscaling as the latter cannot be applied ``as is'' to ML inference clusters, e.g., some autoscaling techniques use CPU utilization as a guide~\cite{k8s_hpa, autopilot, Xue_2022, tiny_autoscaler, cloudscale}, but they do not work well in ML due to the low correlation between latency and CPU utilization in ML inference jobs.

\Figure{}~\ref{fig:motivation_without_autoscaling} shows that without an autoscaler, SLO satisfaction is poor.
Table~\ref{tab:slo_usecases} shows that SLO targets of ML inference jobs are sub-second latency, and are more stringent compared to cold start overheads which can be tens of seconds.

As a result, autoscaling for steady workloads~\cite{serverless_autoscaling} does not fit with ML inference workloads, and even reactive autoscaling~\cite{henge, autopilot, k8s_hpa, autoscale} cannot react quickly enough.

\smallskip
\noindent {\bf Challenges in \sysname{}:} 
{\sysname{}'s goal is autoscaling in a fixed-size cluster where a fixed set of jobs (pre-trained models) is running, and each job is receiving a stream of queries. Query rates vary significantly over time. Each job has an SLO that is a latency target (tail latency or average latency). Further, the cluster administrator may wish to specify a cluster-level objective that targets either maximizing all jobs' SLOs, a heavy hitter job's SLO, or SLO fairness across jobs, etc. }

\begin{table}[t]
    \centering
    \caption{\it SLO Ranges for Use Cases.}
    \vspace*{-0.1cm}
    \vspace*{-0.1cm}
    \adjustbox{max width=\linewidth}{
    \begin{tabular}{c c}
        \toprule 
         {\bf Use Case} & {\bf Latency SLO Requirement}  \\
        \midrule 
         Virtual assistant response & less than 300 ms \cite{netapp_latency_req} \\
         Estimated time of arrival prediction & a few ms \cite{uber_deepeta} \\
         Automated content moderation & a few ms~\cite{xbox_content_moderation} \\
         Fraud detection & less than 300 ms \cite{visa_fraud_detection} \\
        \bottomrule
    \end{tabular}
    }
    \vspace*{-0.2cm}
    \label{tab:slo_usecases}
\end{table}

\sysname{} has to tackle the following challenges: (i) specifying the {\it cluster-level objective} to maximize  (which we call the  {\it utility function}), (ii) {\it right-sized} workload prediction for each job as a function of its current query rate, 
and (iii) making {\it anticipatory} autoscaling decisions that are both {\it non-myopic and fast}.
\sysname{} tackles {the first challenge above} 
by distilling each job's SLO target and current SLO performance into automatically calculating a per-job {\it max utility} value and {\it current utility value}---concepts of utility have been used elsewhere~\cite{henge, zhang2017live, wang2013slo, sloml, wilkes2009utility, wilkes06, cilantro, jockey}. 
Using these, \sysname{} specifies a {\it cluster-wide utility}  function that the autoscaler aims to maximize. \sysname provides a family of utility functions for the cluster administrator to choose from---for either maximizing cluster utility, or fairness across jobs, or minimizing penalty from dropped requests, or combinations thereof. 
This cluster utility formulation can then be input to optimization solvers.

For the second challenge of workload prediction, existing techniques often focus on predicting future workload in a rather \textit{precise} way---this is unable to capture fluctuations in workloads, and thus often underestimates resource needs to meet the SLO. \sysname{} instead uses {\it probabilistic prediction}. 
This ``sloppy'' prediction provides a ``window'' (range)  of resource needs.
This window (distribution) is fed as input to our autoscaler, which can make anticipatory up/down-scaling decisions.  
Our choice of sloppy prediction is also motivated by other works that have shown that so-called  ``accurate'' predictions of systems performance are not so accurate after all~\cite{fu2023blackboxprediction}.

For the third challenge of making autoscaling decisions in fast and non-myopic ways---first, we observe that solving a cluster-wide utility optimization function can be time-consuming (and thus at odds with the need to adapt quickly). For instance, a statistical package like {\tt scipy.optimize}~\cite{scipy} takes 15-20 seconds to solve an objective function (and still outputs a solution quite far from optimal!). 
Adaptation needs to be faster, within seconds because of stringent latency SLOs.
While other areas have used sloppification~\cite{cedar, coral}, \sysname{} is the first to apply it to ML inference, by creating an {\it approximate} but quickly-solvable objective function.

OS-level approaches based on jobs' priorities and preempting resources, are not viable solutions.  Priority preemption is a match with only some of the above listed cluster administrator goals, e.g., it can satisfy heavy hitters' SLOs but neither SLO fairness nor maximizing all jobs' SLOs.

\begin{table}[t]
    \centering
    \caption{
    \it \sysname{} vs. Baselines: Average Lost Cluster Utility. Total 32 Replicas. Top 9 Azure Function~\cite{azure_function_2019}  and  Twitter~\cite{twitter_trace} Traces. }
    \vspace*{-0.1cm}
    \begin{adjustbox}{max width=0.9\linewidth}{
    \begin{tabular}{c c c c c}
        \toprule
        Faireshare & Oneshot & AIAD & \makecell{Mark  \cite{mark, cocktail, barista}} 
        & \sysname{} \\
        \midrule
        \changenew{1.95}{2.42} & \changenew{4.73}{4.83} & \changenew{2.14}{1.96} & \changenew{1.72}{2.02} & {\bf \changenew{0.69}{0.79}} \\
        \bottomrule
    \end{tabular}
    }
    \end{adjustbox}
    \vspace*{-0.2cm}
    \label{tab:compact_results}
\end{table}

\smallskip
\noindent {\bf Benefits over Existing ML Inference Autoscalers:}
Autoscaling techniques for a given \changenew{}{inference} job may be {\it myopic} if they either: (a) give too many resources right away  in a ``one-shot'' way (e.g., Ray Serve and K8s HPA), 
(b) scale up/down incrementally~\cite{infaas},
or (c) individually scale each job up/down with the state-of-the-art techniques~\cite{mark, barista, cocktail}.

Table~\ref{tab:compact_results} shows the average lost utility
(i.e., max possible cluster utility minus actual achieved cluster utility)
by these techniques with production traces~\cite{azure_function_2019, twitter_trace}.
The first existing approach (a) \textit{Oneshot} gives a lot of resources to a needy job all at once, thus overprovisioning resources and starving other jobs.
With the second category (b) \textit{AIAD}, adaption occurs in an additive-increase-additive-decrease way (e.g., \cite{infaas}), but this adaptation pace is too slow. 
The third category (c) \textit{\markpolicy{}}~\cite{mark,barista,cocktail} does better than Oneshot and AIAD, but operates on each job independently and thus misses opportunities for cross-job resource movements. 
\sysname{} addresses these by leveraging the probabilistic prediction to make {\it intelligent customized} autoscaling decisions for each job periodically,
while maximizing the cluster-wide objective.

\begin{figure}[t]
    \centering
    \includegraphics[width=\linewidth]{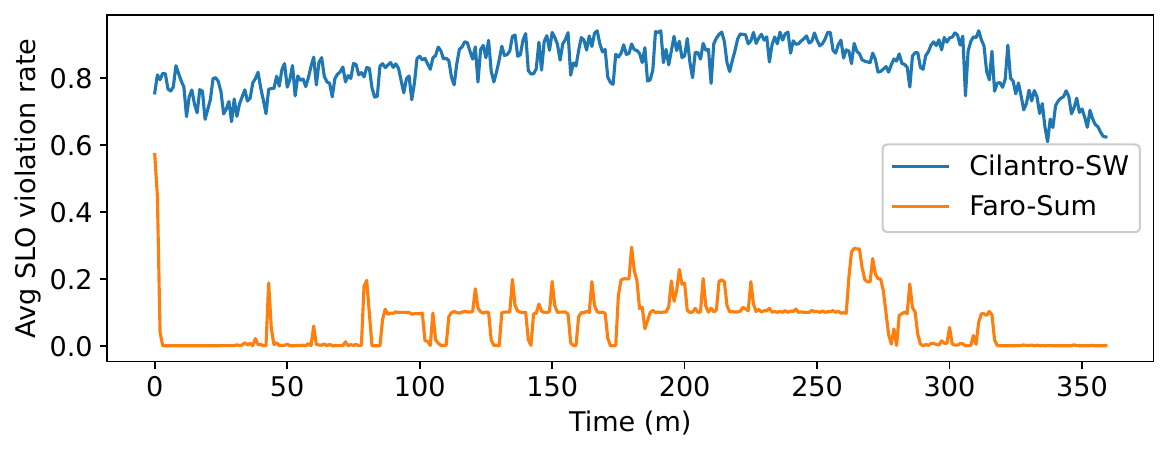}
    \vspace*{-0.8cm}
    \caption{\change{\it Cilantro-SW vs. \sysname{}-Sum: Cluster Size of 32 Replicas. 
    {Top 9 Azure Function Traces~\cite{azure_function_2019}  and  Twitter Trace~\cite{twitter_trace}.}
    }}
    \label{fig:cilantro_comparison}
\end{figure}

\smallskip
\noindent {\bf Benefits over Multi-tenant Utility-based Autoscaler: }
Cilantro \cite{cilantro} uses utilities and ML-based approaches for SLO satisfaction in clusters with microservices, databases, and  ML jobs.
Our experiment in \Figure{}~\ref{fig:cilantro_comparison} shows that for an ML inference job mix with 720 ms latency SLO, Cilantro is unable to react quickly, suffering {average 83.4\%} SLO violation rates. Our \sysname{} system has only {6.9\% average} SLO violation rates.
While Cilantro uses utilities (like \sysname{} and past work~\cite{henge, zhang2017live, wang2013slo, sloml, wilkes2009utility, wilkes06})
and confidence intervals (unlike \sysname{}), its  simple {online} learning {(tree-based binning \cite{kandasamy2020online} and ARMA~\cite{arma})} 
causes it to adapt too slowly to workload changes and spikes.

While this paper is not about optimizing Cilantro, we offer our hypotheses about its performance.
The tree-based binning estimator improves prediction quality initially, but it converges quickly and shows no further improvement with longer training time---even when we ran a 6-hour experiment the prediction quality did not improve much.
We attempted to improve  Cilantro 
in several ways, but they did not help much. Concretely, we found that Cilantro's ARMA model continued being re-trained (fitted) only with the fixed-sized latest request arrival rates, and so early stopping did not move the needle on performance.  
Prior work showed that even with extensive hyperparameter tuning, the prediction quality of regression models like ARMA is inferior to deep learning-based models like \sysname{}'s model \cite{arima_comparison1, arima_comparison2, cocktail}.
\sysname{} achieves SLO satisfaction with $<$ 10 mins of time-series prediction model training (without extensive hyperparameter tuning efforts) and analytic-model-based latency estimation.
Fast autoscale adaptation (for ML inference) can be achieved via  \sysname{}'s sloppified approach. Since we found all  ML inference autoscalers (in the previous paragraph) outperformed Cilantro on SLO satisfaction, we only compare \sysname{} against those baselines in the rest of this paper. 

\smallskip
\noindent {\bf Contributions: }
\sysname's contributions are that it:
(1) Distills user-specified SLOs into utility functions; 
(2) Forecasts workload \changenew{variation}{fluctuation} by   probabilistic prediction; 
(3) Makes fast and close-to-optimal resource allocation decisions for ML jobs by carefully relaxing cluster utility functions;  
(4) Is integrated with Ray Serve framework atop Kubernetes; 
(5) Is evaluated both on a real deployment (cluster) and a deployment-matched simulator, using Twitter \& Azure production traces; 
(6) Reduces both SLO violations by \sloimprovement{}
and lost cluster utility by \utilimprovement{},
vs. state-of-the-art. 

While Faro adapts techniques from elsewhere (e.g., utility, latency estimation, time-series prediction, etc.) our two major contributions are (1) combining these techniques (\Section{} \ref{sec:building_blocks} and \ref{sec:autoscaler}), and (2) using sloppification to enable fast reaction (\Section{} \ref{sec:prob_prediction} and \ref{sec:optimization}). In a sense, sloppification is the philosophy, or the ``glue'', that binds the different techniques. 

\section{\sysname{} Building Blocks}
\label{sec:building_blocks}

\begin{figure}[t]
    \centering
    \includegraphics[width=\linewidth]{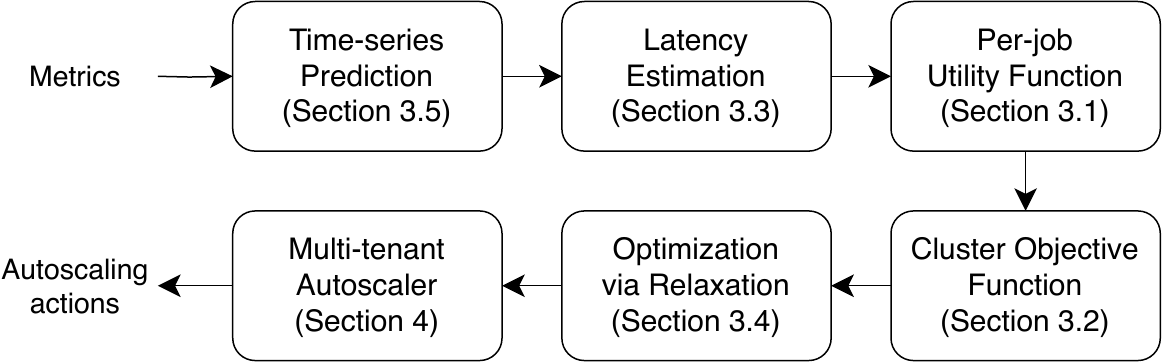}
    \vspace*{-0.5cm}
    \caption{\it Workflow of \sysname{} Autoscaler.}
    \label{fig:workflow}
\end{figure}

\begin{table}[tb!]
\caption{
{
\it {Notations} and Terms Used in the Paper.
}
\vspace*{-0.3cm}
}
\small
\centering
\begin{tabularx}{\linewidth}{lX}
    \toprule
    \changenew{
    \makecell[l]{Job/ \\ Application}
    }{Job ($i$)}
    & One pre-trained model receiving multiple queries \\
    \hline
    $l^i$ & $k$\textsuperscript{th} percentile latency for job $i$ \\
    \hline
    $s^i$ & SLO target latency for job $i$ \\
    \hline
    $x^i$ & Number replicas of job $i$ \\
    \hline
    $\lambda^i$, $p^i$  & For job $i$: Request's  arrival rate, processing time\\
    \hline
    $d^i$ & Drop rate for job  $i$ \\
    \hline
    $\pi^i$, $U^i$, $EU^i$ & For job $i$: Priority, Utility, Effective utility \\
    \hline
    {$Res_{cpu}(x_i)$} & {Required vCPU for $x_i$ replicas} \\
    \hline
    {$Res_{mem}(x_i)$} & {Required memory for $x_i$ replicas} \\
    \hline
    {$ResMax_{cpu}$} & {Total available vCPU in cluster} \\
    \hline
    {$ResMax_{mem}$} & {Total available memory in cluster} \\
    \bottomrule
\end{tabularx}
\vspace*{-0.4cm}
\label{fig:notation}
\end{table}

\Figure{}~\ref{fig:workflow} shows the workflow of \sysname{} for making periodic autoscaling decisions. Metrics are regularly ingested and used to predict both future workloads and resultant latency for each job. The SLO and performance of a job are captured into a {\it utility function}. The utility functions of all jobs are collected and specified in a unified {\it cluster objective function}. Because this is expensive to solve, the objective function is {\it relaxed} and solved via stock solvers. Finally, any remaining {\it multi-tenancy} allocation aspects are solved including minimizing each job's replica count required to meet its SLO.

We first describe  per-job utility functions (\Section{}~\ref{sec:per-job_utility}) and the combined cluster  objective function  (\Section{}~\ref{sec:objective_funcs}). We describe latency estimation  (\Section{}~\ref{sec:latency_estimation}), an input to the relaxed cluster objective  (\Section{}~\ref{sec:optimization}). Lastly we discuss  time-series prediction (\Section{}~\ref{sec:prediction}). These building blocks will then be put together to form our multi-tenant autoscaler in the subsequent section (\Section{}~\ref{sec:autoscaler}). Table~\ref{fig:notation} shows notations used below.

\subsection{Per-job Utility Functions}
\label{sec:utility}
\label{sec:per-job_utility}

The developer of each job merely needs to specify a service level objective (SLO) for the job, containing two pieces: (i) a {\it latency target}, defined as the time between a query being received by the system until an inference output is generated; and (ii) a {\it percentile} $k$, e.g., 99\textsuperscript{th} percentile, 90\textsuperscript{th} percentile, median (50\textsuperscript{th} percentile), etc. Denote the SLO of the $i$\textsuperscript{th} job as $s^i$. If the $k$\textsuperscript{th} percentile latency currently experienced by the job is $l^i$ (Table~\ref{fig:notation}), then \sysname automatically distills the job's SLO into a {\it utility} function: 
\begin{equation*}
    U_{original}(l^i, s^i) = \begin{cases} 
      1 & l^i \leq s^i \\
      0 & l^i > s^i 
    \end{cases}
\end{equation*}
This is a step function. When the latency target is met, the job's utility $U_{original}$ is 1, otherwise it is 0.

\begin{figure}[t]
    \centering
    \begin{subfigure}[t]{0.49\linewidth}
        \includegraphics[width=\linewidth]{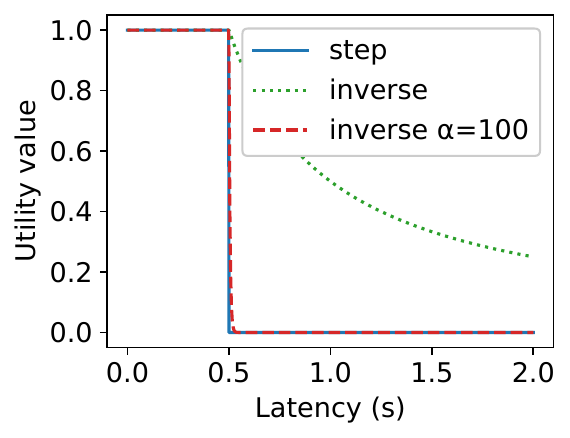}
        \vspace*{-0.5cm}
        \caption{\it Relaxed Utility Function Shapes: Latency SLO Target 0.5 s.}
        \label{fig:utility_function_shape}
    \end{subfigure}
    \begin{subfigure}[t]{0.49\linewidth}
        \includegraphics[width=\linewidth]
        {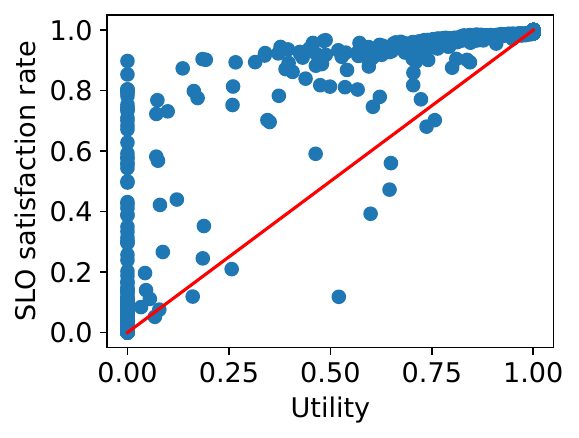}
        \vspace*{-0.5cm}
        \caption{\it Correlation between SLO Satisfaction and Utility with 99\textsuperscript{th} Percentile Latency.
        }
        \label{fig:utility_correlation}
    \end{subfigure}
    \vspace*{-0.3cm}
    \caption{\it Utility Functions: Original and Relaxed, and their relation to SLO Satisfaction. 
    }
\end{figure}

While utility functions have been used to capture SLOs~\cite{henge,cilantro,wilkes2009utility}, 
a major drawback of $U_{original}$ is that
it has a \textit{plateau}---
plateaus~\cite{ai_textbook} are flat regions in the search space where nearby local search points return the same value, which makes it hard for optimization solvers to find global extrema. 
Optimization formulations involving plateau functions require computationally-heavy optimization operations  (e.g., evolutionary algorithm~\cite{de}) and yet produce suboptimal results~\cite{FRIEDRICH2010854, pascanu2014saddle, dauphin2014identifying}.
This makes $U_{original}$ impractical to use.

Hence we also write  {\it relaxed} variants of the original utility that is non-plateau (except the global maxima), 
as follows: 
\vspace*{-0.1cm}
\begin{equation}
    U(l^i, s^i) = \min((\frac{s^i}{l^i})^\alpha, 1) \label{eqn:inverse_util}
    \vspace*{-0.05cm}
\end{equation}

\sysname{} automatically derives this relaxed utility equation from the SLO (latency, percentile). 
\Figure{}~\ref{fig:utility_function_shape} shows that as $\alpha$ increases (to $\infty$), this relaxed utility $U$ approaches
original utility $U_{original}$. 
{Overall, utility values are good proxies for SLOs.
\Figure{}~\ref{fig:utility_correlation} plots, for a trace-driven job, {\it SLO satisfaction rates} defined as {\it fractions of requests satisfied within latency SLO}. The plot confirms that utility values are {\it lower bounds} on the SLO satisfaction rates. \sysname can thus use these pessimistic (utility) estimates for resource allocation decisions.  
}
\sysname{} does not {\it require} jobs to be amenable to a weakened form of SLO (utility)---instead, utility is \sysname's way of solving the SLO problem.
We use both SLO satisfaction rates and utilities as our main experimental metrics.

\subsection{Cluster Objective Functions}
\label{sec:objective_funcs}

In a constrained cluster running multiple jobs (each job \changenew{being}{\bjt{is}} a model receiving inference queries), cluster administrators may have different goals of cluster-wide objectives, including (but not limited to): SLO fairness, total SLO satisfaction, penalizing jobs dropping requests, etc.  
\sysname{} provides a family of {\it cluster objective functions} to choose from. 

While \sysname can handle arbitrary cluster objectives, for concreteness we present below five concrete cluster objectives for utility maximization, fairness, and drop request penalty.

\noindent{\bf 1. \sysname{}-Sum: } Maximize cluster-wide satisfaction of SLOs, i.e., the sum of job utilities, Formally, the goal is to: 
\begin{equation*}
    \maximize{} \sum_i \pi^i U^i
\end{equation*}
Coefficients $\pi^i$s can reflect job $i$'s priority ({1 by default}). 

\noindent{\bf 2. \sysname{}-Fair: } Maximize  fairness, i.e., minimize the difference between maximum and minimum utility across all jobs: 
\begin{equation*}
\minimize{( \max_i U^i - \min_i U^i )}
\end{equation*}
%

\noindent{\bf 3. \sysname{}-FairSum: } Hybrid of \sysname-Sum and \sysname-Fair: 
 \vspace*{-0.1cm}
\begin{align*}
\maximize{\left[ \sum_i \pi^i U^i - \gamma\ (\max_i U^i - \min_i U^i) \right]}
\end{align*}
Parameter $\gamma$ determines the relative weight of fairness. We recommend setting it to job count, normalizing both terms. 

\begin{table}[t]
    \centering
    \small
    \caption{\it Penalty Functions in \sysname{}: based on AWS SLAs~\cite{aws_sla, ibmcloud_sla}.
    }
    \vspace*{-0.3cm}
    \begin{adjustbox}{max width=\linewidth}
    \begin{tabular}{c c}
        \toprule
        Availability Percentage & Service Credit Percentage \changenew{}{(Penalty)} \\
        \midrule
        \changenew{}{Availability $\geq$ 99.0\%}  & \changenew{}{0\%} \\
        95.0\% $\geq$ Availability $<$ 99.0\%  & 25\% \\
        90.0\% $\geq$ Availability $<$ 95.0\%  & 50\% \\
        Availability $<$ 90.0\%  & 100\% \\
        \bottomrule
    \end{tabular}
    \end{adjustbox}
    \vspace*{-0.2cm}
    \label{tab:sla_example}
\end{table}

\noindent{\bf 4. \sysname{}-PenaltySum: } 
When a cluster with limited resources is overloaded under  incoming requests, some requests may need to be explicitly dropped in order to:
(i) satisfy the SLO for remaining queries, and (ii) avoid OOM (Out of Memory) errors. Dropping queries should naturally incur a penalty. To calculate this penalty \sysname{} uses a {\it penalty multiplier}, structured similar to   credit functions commonly used by cloud providers today for dropped requests~\cite{aws_sla, ibmcloud_sla}---Table~\ref{tab:sla_example} shows the penalty function, \changenew{}{$penalty(availability = 1 - d^i(\text{drop rate}))$,} we use,  borrowed from AWS.
\sysname uses this to calculate the {\it Effective Utility (EU)} of the job:
\begin{equation}
    \changenew{
    EU^i = penalty(d^i) \times U^i(\text{non-dropped requests})
    }{
    EU^i = \phi(d^i) \times U^i(\text{non-dropped requests})
    }
    \label{eqn:eu}
\end{equation}
where $\phi(d^i) = 1 - penalty(1-d^i)$.
This allows us to define \sysname-PenaltySum's cluster objective: 
\begin{equation*}
    \maximize{\sum_i \pi^i EU^i}
\end{equation*}

\noindent{\bf 5. \sysname{}-PenaltyFairSum: } This combines \sysname-FairSum and \Equation{}~\ref{eqn:eu} to capture both fairness and penalty: 
\begin{equation*}
\maximize{\left[ \sum_i \pi^i EU^i - \gamma\ (\max_i EU^i - \min_i EU^i) \right]}
\end{equation*}

\subsection{Latency Estimation}
\label{sec:latency_estimation}

Our relaxed cluster optimization (discussed soon) uses  \sysname's two techniques for latency estimation: 

\smallskip
\noindent {\bf I. Upper-Bound Estimator:} This is a pessimistic estimate. 
If $\kappa$ requests arrive simultaneously, with  $N$ replicas and per-request processing time of $c$,  the completion time is $\frac{c\times\kappa}{N}$. 

\smallskip
\noindent {\bf II. Queuing Theory: } 
It is known that  ML inference workloads show Poisson arrival patterns~\cite{deeprecsys, mlperf_inference} and request processing times show low variation~\cite{swayam}. Thus the pessimistic approach may be too conservative. Our second approach is to leverage the {\it M/D/c queuing model}~\cite{mdc_queue, shortle2018fundamentals}, which assumes Poisson arrivals and constant processing time.
The estimate of the $k$\textsuperscript{th} percentile latency is: 
\begin{equation*}
    \begin{cases}
        latency_{M/D/c}(k, p, \lambda, N) & \rho = \frac{p\times\lambda}{N} < 1 \\
        \infty & otherwise
    \end{cases}
\end{equation*}

In practice, we observed that the M/D/c model estimates fewer replicas than the upper-bound model. 
Suppose $p$ = 150 ms, $\lambda$ = 40 req / s, and $s$ = 600 ms (See Table~\ref{fig:notation} for notations).
The upper-bound model estimates 10 replicas, but the M/D/c model estimates 8 replicas with 99.99\textsuperscript{th} percentile latency.  
This can avoid overprovisioning of resources to meet latency SLOs, an asset when cluster resources are limited.

To expedite the M/D/c queue latency estimation, we adopt common engineering approximations~\cite{mdc_approximation} by  treating M/D/c queue waiting time $\approx$ $\frac{1}{2}$ M/M/c queue~\cite{shortle2018fundamentals} waiting time.

\subsection{Cluster Optimization via Relaxation}
\label{sec:optimization}

\begin{figure}[t]
    \centering
    \includegraphics[width=0.8\linewidth]{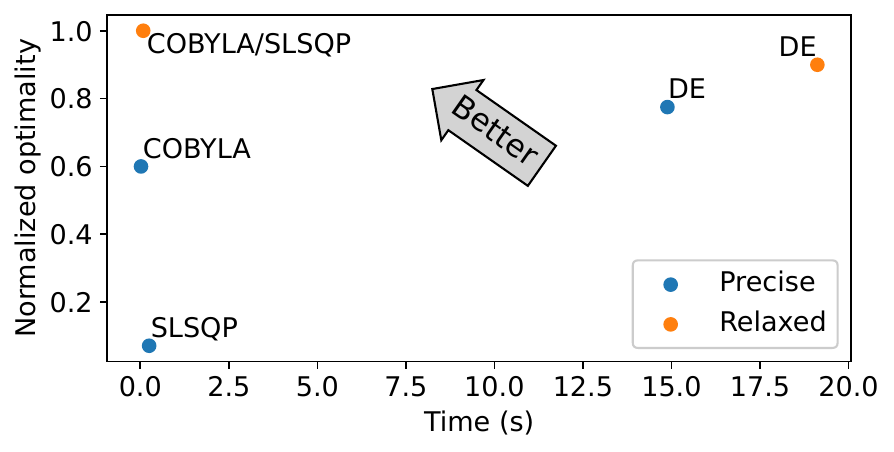}
    \vspace*{-0.1cm}
    \vspace*{-0.1cm}
    \vspace*{-0.1cm}
    \caption{\change{\it Precise vs. Relaxed Solvers in  \sysname. 
    {10 Jobs  (Azure/Twitter traces),  
    40 Total Replicas. 4-core CPU machine.}}
    }
    \label{fig:optimization_solutions}
\end{figure}

Given the latency estimation from \Section{}~\ref{sec:latency_estimation}, we now rewrite the cluster objective functions of \Section{}~\ref{sec:objective_funcs} into precise forms. Then we discuss why and how to relax them.

\smallskip
\noindent{\bf Precise Optimization: } For each of the five cluster objectives listed in \Section{}~\ref{sec:objective_funcs}---\sysname-Sum, \sysname-Fair, \sysname-FairSum, \sysname-PenaltySum, \sysname-PenaltyFairSum---we have precise optimization formulations. 
For concreteness and due to brevity reasons, we present one of the five. For \sysname{}-PenaltySum, \sysname{} calculates utility 
using  the step per-job utility function ($U_{original}$ from \Section{}~\ref{sec:per-job_utility}) and  via  99\textsuperscript{th} percentile M/D/c queue model from  \Section{}~\ref{sec:latency_estimation}.  This gives the following precise cluster optimization problem specification: 
\begin{align}
    \argmax_{x^i, d^i} &{\sum_i \pi^i EU_{original}^i(\lambda^i, x^i, s^i, p^i)} \label{eqn:precise}\\
     = & \sum_i \pi^i {\phi(d^i)} U_{original}^i(\lambda^i(1-d^i), x^i, s^i, p^i) \tag*{} \\
\text{\bf s.t.} \quad & x_i \geq 1, 0 \leq d_i \leq 1 \tag*{(min. \# replicas, drop rate)} \\
    & \sum_i Res_{cpu}(x_i) \leq ResMax_{cpu} \tag*{(vCPU limit)} \\
    & \sum_i Res_{mem}(x_i) \leq ResMax_{mem} \tag*{(memory limit)} 
\end{align}
\sysname{}-Fair, \sysname-Sum{}, \sysname-FairSum{}, and  \sysname{}-PenaltyFairSum, 
each has similar precise formulations. 

\smallskip
\noindent{\bf Difficulties of Solving the Precise Optimization: } Solving this formulation precisely {\it and} quickly is hard. 
\Figure{}~\ref{fig:optimization_solutions} shows data from solving a snapshot from production traces~\cite{azure_function_2019, twitter_trace} by using constrained nonlinear local optimization algorithms: (i) SLSQP~\cite{slsqp}, (ii) COBYLA~\cite{cobyla}, and (iii) an evolutionary algorithm (Differential Evolution or DE~\cite{de}). All three are initiated in the \texttt{scipy.optimize}~\cite{scipy} package. 

We observe that SLSQP and COBYLA solve the problem fast (within a second), but generate suboptimal solutions because of the plateaus arising from $U_{original}$ (described in \Section{}~\ref{sec:utility}). 
Differential Evolution (DE) is able to escape  plateaus, yet even it takes 15 seconds
while still generating suboptimal results.
This implies that solving the precise formulation may neither be fast nor produce optimal solutions.

\begin{figure}[t]
    \centering
    \includegraphics[width=\linewidth]
    {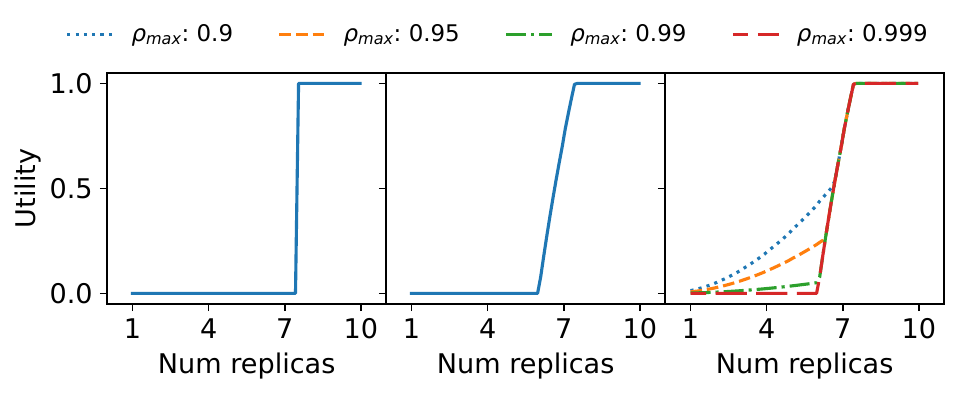}
    \vspace*{-0.6cm}
    \caption{{\it
    {{
    {Relaxation Stages}: (Left) Precise Objective from Job's SLO, (Middle) Relaxation via Inverse (\Section{}~\ref{sec:per-job_utility}), (Right) Second Relaxation via M/D/c queueing.
    }}}
    }
    \label{fig:relaxation_steps}
\end{figure}

\begin{figure}
    \centering
    \begin{subfigure}{0.49\linewidth}
        \includegraphics[width=\linewidth]{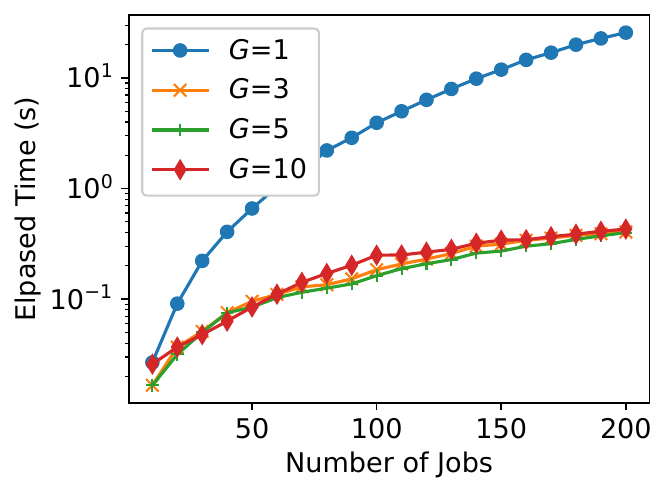}
        \vspace*{-0.5cm}
        \caption{\changenew{}{\it Optimization Time ($G$ groups)}}
        \label{fig:hierarchical:time}
    \end{subfigure}
    \begin{subfigure}{0.49\linewidth}
        \includegraphics[width=\linewidth]{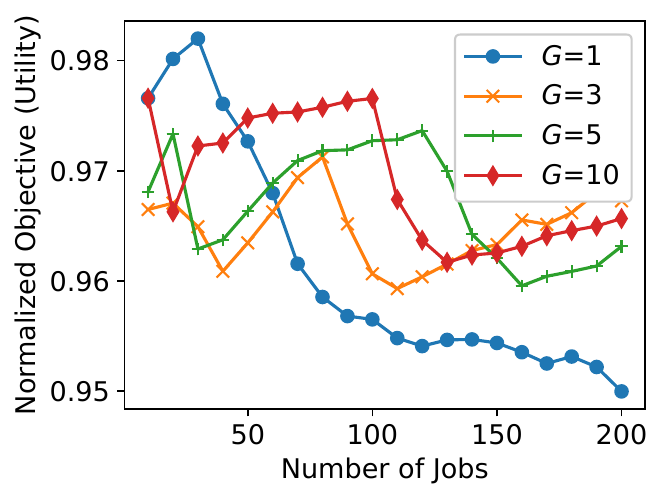}
        \vspace*{-0.5cm}
        \caption{\changenew{}{Normalized Objective Values}}
        \label{fig:hierarchical:util}
    \end{subfigure}
    \vspace*{-0.2cm}
    \caption{{\it Hierarchical Optimization. 70-sample Average. Azure Function~\cite{azure_function_2019} \& Twitter~\cite{twitter_trace}. COBYLA, 1000 iterations. 
    }}
    \label{fig:hierarchical}
\end{figure}

\begin{figure*}[t]
    \centering
    \begin{subfigure}[b]{0.33\linewidth}
        \includegraphics[width=\linewidth]
        {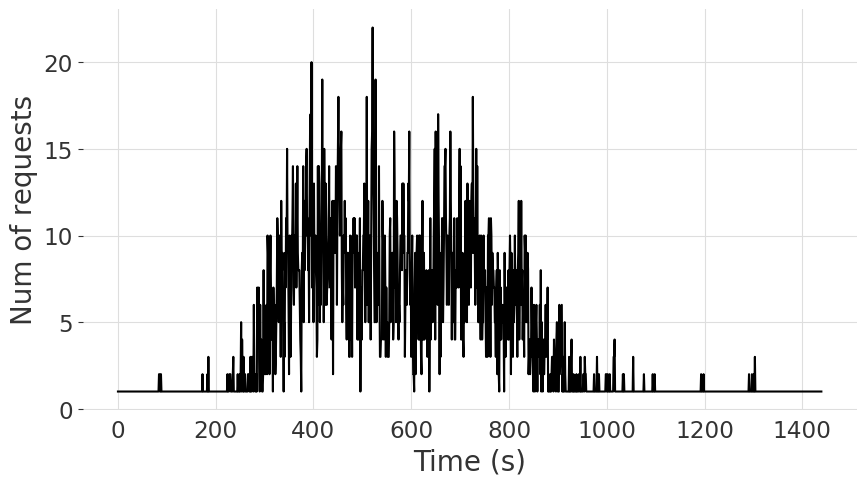}
        \caption{Azure Function 1-day Trace (Sample).}
        \label{fig:sample_workload}
    \end{subfigure}
    \begin{subfigure}[b]{0.33\linewidth}
        \includegraphics[width=\linewidth]{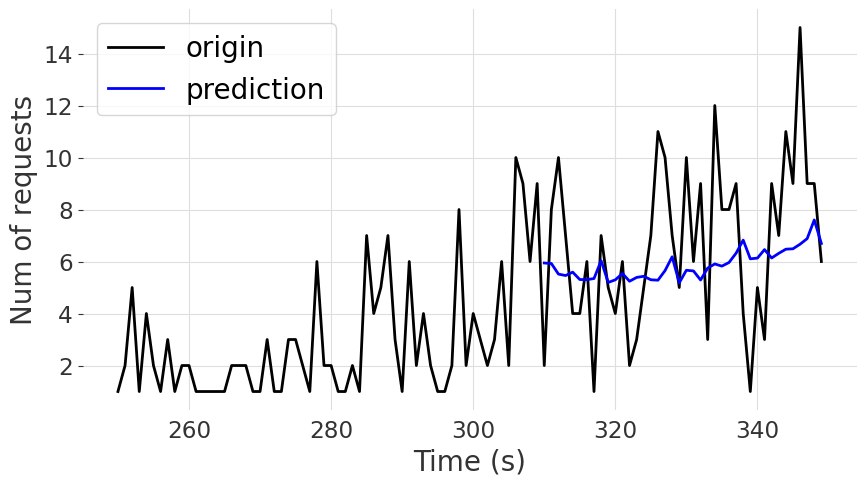}
        \caption{N-HiTS (RMSE) {Prediction}.}
        \label{fig:nhits_rmse_pred_sample2}
    \end{subfigure}
    \begin{subfigure}[b]{0.33\linewidth}
        \includegraphics[width=\linewidth]{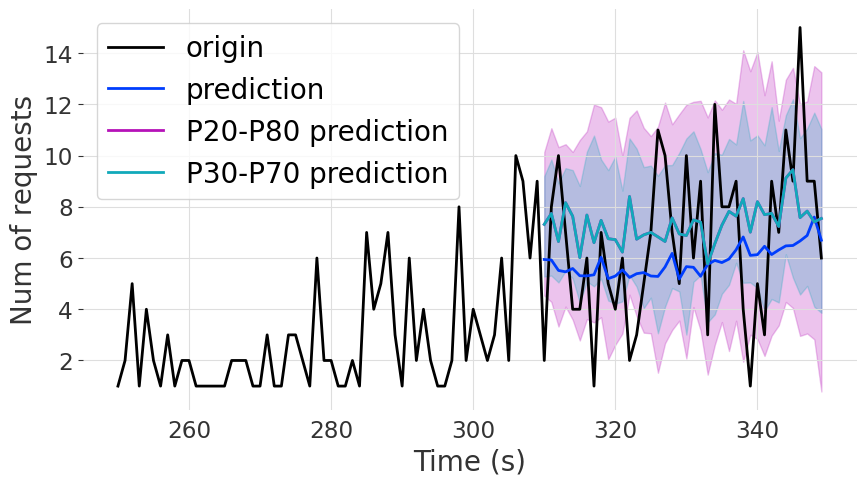}
        \caption{
        {\sysname's N-HiTS (Gaussian) Prediction}.}
        \label{fig:nhits_gaussian_pred_sample2}
    \end{subfigure}
    \caption{\it Azure Function 1-day Trace \cite{azure_function_2019} Sample: N-HiTS Prediction and \sysname's Probabilistic Prediction. }
    \label{fig:probabilistic_prediction}
\end{figure*}

\smallskip
\noindent{\bf Relaxing the Optimization Function: } To derive fast and good quality solutions, 
we relax  precise optimization (e.g., \Equation{}~\ref{eqn:precise}) by converting the plateau functions ({for} per-job utility function,  M/D/c queue latency estimation, and penalty multiplier)  into non-plateau functions. 
\Figure{}~\ref{fig:optimization_solutions} shows that our relaxed variant is fast and produces good-quality solutions. 

\Figure{}~\ref{fig:relaxation_steps} depicts a concrete example of the two stages  of \sysname's relaxation of utility for a given job. 
The {left} plot, derived directly from the SLO, is relaxed in the middle plot using the inverse utility function  (\Section{}~\ref{sec:per-job_utility}).
However, this still has a plateau---if a job receives requests more than its processing capability ($\rho=\frac{p\times\lambda}{N}>1$), its queue size keeps growing (unstable queue). In such a case, we estimate latency as $\infty$ (\Section{}~\ref{sec:latency_estimation}) but this does not differentiate \textit{how fast} the queue size grows. This creates a plateau in {the middle plot}.

Hence we next relax the M/D/c queue latency estimation to produce the right plot in \Figure{}~\ref{fig:relaxation_steps}.
Concretely, \sysname{} estimates latency with an unstable queue as penalized latency under a stable queue. 
{We thus define the 
penalty proportional to the queue size growth rate ($\propto \lambda$), as the following}:
\begin{equation*}
    \begin{cases}
        latency_{M/D/c}(k, p, \lambda, N) & \rho \leq \rho_{max} \\
        \frac{\lambda}{\lambda^{\rho_{max}}} \times latency_{M/D/c}(k, p, \lambda^{\rho_{max}}, N) & otherwise
    \end{cases}
\end{equation*}
Here, $\rho_{max}$ ($= p\lambda^{\rho_{max}}/N < 1$) is a knob determining how close original and relaxed estimations are. 
Values of $\rho_{max}$ near 1.0 
accurately {model}  actual queue behavior, but 
may create a plateau.
\sysname{} sets $\rho_{max} = 0.95$ so as to remove the  plateau but still stay close to estimated latency. 

{
Finally, we note that relaxing the penalty multiplier (Table~\ref{tab:sla_example} in \Section{}~\ref{sec:objective_funcs}) into a piece-wise linear function also creates a plateau-free optimization function.
}

Putting together all these relaxations, 
all three optimizers find (nearly-)optimal solutions (orange dots of \Figure{}~\ref{fig:optimization_solutions}). 
The local optimizers (COBYLA and SLSQP) {solve} the problem within a sub-second, but the evolutionary algorithm (DE) takes longer ({20} seconds) as 
it explores a larger state space after the relaxation. 
{Hence \sysname{} uses COBYLA by default.  }

\smallskip
\noindent {\bf Hierarchical Optimization: }
With more jobs, the optimization variables rise too. To scale the solving, we use a 
\textit{hierarchical optimization}. \sysname{}  creates $G$ groups,  assigning each job to a random group. Group $g_i$'s aggregated job arrival rate and processing time are then: $\lambda_{g_i} = \sum_{j\in g_i} \lambda^j$, $p_{g_i} = \sum_{i\in {g_i}} p^j/|g_i|$. 
{We plug this into \Equation{}~\ref{eqn:precise} as $Res_{cpu}(g_j) = \sum_{i \in g_j} Res_{cpu}(x_i)$, $Res_{mem}(g_j) = \sum_{i \in g_j} Res_{mem}(x_i)$, $\sum_i Res_{cpu}(g_i) \leq ResMax_{cpu}$ and $\sum_i Res_{mem}(g_i) \leq ResMax_{mem}$ }
(Table~\ref{fig:notation} has notations).

\Figure{}~\ref{fig:hierarchical:time} shows 
just a few groups speed up the base \sysname{} solution ($G=1$) by 64$\times$, improving utility by $2\%$ at $>$ 50 jobs. At $<$ 50 jobs, aggregated $\lambda_{g_i}$ and $p_{g_i}$ degrades utility. For simplicity of design we use $G=10$ as default in the rest of the paper. 



\subsection{Probabilistic Time-series Prediction}
\label{sec:prediction}

We describe how \sysname{} predicts a job's future workload. 

\subsubsection{Time Series Prediction Model: N-Hits}

Prior works for proactive autoscaling in ML inference 
employed prediction algorithms like linear regression~\cite{swayam}, Prophet~\cite{barista}, neural network based prediction (e.g., LSTM~\cite{mark}, DeepAR~\cite{cocktail}).
{We implemented 
LSTM and DeepAR, but 
they had lower prediction quality (RMSE: 123.95 and 122.38) vs.  \sysname's approach (RMSE: 116.24) 
(the same setup as \Section{}~\ref{sec:eval}). 
They also incurred 2--3$\times$ higher inference latency over \sysname{}'s predictor.}

At its base \sysname uses a N-HiTS~\cite{nhits}-based prediction model to predict future arrival rates for proactive autoscaling. N-HiTS is a state-of-the-art neural network based time series prediction model that incorporates hierarchical interpolation and multi-rate data sampling to reduce computational complexity and to address prediction volatility.

\subsubsection{Probabilistic Prediction}
\label{sec:prob_prediction}

The base time-series prediction approach is 
{\it too precise}, and thus unable to capture workload fluctuation.
For instance, it is common to train a time-series prediction model using mean absolute error (MAE) or root-mean-square error (RMSE) as a loss function \cite{cocktail, nhits}.
However, the trained model with MAE or RMSE does not capture the fluctuation in a dynamic input workload.

\Figure{}~\ref{fig:sample_workload} shows a portion of a 1-day arrival trace from  
Azure Function trace~\cite{azure_function_2019}. 
We trained the N-HiTS model to predict arrival rates over next {(say)} 40 time units, by using arrival rates over the past {(say)} 60 time units (RMSE loss function).
\Figure{}~\ref{fig:nhits_rmse_pred_sample2} shows prediction results. 
The blue line shows the damped (weighted) average, however it is unable to capture any of the natural fluctuation in the workload.

Capturing fluctuation is critical for SLO satisfaction. To minimize SLO violations, the autoscaling decision should ideally handle 
the {\it maximum expected} arrival rate over the prediction window. 
As \Figure{}~\ref{fig:nhits_rmse_pred_sample2} shows, 
the ground-truth maximum arrival rate over the prediction window is over 2$\times$ higher than  predicted arrival rate.

Prior works address this underestimation problem by threshold based overprovisioning~\cite{swayam} or reactive upscaling~\cite{cocktail, mark} when SLO violations are observed.
The former approach 
is susceptible to misconfiguration. 
In the latter, 
SLO violations still occur 
due to non-negligible upscaling overhead.

\sysname addresses this challenge by leveraging probabilistic time-series prediction~\cite{prob_forecast}. 
Instead of predicting {\it single-valued future arrival rates}, a probabilistic prediction technique predicts the {\it probability distributions} of future arrival rates, i.e.,  parameters of the distribution for future arrivals.
Concretely \sysname incorporates the Gaussian distribution model into N-HiTS. 
\Figure{}~
\ref{fig:nhits_gaussian_pred_sample2} shows \sysname's  prediction results---
we generate 100 prediction samples and plot the range between the min-max samples, 20\textsuperscript{th}-80\textsuperscript{th} percentiles, and 30\textsuperscript{th}-70\textsuperscript{th} percentiles.
The generated sample ranges 
cover the ground truth fluctuation, and \sysname{} captures the workload fluctuation. 

\section{\sysname{} Autoscaler Design}
\label{sec:autoscaler}

Using the building blocks just described, we now detail how \sysname{}'s multi-tenant autoscaler makes decisions in a centralized manner. 
{These decisions {\it implicitly} give more (or less) resources to individual jobs, and thus implicitly move resources (vCPUs {and memory}) from one job to another (since the cluster is capped in size).} 
\sysname{} periodically collects metrics and periodically invokes the \sysname{} autoscaler  (we discuss frequencies later). 
Whenever \sysname's autoscaler is invoked, it executes in  three stages: (1) {\it Per-job Autoscaling Formulation}, (2) 
{\it Multi-tenant Autoscaling}, and (3) {\it Multi-Job Shrinking}.

\subsection{Stage 1: Per-job Autoscaling Formulation}
\label{sec:per-job_problem}

We specify the per-job optimization problem which relates its replicas to SLO satisfaction.
Concretely, \sysname{} fetches metrics for each job: average per-request replica processing time ($p$) and past arrival rates. Over the next 
$w$ time units, future loads 
($\Lambda_{pred}$ = $\{\lambda^1_{pred},\dots,\lambda^w_{pred}\}$) are  predicted via the {probabilistic time-series prediction} 
of \Section{}~\ref{sec:prediction}.
For given future loads, processing time, and SLO target ($s$),
\sysname{} derives a per-job objective function with respect to the number of replicas ($x$). It uses the latency estimation ($L$) of \Section{}~\ref{sec:latency_estimation} and  {the} per-job utility function ($U$) of \Section{}~\ref{sec:per-job_utility}. This per-job objective is:  
\begin{equation*}
    \maximize_{x} \frac{1}{w} \sum_k^w U(L(\lambda_{pred}^k, p, x), s).
\end{equation*}
A cold-start penalty for new replica creation is typically 10s of seconds to a few minutes (up to 70 s in our runs).
\sysname{} incorporates this cold-start penalty into the optimization by planning for replica startup after the cold start. 

\algtext*{EndWhile}
\algtext*{EndFor}
\algtext*{EndIf}
\algtext*{EndFunction}

\subsection{Stage 2: Multi-tenant Autoscaling} \label{sec:multitenant_autoscaling} 
Merely solving the per-job SLO optimization problem from \Section{}~\ref{sec:per-job_problem} may exceed the total cluster resources.
Hence, \sysname{} formulates the multi-tenant autoscaling problem using the approach from  \Section{}~\ref{sec:optimization}. 
\sysname{} specifies this by using the cluster objective function in \Section{}~\ref{sec:objective_funcs} which contains the per-job autoscaling functions from  \Section{}~\ref{sec:per-job_problem}.

\sysname{} solves the multi-tenant autoscaling objective function by using the nonlinear programming solver called COBYLA \cite{cobyla}, using the total cluster resources 
\changenew{}{(vCPU and memory)} 
as a constraint (used resources $\leq$  total cluster resources). 
This solving is usually fast, {e.g., less than 1 second.}
The solution is post-processed by converting it into integers for replica counts, staying within the cluster size. 

\subsection{Stage 3: Shrinking}
\label{sec:shrinking}

While \sysname{}'s multi-tenant autoscaler 
scales jobs to optimize the cluster objective {within the total cluster resources}, 
more resources can be allocated to a job than it needs to meet its SLO for future load.
This especially happens when resources are sufficient for all jobs to satisfy their SLOs.

We subsequently shrink job replica counts
(thus improving resource utilization) 
by {\it post-processing} the previous stage output.
\sysname{} iteratively shrinks (reduces) a replica count for (only) a job with (predicted) utility 1, until the cluster utility changes---thereupon, iterative shrinking is stopped for that job, and the next job is tried. 
(Note this criterion uses cluster utility instead of ``any job's utility'' since \sysname's goal is achieving  cluster objective as specified by the cluster administrator.) 
The result of Stage 3 is {\it right-sized} allocation of resources for the collection of jobs, just sufficient to meet their respective SLOs.
Naturally, Stage 3 is inapplicable if no jobs are meeting their SLOs (i.e., all jobs have utility $< 1$).
Stage 3's decisions are executed automatically by \sysname{}.

\subsection{Selecting an Autoscaling Frequency: \sysname{}'s Hybrid Autoscaler}
\label{sec:hybrid autoscaler}

Too frequent autoscaling makes some resources unusable during the autoscaling operations (e.g., reallocating resources from one job to another job).
If the autoscaling interval is too large, multiple concerns arise: (i) prediction quality degradation,  
(ii) suboptimal autoscaling decisions due to dynamic workload changes, and 
(iii) failure to handle unpredictable workload changes (e.g., spike).

To address this dilemma, \sysname{} contains a {\it hybrid autoscaler} that uses an intelligent combination of a {\it long-term predictive}  autoscaler and a {\it short-term reactive} autoscaler.
\sysname{} performs long-term predictive autoscaling (every 5 min) by using the multi-tenant autoscaler.
\sysname{} triggers its short-term reactive autoscaling more frequently (10 s)
in order to react to unpredictable workload changes.
In this short-term autoscaler, \sysname{}  scales up jobs in an {\it additive} manner and only if SLO violations are observed.
Since the long-term autoscaler determines the current number of replicas, \sysname{} does not use its short-term autoscaler for downscale operations.

\section{Implementation}
\label{sec:impl}

\begin{figure}[t]
    \centering
    \includegraphics[width=\linewidth]{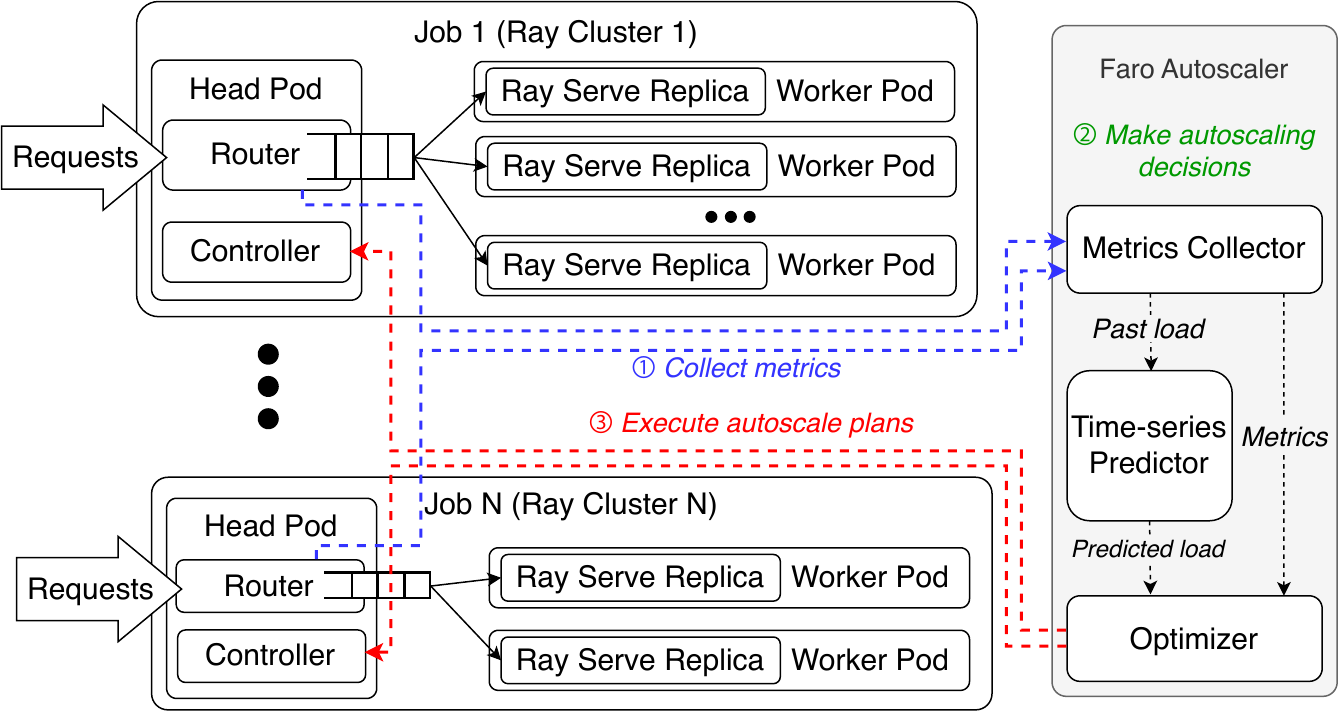}
    \vspace*{-0.2cm}
    \vspace*{-0.2cm}
    \caption{\it \sysname{} Integration with Ray Serve | Kubernetes.}
    \label{fig:architecture}
\end{figure}

\Figure{}~\ref{fig:architecture} shows how  \sysname{} is integrated with  Ray Serve~\cite{ray_serve} v2.0.0 running on a Kubernetes \cite{kubernetes} cluster. 
In Ray terminology, a {\it Ray cluster } refers to a subcluster (a subset of nodes in the cluster).
Atop Kubernetes, a subcluster consists of its controller server (head pod) and multiple worker servers (worker pods).
Ray Serve deploys ML inference jobs on a subcluster (Ray cluster) by running Ray Serve Replicas that perform inference tasks on worker pods, and running a Router that accepts and routes incoming requests to Ray Serve Replicas, on the head pod. 

We allocate a subcluster to a single ML inference job. This isolates each job's congestion at its Ray Router, and  
prevents resource fragmentation as we configure (subcluster-specific) worker pod resource allocation size appropriate to a single Ray Serve Replica.
We modify Ray Router to:
(i) collect metrics (arrival rates, average per-request replica processing time, and average per-request latency), all {continually};  
(ii) export these metrics on API request;
(iii) explicitly drop requests as instructed by \sysname{} autoscaler; and 
(iv)  tail drop~\cite{comer2000internetworking}  requests when the queue length at a Router (for a job) reaches a specified threshold (50; values in range $[20, 100]$ behaved similarly), sending the client an HTTP 503 response. 

\sysname{} autoscaler  (\Section{}~\ref{sec:autoscaler}) runs as a separate Kubernetes pod and periodically scales all Ray Serve jobs. 
Our scaling period of 5 mins is chosen to be larger than autoscaling \changenew{}{(cold-start)} overhead (1 min) but frequent enough to capture workload changes.
The prediction window is 7 mins, so it 
overlaps with the next autoscaling cycle and covers cold start overhead.
\sysname{} changes the number of Ray Serve replicas by using Ray Serve Python API, and drops requests by sending control requests to Routers \changenew{}{based on autoscaling decisions}. 

We implement \sysname{}'s autoscaler in Python and its time-series predictor (\Section{}~\ref{sec:prediction}) using Darts~\cite{darts} v0.22. We train it with a 15-min arrival rate history to predict the next 7-min window.
For optimization solver (\Section{}~\ref{sec:optimization}), 
\sysname{} uses  
COBYLA~\cite{cobyla} in \texttt{scipy.optimize}~\cite{scipy} with the initial variable change of 2. We accelerate the objective function calculation via Numba~\cite{numba}.

\section{Evaluation}
\label{sec:eval}

Our evaluation uses both cluster deployments and simulations. We address the following research questions:
%
{1. How does \sysname{} compare to baseline ML autoscalers? 
2. Do \sysname{} variants (\Section{}~\ref{sec:objective_funcs}) optimize  cluster utility and fairness? 
What tradeoffs are entailed? 
3. How does \sysname{} perform when the cluster is slightly or severely constrained to meet SLOs of all jobs? 
4. Which \sysname components do help the most? 
5. How does \sysname{} work with large-scale workloads?
}

\smallskip
\noindent {\bf Workloads: } 
We generate workloads by using   Azure function traces~\cite{azure_function_2019} and  Twitter traces~\cite{twitter_trace}. Our baseline systems used these same traces 
{\cite{barista, cocktail, clockwork, infaas}.}
We create a diverse set of 10 jobs~\cite{swayam}: 9 workloads using arrival patterns of top 9 Azure function traces (invocation counts), and a 10$^{th}$  workload using query arrival pattern of Twitter trace. 
We use the first 11 days of traces and re-scale them to inject between 1--1600 requests per minute. 
Days 1-10 are used for training the time-series predictor, and Day 11 is used for evaluation.

For cluster deployments, we reduce original traces by splitting them into 4-minute windows and averaging them to reduce experiment time while retaining the temporal patterns. 
The load generator uses Poisson distribution, similar to \cite{mark, swayam, deeprecsys, infaas}. Dropped requests  (\Section{}~\ref{sec:impl}) are marked as failed by the load generator and not resent. 


\begin{table}[t]
    \small
    \caption{\it Baseline Policies Used in Evaluation. \changenew{}{*No Downscale.}
    }
    \vspace*{-0.2cm}
    \centering
    \begin{adjustbox}{max width=\linewidth}
    \begin{tabular}{c c}
        \toprule
         Policy & Systems \\
         \midrule
         FairShare & \changenew{}{Clipper~\cite{clipper}, TensorFlow-Serving~\cite{tf_serving}} \\
         Oneshot & \changenew{}{K8s HPA~\cite{k8s_hpa}, Henge~\cite{henge}, Ray Serve Autoscaler~\cite{ray_serve_autoscaler}}\\
         AIAD & \changenew{}{INFaaS*~\cite{infaas}}
         \\
         \markpolicy{} & \changenew{}{Mark~\cite{mark}, Barista~\cite{barista}, Cocktail*~\cite{cocktail}}
         \\
         \bottomrule
    \end{tabular}
    \end{adjustbox}
    \vspace*{-0.1cm}
    \vspace*{-0.2cm}
    \vspace*{-0.1cm}
    \label{tab:baseline_policies_and_systems}
\end{table}

\begin{figure*}[ht]
    \centering
    \begin{subfigure}{0.33\linewidth}
        \includegraphics[width=\linewidth]
        {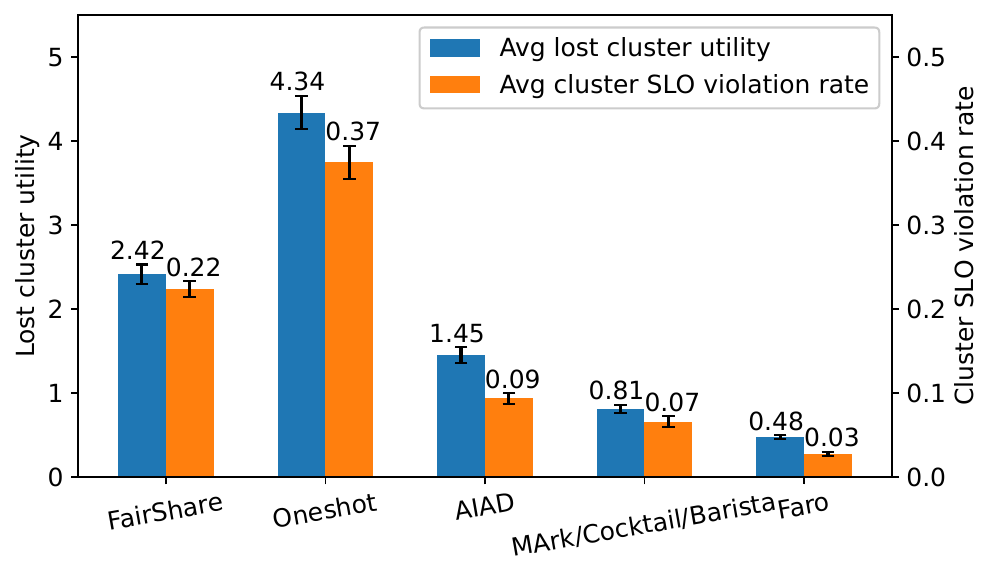}
        \vspace*{-0.5cm}
        \caption{
        Right-sized (RS) Cluster.}
        \label{fig:36_avg_lost_util_slo_plot}
    \end{subfigure}
    \begin{subfigure}{0.33\linewidth}
        \includegraphics[width=\linewidth]{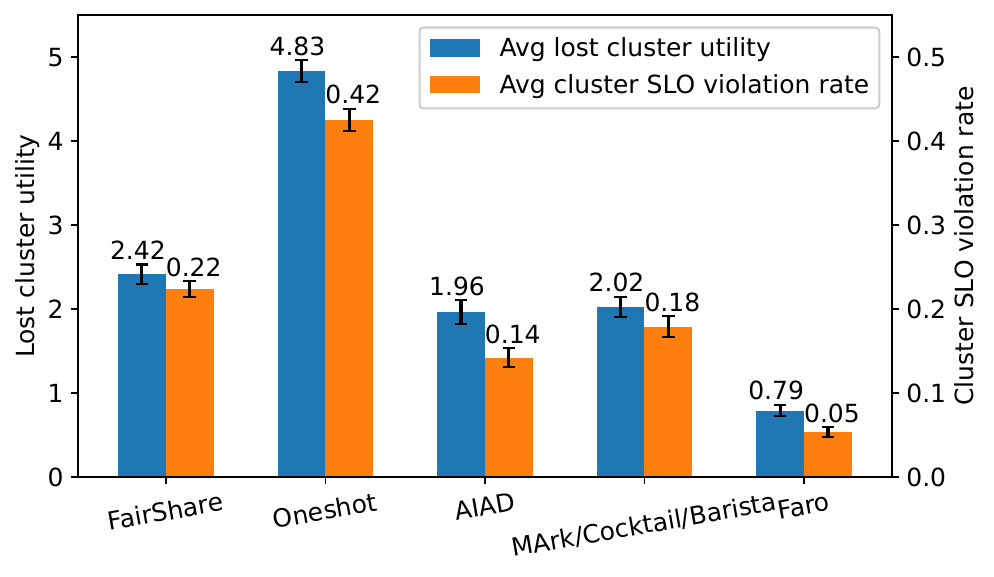}
        \vspace*{-0.5cm}
        \caption{
        Slightly \Oversubscribed{} (\SO{}) Cluster.}
        \label{fig:32_avg_lost_util_slo_plot}
    \end{subfigure}
    \begin{subfigure}{0.33\linewidth}
        \includegraphics[width=\linewidth]
        {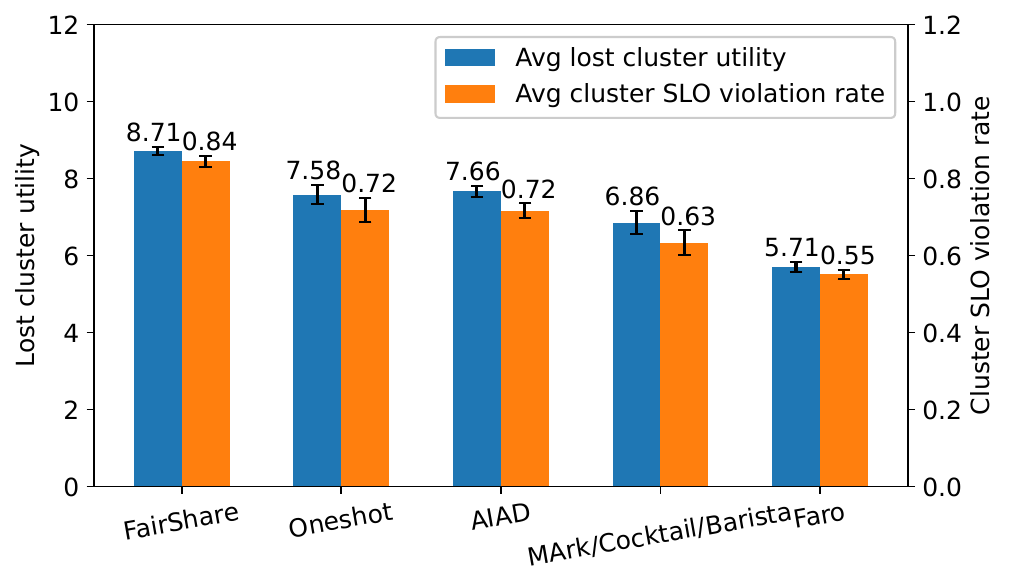}
        \vspace*{-0.5cm}
        \caption{
        Heavily \Oversubscribed{} (\HO) Cluster.
        }
        \label{fig:16_avg_lost_util_slo_plot}
    \end{subfigure}
    \vspace*{-0.5cm}
    \caption{\change{\it Lost Cluster Utility and SLO Violation Rate: Lower is Better for Both Metrics.
    {Cluster sizes (total replicas): 36 (RS), 32 (\SO{}), 16 (\HO{}). Uses \sysname{}-FairSum (RS and \SO{}) and \sysname{}-Sum (\HO{}).}
    \changenew{}{Error Bars: Standard Deviation (SD).}
    }}
    \label{fig:avg_lost_util_slo_plot}
\end{figure*}

\smallskip
\noindent {\bf Experiment Setup: } 
We run our evaluation on {\evalenvfullname{}}.
We deploy a Kubernetes cluster that consists of two {cx2-32x64} {VM instances each} with 32 vCPU and 64 GB memory. We use Kubernetes resource quota~\cite{k8s_resource_quota} to change the cluster size (i.e., total replicas). 
We deploy jobs starting with 1 Ray Serve replica running   ResNet34~\cite{resnet} on  PyTorch~\cite{pytorch} runtime.
We allocate 1 vCPU and 1 GB memory for each Ray worker pod (1 Ray Serve replica). 
Ray head pod runs with 1 vCPU and 4 GB memory. 
\sysname{} autoscaler gets 4 vCPU and 8 GB memory.
To evaluate generality, we do not modify any of \sysname's internal parameters.

\smallskip
\noindent {\bf Baselines: } 
We use {four} 
baselines (Table~\ref{tab:baseline_policies_and_systems}): 

{
\noindent 1. {{\bf FairShare}: No autoscaling. Cluster resources (replicas) are equally shared among jobs{: $\lfloor \frac{total\:resource}{num \:jobs} \rfloor$} }

\noindent 2. {\bf Oneshot}: Reactive heuristic  that allocates a replica count proportional to $\frac{latency}{SLO}$.

\noindent 3. {\bf Additive-increase-additive-decrease (AIAD): } Reactive 
heuristic 
that increments/decrements a replica count when current latency is respectively greater/less than SLO target. 
 
\noindent 4. {\bf \markpolicy{}:} Proactive policy that calculates a replica count
based on the maximum throughput of each replica independently. 
We implemented this in our codebase.}

For Oneshot and AIAD, when a job has stayed overloaded/ underloaded for a threshold length of time, it is marked for scale-up/down.  
We use suggested values of 30 s and 5 mins respectively~\cite{ray_serve_autoscaler,scale_or_not}---this is aggressive scale-up and conservative scale-down. 
For fair comparison we use the same trigger thresholds for \sysname{}'s short-term autoscaler (\Section{}~\ref{sec:hybrid autoscaler}).

\smallskip
\noindent {\bf Metrics: }
We track the actual SLO satisfaction rate, i.e., whether each job's 99\textsuperscript{th} percentile latency (measurements taken every minute) meets the SLO target. If a request is dropped, we assign it $\infty$ latency (this affects the 99th percentile latency). We also plot the {\it job's utility} by plugging latency into the inverse utility function from \Section{}~\ref{sec:per-job_utility} (\Equation{}~\ref{eqn:inverse_util}). The {\it cluster utility} is then just the sum of the utilities of all jobs in the cluster. 

Our main metric is a job's {\it SLO violation rate}  defined as the ratio between number of requests that violate latency SLO requirement, to total number of incoming requests (including dropped requests).
Similarly  {\it cluster SLO violation rate} is defined as average of SLO violation rates of all jobs. 
For \sysname{}-Penalty variants, we  measure {\it effective utility} (\Equation{}~\ref{eqn:eu} in \Section{}~\ref{sec:objective_funcs}). 
Since utility values are usually close to max possible values, we  instead plot the inverse metric, i.e.,:
\begin{equation}
    \textit{(Lost Utility)} = \textit{(Max Utility)} - \textit{(Actual Utility)}.
    \label{eq:lost_utility}
\end{equation}
%
%
Lower lost values are preferable, for both job and cluster.

We set  target latency SLO to 720 ms, $4\times$ of average per-request  processing time (180 ms); 
changing this did not affect the relative performance of \sysname vs. baselines.
We calculate average and standard deviation (SD) of metrics over 5 trial runs for all systems.

\smallskip
\noindent {\bf Deployment Cluster Sizes:} 
We wish to study performance in clusters that are \oversubscribed{} vs. \undersubscribed{} vs. right-sized to meet all jobs' latency SLOs.
\Oversubscribed{} clusters have motivated other problems in ML systems~\cite{polca,asplos2019li, shockwave} as well as other domains~\cite{tetris, vl2, towards2021baset, cake}.
For our workloads, a cluster with a total of 36 total available replicas (across jobs) is {\it right-sized}. 
Any cluster size below 36 is {\it \oversubscribed{}} \changenew{}{(has insufficient resources)}, 
and any cluster size above 36 is {\it \undersubscribed{}}
(has more than sufficient resources). 

\subsection{Comparison with  Baselines}
\label{sec:eval_comparison}

\begin{figure}[t]
    \centering
    \includegraphics[width=\linewidth]{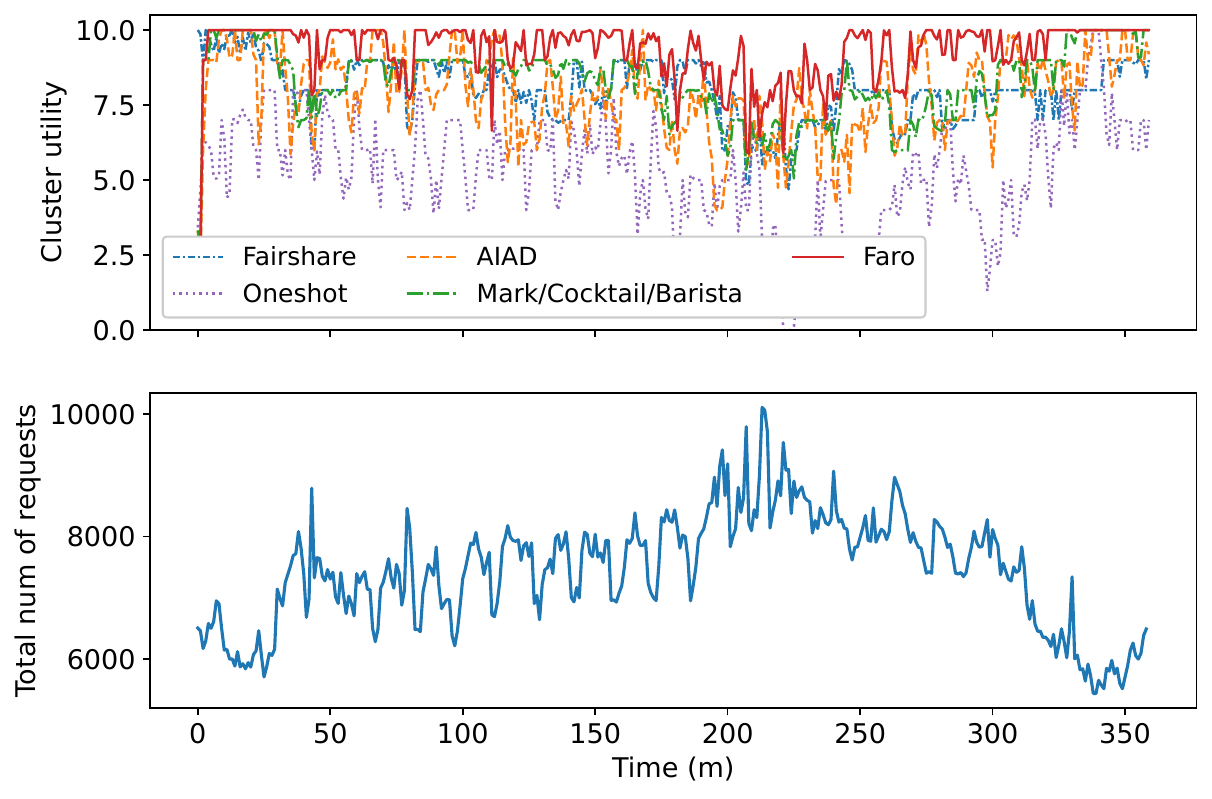}
    \vspace*{-0.2cm}
    \caption{\it Cluster Utility (above) and (below) Workload (Total Number of Requests): Cluster Size of 32 Replicas.}
    \label{fig:avg_utilty_timeline}
\end{figure}

\begin{figure*}[t]
    \centering
    \begin{subfigure}{0.33\linewidth}
        \includegraphics[width=\linewidth] {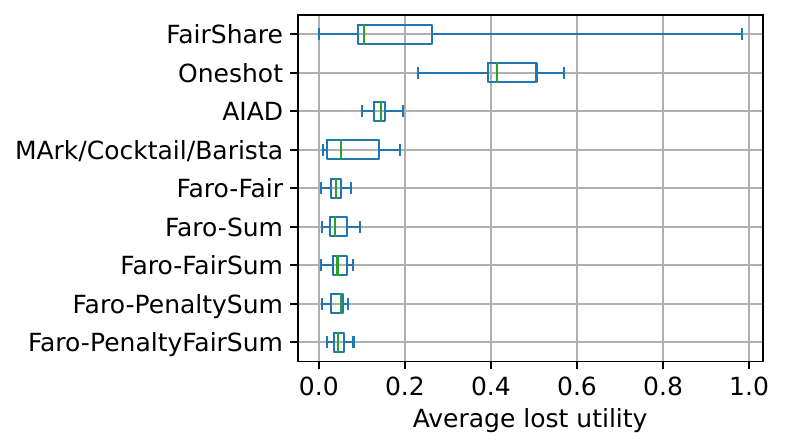}
        \vspace*{-0.5cm}
        \caption{Right-sized (RS) Cluster.}
        \label{fig:36_lost_util_box_plot}
    \end{subfigure}
    \begin{subfigure}{0.33\linewidth}
        \includegraphics[width=\linewidth]
        {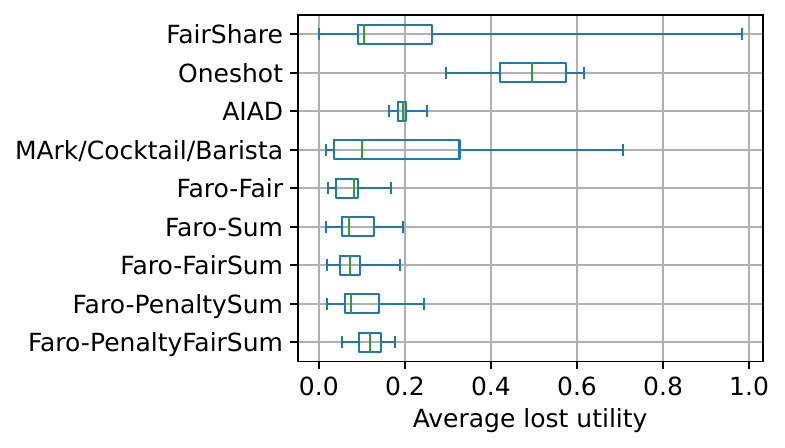}
        \vspace*{-0.5cm}
        \caption{Slightly \Oversubscribed{} (\SO{}) Cluster.}
        \label{fig:32_lost_util_box_plot}
    \end{subfigure}
    \begin{subfigure}{0.33\linewidth}
        \includegraphics[width=\linewidth]
        {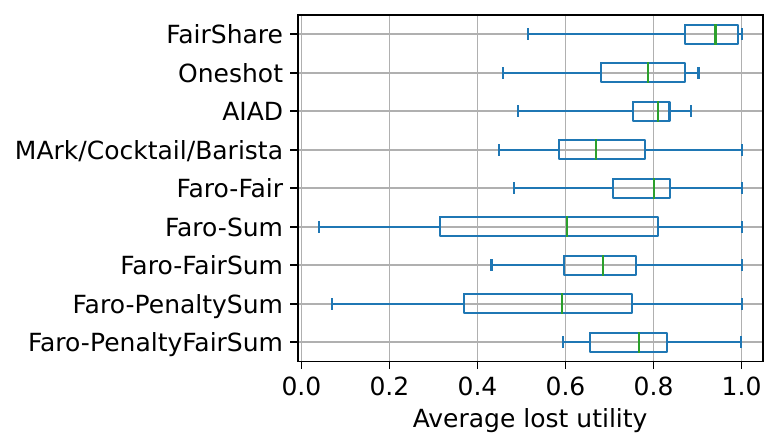}
        \vspace*{-0.5cm}
        \caption{Heavily \Oversubscribed{} (\HO{}) Cluster.}
        \label{fig:16_lost_util_box_plot}
    \end{subfigure}
    \vspace*{-0.5cm}
    \caption{\change{\it \sysname{} Variants vs. Baselines: Utility.
    {Cluster sizes (total replicas): 36 (RS), 32 (\SO{}), 16 (\HO{}).}
    }}
    \label{fig:box_avg_lost}
    \vspace*{-0.2cm}
\end{figure*}

\begin{figure*}[t]
    \begin{subfigure}{0.33\linewidth}
        \includegraphics[width=\linewidth]{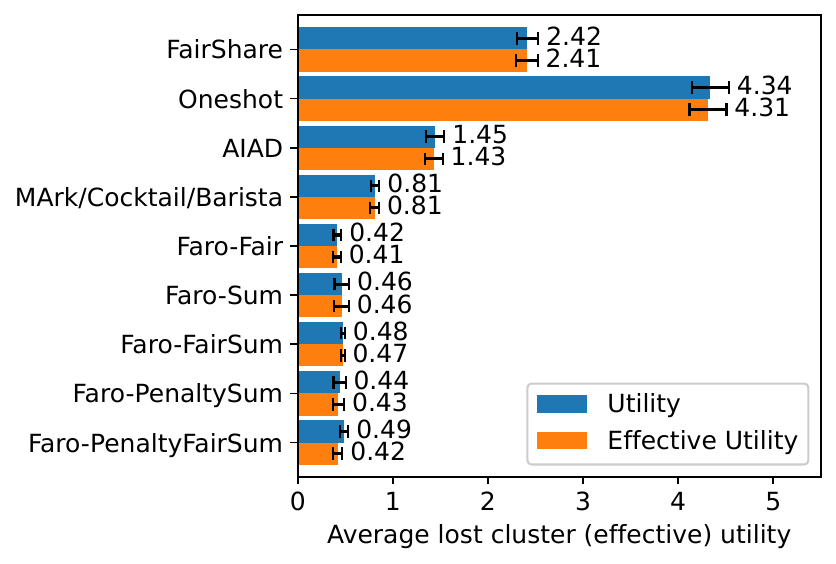}
        \vspace*{-0.5cm}
        \caption{Right-sized (RS) Cluster.}
        \label{fig:36_lost_effective_util}
    \end{subfigure}
    \begin{subfigure}{0.33\linewidth}
        \includegraphics[width=\linewidth]{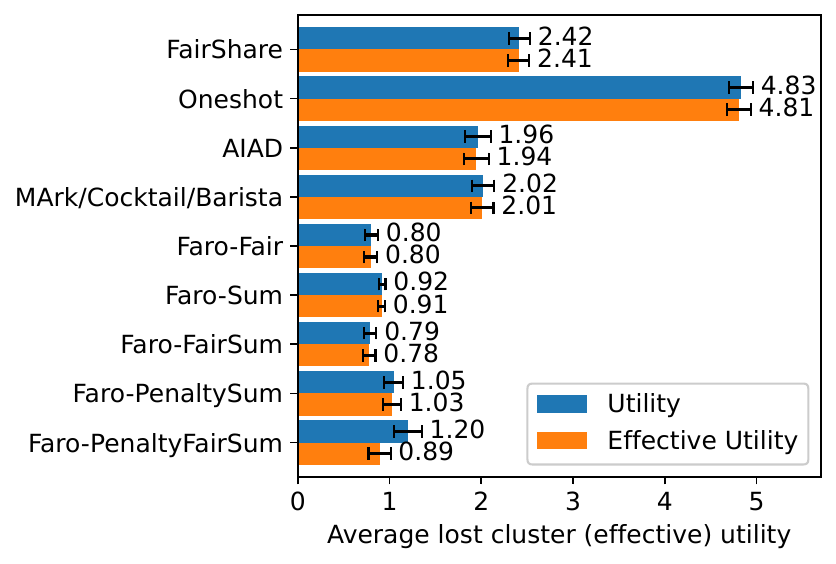}
        \vspace*{-0.5cm}
        \caption{Slightly \Oversubscribed{} (\SO{}) Cluster.}
        \label{fig:32_lost_effective_util}
    \end{subfigure}
    \begin{subfigure}{0.33\linewidth}
        \includegraphics[width=\linewidth]{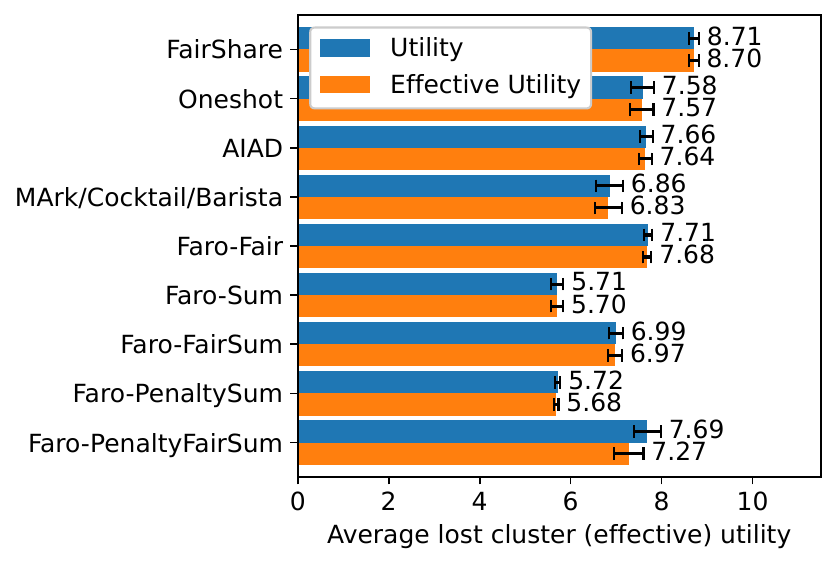}
        \vspace*{-0.5cm}
        \caption{Heavily \Oversubscribed{} (\HO{}) Cluster.
        }
        \label{fig:16_lost_effective_util}
    \end{subfigure}
    \vspace*{-0.5cm}
    \caption{\change{\it \sysname{} Variants vs. Baselines: Utility Metrics.
    Cluster sizes (total replicas): 36 (RS), 32 (\SO{}), 16 (\HO{}). \changenew{}{Error Bars: SD.}
    }}
    \vspace*{-0.2cm}
    \label{fig:lost_effective_util}
\end{figure*}


\noindent {\bf \sysname{} vs. Baselines: } 
{Since cluster utility stays very high, we plot the inverse metric, i.e.,  {\it lost cluster utility} \changenew{defined as: Max possible cluster utility minus actual achieved cluster utility (lower is better)}{(\Equation{}~\ref{eq:lost_utility})}.
\Figure{}~\ref{fig:36_avg_lost_util_slo_plot} shows that in the right-sized cluster with  36 total replicas (just sufficient to satisfy all SLOs), 
\sysname{} outperforms  baselines, lowering cluster SLO violation rate (actual queries missing SLOs) by 
\changenew{1.7$\times$--12$\times$}{2.3$\times$--12.3$\times$}.

We also show the (distilled) lost cluster utility, and it is also lowered by \changenew{1.2$\times$--8.5$\times$}{1.7$\times$--9$\times$}.
Constraining the cluster by dropping it from 36 to 32 total replicas in \Figure{}~\ref{fig:32_avg_lost_util_slo_plot}, degrades \markpolicy{} significantly by increasing its SLO violation rate and lost cluster utility by 157\% and 149\%, respectively.
\sysname{} degradation is lower: 67\% rise in SLO violation rate and 65\% increase in lost cluster utility.

In this \oversubscribed{} cluster compared to the four baselines, \sysname{} lowers cluster SLO violation rate by a range 2.8$\times$--8.4$\times$ and lost cluster utility by 2.5$\times$--6.1$\times$.
Even when the cluster is heavily \oversubscribed{} (16 total replicas), \sysname{} still outperforms all baselines: 1.1$\times$--1.5$\times$ lower cluster SLO violation rate and 1.2$\times$--1.5$\times$ lower lost cluster utility.

\smallskip
\noindent {\bf Why do Baselines Perform Poorly?} Oneshot has the lowest SLO satisfaction: \changenew{58}{42}\% requests at \oversubscribed{} cluster size of 32 total replicas. 
Its aggressive up-scaling of overloaded jobs and its delayed down-scaling (after 5 mins),  prevent efficient resource sharing among jobs.

AIAD achieves \changenew{2.6}{3}$\times$ 
less SLO violations than Oneshot at the cluster size of 32.
Essentially AIAD's conservative and reactive upscaling 
reduces resource underutilization, improving resource sharing. 
But this also means jobs with dynamically changing workloads are not scaled up quickly enough, 
causing 2.8$\times$ more cluster SLO violations than \sysname{}. 

Recall that \markpolicy{} is aimed at meeting a SLO for a single job, and when there are sufficient (infinite) available resources. However, \Figure{}~\ref{fig:36_avg_lost_util_slo_plot} shows that in spite of their sophisticated techniques, their SLO satisfaction at 36 replicas (sufficient cluster size) is comparable to a simpler policy like AIAD. This shows that merely generalizing single-job autoscaling policies to a multi-tenant cluster is insufficient and impractical. \sysname's techniques are needed. 

}

\smallskip
\noindent {\bf Timeline:} \Figure{}~\ref{fig:avg_utilty_timeline} shows a timeline of cluster utility (with workload shown below it). \sysname{} provides max cluster utility (10) for longer periods compared to baselines. Even when the cluster is overloaded ($t$ = 200--230 m) \sysname{}  provides the highest utility vs. baselines. Only during fleeting load spike instances (e.g., $t$ = 45 min, 80 min, etc.), there is a lot of fluctuation among all techniques. 
Nevertheless \sysname{} is able to recover quickly using its short-term autoscaler (\Section{}~\ref{sec:hybrid autoscaler}).

\begin{figure}[t]
    \centering
    \includegraphics[width=0.8\linewidth]{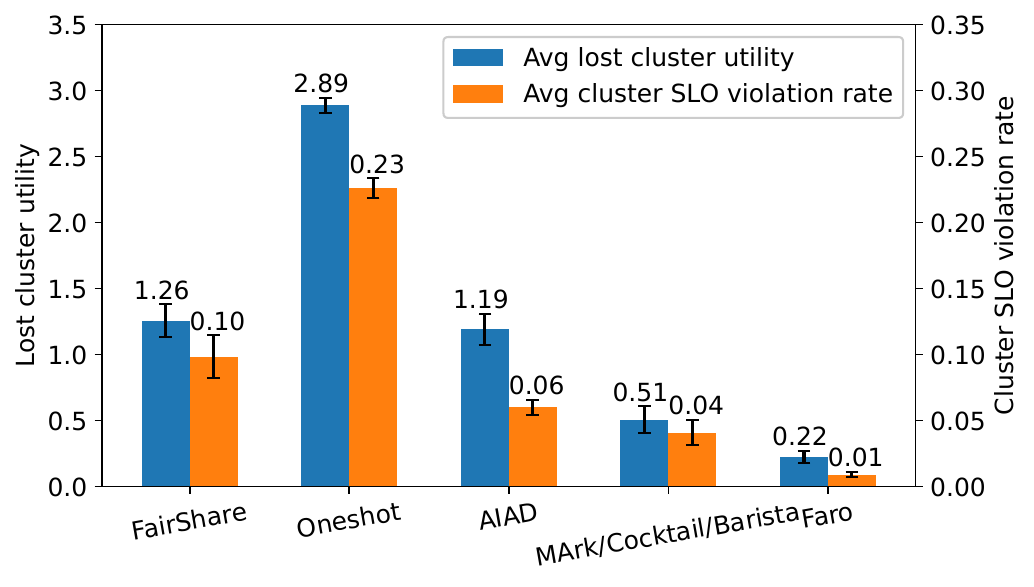}
    \vspace*{-0.4cm}
    \caption{\change{\it Mixed Workloads: Multiple jobs of Multiple Models (ResNet18 and ResNet34).
    Uses \sysname{}-FairSum.
    \changenew{}{Error Bars: SD.}
    }}
    \label{fig:mixed_avg_lost_util_slo_plot}
    \vspace*{-0.4cm}
\end{figure}

\subsection{Faro Variants}
\label{sec:eval_variants}

We compare \sysname{} variants using  \Section{}~\ref{sec:eval_comparison}'s configurations.

\noindent {\bf Fairness: }
\Figure{}~\ref{fig:box_avg_lost} 
shows box plots of lost job utility, measured across jobs. {\it Tighter (smaller) whiskers indicate higher fairness across jobs.  }  
Each box ranges from  25\textsuperscript{th} percentile to  75\textsuperscript{th} percentile, and the middle line shows the median. Whiskers show minimum and maximum utility. 

{
Counterintuitively FairShare is unfair. Since the required resources to satisfy an SLO vary over time, static fair allocation is ineffective. 
}
Oneshot exhibits unfairness as well as poor SLO satisfaction---a single overloaded job takes most of the available resources, preventing others from up-scaling.
AIAD is better at fairness as its cautious autoscaling does not over-allocate, and thus it retains resources for needy jobs.

Although \markpolicy{} has higher cluster utility than AIAD and Oneshot, it is unfair especially in a slightly \oversubscribed{} cluster (\Figure{}~\ref{fig:32_lost_util_box_plot}), where max lost utility is \changenew{20}{7}$\times$ worse than median lost utility. This behavior is due to its independent autoscaling decisions which cause 
some jobs to stay at a low SLO satisfaction over the entire run.  

\sysname{}'s fairness variants have ``tighter'' boxplots,  
indicating smaller utility differences among jobs, and thus better fairness. Not surprisingly, \sysname{} variants that explicitly include fairness (e.g., all \sysname{}-*Fair* variants) exhibit better fairness than other variants  (\sysname{}-Sum and \sysname{}-PenaltySum).

\smallskip
\noindent {\bf Cluster Utility and Penalty: } 
\Figure{}~\ref{fig:lost_effective_util} plots both 
lost cluster utility, as well as {\it effective utility} which includes a penalty from dropped requests (Section~\ref{sec:objective_funcs}, Table~\ref{tab:sla_example}).
We observe that all \sysname{} variants outperform baselines \changenew{}{at RS and \SO{} clusters}.
Second,  cluster utilities of \sysname{}  variants are similar. Interestingly, \sysname{}-Fair, focusing only the fairness, also achieves high cluster utility, since  when most jobs are able to  achieve max utility (i.e., 1.0), \sysname{}-Fair finds a solution that minimizes the utility difference, i.e., {\it ``rich jobs pull poorer jobs upwards''} 
when cluster resources are sufficient.
Third, we observe that while \sysname-Penalty* variants reduce job workloads by {\it explicitly} dropping requests (based on \Equation{}~\ref{eqn:precise}), 
in a right-sized cluster these variants improve neither cluster utility nor effective utility vs. non-Penalty variants

\subsection{Mixed Workloads}

We evaluate a mixed workload with 50\% jobs as ResNet18 and 50\% as ResNet34, in a right-sized cluster.  Latency SLO is set for ResNet18  (400 ms)  similar to ResNet34: 4$\times$   average per-request processing  (100 ms).
\Figure{}~\ref{fig:mixed_avg_lost_util_slo_plot} shows that compared to baselines, \sysname{} lowers cluster SLO violation rates by 4$\times$--23$\times$ and lowers lost cluster utility by 2.3$\times$--13.1$\times$.

\begin{table*}[t]
    \centering
    \caption{\changenew{}{\it Matching  Simulator and Cluster Deployment:
    Utility (SD). Cluster sizes (total replicas): 36 (RS), 32 (\SO{}), 16 (\HO{}).
    }}
    \vspace*{-0.2cm}
    \begin{adjustbox}{max width=\linewidth}
    \changenew{}{\begin{tabular}{c c c c c c c c c c c}
        \toprule
        Size & Environment & Rank \#1 & Rank \#2 & Rank \#3 & Rank \#4 & Rank \#5 & Rank \#6 & Rank \#7 & Rank \#8 & Rank \#9 \\
        \midrule
         \multirow[b]{2}{*}{RS} & Cluster & \makecell{Faro-Fair\\ 0.42 (0.04)} & \makecell{Faro-PenaltySum \\ 0.44 (0.06)} & \makecell{Faro-Sum \\ 0.46 (0.08)} & \makecell{Faro-FairSum \\ 0.48 (0.02)} & \makecell{Faro-PenaltyFairSum \\ 0.49 (0.04)} & \makecell{MArk/Cocktail/Barista  \\ 0.81 (0.05)} & \makecell{AIAD     \\ 1.45 (0.10)} & \makecell{FairShare\\ 2.42 (0.11)} & \makecell{Oneshot  \\ 4.34 (0.20)} \\
         & Simulation & \makecell{Faro-PenaltySum \\ 0.37 (0.02)} &  \makecell{Faro-Sum \\ 0.37 (0.02)} & \makecell{Faro-FairSum \\ 0.37 (0.02)} & \makecell{Faro-Fair \\ 0.38 (0.02)} & \makecell{Faro-PenaltyFairSum \\ 0.44 (0.10)} & \makecell{MArk/Cocktail/Barista  \\ 0.64 (0.07)} & \makecell{AIAD  \\ 1.39 (0.02)} & \makecell{FairShare \\ 2.08 (0.02)} & \makecell{Oneshot \\ 3.87 (0.03)}   \\
        \midrule
        \multirow[b]{2}{*}{\SO{}} & Cluster & \makecell{Faro-FairSum \\ 0.79 (0.07)} & \makecell{Faro-Fair \\ 0.80 (0.07)} & \makecell{Faro-Sum \\ 0.92 (0.04)} & \makecell{Faro-PenaltySum \\ 1.05 (0.10)} & \makecell{Faro-PenaltyFairSum \\ 1.20 (0.15)} & \makecell{AIAD \\ 1.96 (0.14)} & \makecell{MArk/Cocktail/Barista \\ 2.02 (0.12)} & \makecell{FairShare \\ 2.42 (0.11)} & \makecell{Oneshot \\ 4.83 (0.13)} \\
         & Simulation & \makecell{Faro-FairSum \\ 0.78 (0.01)} & \makecell{Faro-Fair \\ 0.80 (0.03)} & \makecell{Faro-Sum \\ 0.91 (0.03)} & \makecell{Faro-PenaltySum \\ 0.94 (0.02)} & \makecell{Faro-PenaltyFairSum \\ 1.14 (0.12)} & \makecell{AIAD \\ 1.71 (0.04)} & \makecell{MArk/Cocktail/Barista \\ 1.77 (0.08)} & \makecell{FairShare \\ 2.08 (0.02)} & \makecell{Oneshot \\ 4.37 (0.16) }  \\
         \midrule
         \multirow[b]{2}{*}{\HO{}} & Cluster & \makecell{Faro-Sum \\ 5.71 (0.13)} & \makecell{Faro-PenaltySum \\ 5.72 (0.05)} & \makecell{MArk/Cocktail/Barista \\ 6.86 (0.30)} & \makecell{Faro-FairSum \\ 6.99 (0.16)} & \makecell{Oneshot \\ 7.58 (0.25)} & \makecell{AIAD \\ 7.66 (0.14)} & \makecell{Faro-PenaltyFairSum \\ 7.69 (0.30)} & \makecell{Faro-Fair \\ 7.71 (0.09)} & \makecell{FairShare \\ 8.71 (0.11)} \\
         & Simulation & \makecell{Faro-Sum \\ 6.01 (0.07)} & \makecell{Faro-PenaltySum \\ 6.09 (0.04)} & \makecell{MArk/Cocktail/Barista \\ 7.43 (0.27)} & \makecell{Faro-FairSum \\ 7.85 (0.03)} & \makecell{Oneshot \\ 8.05 (0.13)} & \makecell{AIAD \\ 8.15 (0.08)} & \makecell{Faro-PenaltyFairSum \\ 8.15 (0.11)} & \makecell{Faro-Fair \\ 8.26 (0.04)} & \makecell{FairShare \\ 9.22 (0.00)}  \\
        \bottomrule
    \end{tabular}
    }
    \end{adjustbox}
    \label{tab:sim_real_comparison}
\end{table*}

\subsection{Matched Simulation} 
\label{sec:simres}

We wrote a custom {\it matched} simulator to simulate the behavior of Ray Serve over Kubernetes. The simulator's architecture allows us to reuse the (cluster deployment) code for  \sysname{} and all baselines, and inject the same workloads as \Section{}~\ref{sec:eval_comparison}--\ref{sec:eval_variants}. 
We carefully architected the simulator in such a way that its behavior {\it matches} the cluster deployment results at all cluster sizes of 36, 32, and 16 total replicas. 
Table~\ref{tab:sim_real_comparison} shows that our simulator is able to predict cluster utility values reasonably, with 9.6\% average difference (range: 0.77\%-22.76\%). 
We use the Kendall-Tau metric~\cite{kendall_tau} to compare rankings between cluster vs. simulation: 0 indicates identical, and 1 indicates complete divergence. The Kendall-Tau metric is zero for each of \SO{} and \HO{}, and for RS it is only 0.083.

Thus we can {\it extrapolate} the simulator to both: (1) much larger scales in the cluster ({\it \undersubscribed{}}), and (2) much smaller scales in the cluster ({\it \oversubscribed{}}).

\begin{figure}[t]
    \centering
    \includegraphics[width=0.95\linewidth]{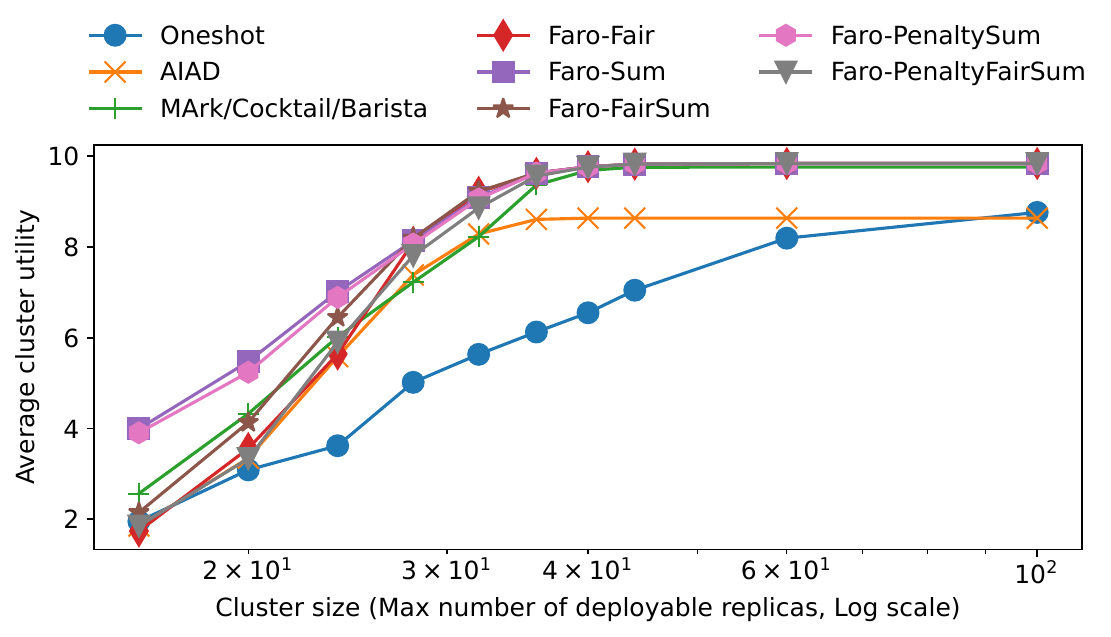}
    \vspace*{-0.2cm}
    \caption{\change{\it Matched Simulation: \Oversubscribed{} ($<36$ replicas) to \Undersubscribed{} Cluster ($\geq 36$ replicas).
    }}
    \label{fig:curve_sim}
    \vspace*{-0.3cm}
\end{figure}

\smallskip
\noindent {\bf From \UnderToOver{} Cluster: }
\Figure{}~\ref{fig:curve_sim} shows that (as expected) at right-sized and \undersubscribed{} clusters (cluster size $\geq$ 36)  the  previously-observed dominance of \sysname  vs. baselines holds. All \sysname{} variants and \markpolicy{} achieve cluster utility close to max (10), but other baselines do not.  
In constrained cluster  ($\leq$ 32 replicas)  \sysname{}  outperforms \markpolicy{} and others. In small clusters, \sysname{}-Sum and \sysname{}-PenaltySum outperform the  3 \sysname{}-*Fair* variants, since   
``equitably'' dividing resources lowers cluster utility. 
%
\sysname{}-Sum beats \sysname{}-PenaltySum, likely since explicitly dropping requests (via our optimization formulation) is overshadowed by underlying queues getting naturally full (due to high relative load) and implicitly dropping requests. 

\begin{figure}[t]
    \centering
    \includegraphics[width=0.9\linewidth]{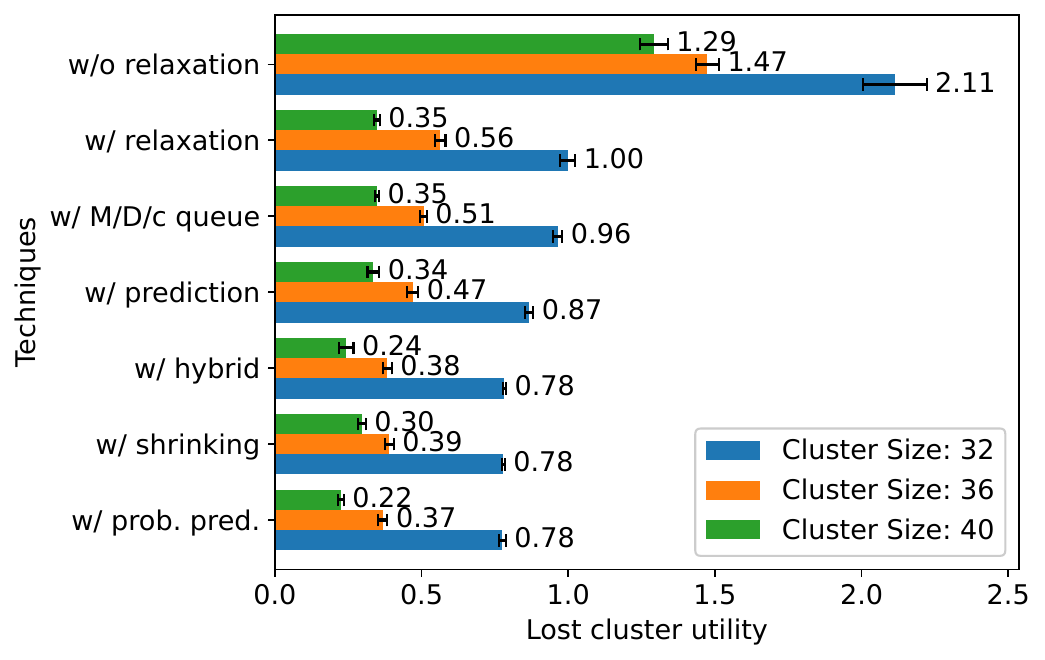}
    \vspace*{-0.3cm}
    \caption{\change{\it Ablation Study: Lost Cluster Utility. Uses Faro-FairSum. (Cluster {size} 36     is right-sized.) \changenew{}{Error Bars: SD.}}}
    \label{fig:ablation_study}
\end{figure}

\smallskip
\noindent {\bf Ablation Study: }
\Figure{}~\ref{fig:ablation_study} shows an ablation study.
Lost cluster utility is lowered 2.1$\times$--3.7$\times$ by Relaxation which makes the optimizer find reasonable  solutions quickly (\Section{}~\ref{sec:optimization}), 
up to 1.1$\times$
by M/D/c latency estimation which helps find right-sized number of replicas
(\Section{}~\ref{sec:latency_estimation}),
up to 1.1$\times$ by Time-series prediction which  minimizes cold start impact (\Section{}~\ref{sec:prediction}),
and up to {1.42}$\times$ by Hybrid autoscaler which handles prediction errors (\Section{}~\ref{sec:hybrid autoscaler}).  
Shrinking (\Section{}~\ref{sec:shrinking}) 
increased lost cluster utility by up to 1.25$\times$ due to overtight resource allocation.
Probabilistic time-series prediction (\Section{}~\ref{sec:prob_prediction}) handles this overtightness, lowering lost cluster utility by up to 
1.36$\times$.

\begin{table}[t]
    \centering
    \caption{{\it Large-scale Workloads: Average Lost Cluster Utility (Utility) and SLO Violation Rates (SLO) with SD.
    }}
    \vspace*{-0.2cm}
    \begin{adjustbox}{max width=\linewidth}{
    {\begin{tabular}{c c c c c c c}
        \toprule
        \makecell{Num Jobs \\ (Cluster Size)} & Metric & Fairshare & Oneshot & AIAD & Mark
        & \makecell{\textbf{\sysname{}} \\ \textbf{FairSum}} \\
        \midrule
        \multirow[b]{2}{*}{\makecell{\lbrack{}Cluster\rbrack{}
        \\ 20 Jobs \\ (70 Replicas)}} 
        & Utility & \makecell{3.48 \\ (0.25)} & \makecell{8.67 \\ (0.27)} & \makecell{2.37 \\ (0.16)} & \makecell{1.77 \\ (0.21)} & {\bf \makecell{0.63 \\ (0.10)}} \\
        & SLO & \makecell{0.14 \\ (0.01)} & \makecell{0.37 \\ (0.01)} & \makecell{0.07 \\ (0.00)} & \makecell{0.08 \\ (0.01)} & \bf \makecell{0.02 \\ (0.00)} \\
        \midrule
        \multirow[b]{2}{*}{\makecell{\lbrack{}Simulation\rbrack{} \\ 100 Jobs \\ (320 Replicas)}} 
        & Utility & \makecell{20.82 \\ (0.08)} & \makecell{53.37 \\ (0.63)} & \makecell{16.72 \\ (0.16)} & \makecell{16.24 \\ (0.13)} & \bf \makecell{7.83 \\ (0.07)} \\
        & SLO & \makecell{0.16 \\ (0.00)} & \makecell{0.48 \\ (0.01)} & \makecell{0.09 \\ (0.00)} & \makecell{0.13 \\ (0.00)} & {\bf \makecell{0.03 \\ (0.00)}} \\
        \bottomrule
    \end{tabular}
    }}
    \end{adjustbox}
    \vspace*{-0.2cm}
    \label{tab:large_scale_util_slo}
\end{table}

\subsection{Large-scale Workloads}

We use a large-sized Kubernetes cluster with 32 
{IBM Cloud VPC cx2-4x8} VMs (each 4 vCPU 
and 8 GB), and 20 jobs (workloads duplicated). We also run a larger simulation with 100 jobs over 320 replicas, equivalent to a cluster with 344 vCPUs and 408 GB. 
Table~\ref{tab:large_scale_util_slo}
shows  even at scale \sysname{} lowers SLO violation rates by 3$\times$--18.5$\times$ and lost cluster utility by 2.07$\times$--13.76$\times$, vs. baselines.



\section{Discussion and Limitations}

\noindent{\bf Beyond On-Premises Clusters: } Limited clusters  also arise beyond on-premises clusters. A common example is deployment on a public cloud wherein developers prefer a VM instance type but have a budget limit (\$ per hour) they can spend. Another example is an edge deployment that has user {\it quotas}.  \sysname{} is also applicable in these scenarios. 

\smallskip
\noindent{\bf CPU/GPU mixes and Heterogeneity: } Due to low prices and high availability of CPU resources, enterprises still prefer to run inference workloads on homogeneous CPUs. Admitting heterogeneous mixes of accelerators (GPUs) with CPUs is an open problem, with \sysname{} representing a first step. 

\smallskip
\noindent{\bf Orthogonal Techniques (Batching, Memory): } \sysname{} is orthogonal to, and thus combinable with:  Ray's and Kubernetes' fault-tolerance mechanisms, intelligent request batching~\cite{batch, clipper}, hyperparameter tuning~\cite{optuna, ray_tune},  and different types of ML parallelism (pipeline, model) for large models~\cite{deepspeed}. 

\smallskip
\noindent {\bf Heterogeneity:} Homogeneous clusters are common due to ease of manageability, and \sysname{} focuses on them; heterogeneity is an open challenge. 
\sysname{} is applicable to ML pipelines that make chained calls to multiple ML jobs, if the application SLO can be split into sub-SLOs for each called model, e.g., proportionally: for a chain with two model calls, if one model takes 2$\times$ other, if the SLO is split as 66\%-33\%. 

\smallskip
\noindent{\bf Decentralized Approaches: } Decentralizing \sysname is not essential but could be an interesting future direction~\cite{sparrow, ownership}.

\smallskip
\noindent{\bf Admission Control: } 
While ML job arrivals/departures are rare, it remains an open question whether admission control decisions can be designed to guarantee SLO satisfaction, perhaps with some workload assumptions. 

\smallskip
\noindent{\bf Beyond ML Inference: } While we have focused on ML inference clusters, it is possible that \sysname{}'s techniques could be extended to other domains, e.g., microservices and batch processing jobs. Our use of ML inference-specific techniques, e.g., M/D/c queueing (\Section{}~\ref{sec:latency_estimation}), will need to be adapted, e.g., via M/M/c or G/G/c queuing, and utility functions for batch processing, e.g., ML training workloads~\cite{cilantro, jockey, wilkes2009utility, shockwave}.

\section{Related Work}

ML inference autoscaling either makes per-job independent scaling decisions, or assumes infinite (cloud) resources. 
Swayam \cite{swayam}  proactively autoscales to minimize resource utilization under SLOs, relying on latency estimation.
However, Swayam cannot work with constrained clusters or multi-tenant setups. 
Swayam's customized decentralized architecture prevented us from reusing it for the Ray Serve | Kubernetes stack.
Mark~\cite{mark} and Barista~\cite{barista} use SLO-aware autoscaling heuristics to minimize cost on various cloud resources, e.g., VM instance types, serverless, etc. 
Greedy heuristics iteratively pick a cloud resource type with the lowest cost-per-request, until job SLOs are met. 
Cost-per-request is based on max throughputs of each instance type. \sysname{} solves a different problem in constrained clusters.
Because of their orthogonality, an open direction is to combine \sysname{} with Mark/Barista's heterogeneity awareness to realize a constrained cluster version of \sysname{} in public clouds.
Cocktail~\cite{cocktail} is an ensemble model-serving system to meet request accuracy and latency requirements. Cocktail \textit{only} scales up jobs (proactively). 
Yet an upscaled job never relinquishes its resources,  even if the workload drops, making it non-adaptable under changing workloads. 
INFaaS~\cite{infaas} auto-selects cost-efficient configurations (e.g., ML model, hardware) based on user needs. It reactively autoscales  {\it additively} by monitoring job throughput, hence our AIAD baseline captures it.

Data center scheduling work~\cite{apollo, tetris, quasar, paragon} focuses on placement but does not deal with autoscaling to meet (human-facing) latency SLOs.
Some recent works~\cite{jockey, morpheus, cilantro, autopilot} tackle SLO-aware autoscaling in fixed clusters~\cite{jockey, morpheus, cilantro, autopilot}. 
Of these, Autopilot and Jockey~\cite{jockey} only autoscale individual jobs, like Mark~\cite{mark} and K8s HPA~\cite{k8s_hpa}, which we have compared against~\cite{mark, barista, cocktail}.
Morpheus~\cite{morpheus} supports only periodic data-analytic jobs. 
Cilantro~\cite{cilantro} maximizes global objectives but converges too slowly (\Figure{}~\ref{fig:cilantro_comparison}). 
ML training scheduling work~\cite{gavel, gandiva, gandiva_fair, pollux, shockwave, themis, slaq}  support neither horizontal autoscaling~\cite{gavel, gandiva, gandiva_fair, themis}, nor stringent latency SLOs (ms) for inference~\cite{pollux, shockwave, slaq}. 

\section{Summary}

We presented \sysname{}, an SLO-aware autoscaling system for ML inference services in a constrained containerized cluster. \sysname{} adopts a ``sloppified but fast'' approach, e.g., w.r.t. objectives, workload prediction, etc.
Deployment and simulation results show that against state-of-the-art systems which focus on single jobs only, \sysname{}: (1)  lowers SLO violations by \sloimprovement{} when the resources are sufficient, 
and (2) lowers SLO violations by  
\sloimprovementho{},
in constrained clusters.
Faro's code is openly available at the following link: \url{https://dprg.cs.uiuc.edu/
traces/go.php?id=40}

\begin{acks}
This work was supported in part by the following grants: IIDAI (IBM-Illinois Discovery Accelerator Institute) Grant 107275, NSF IIS Grant 1909577, NSF CNS Grant 1908888, and a gift from Microsoft.
\end{acks}


\balance

\bibliographystyle{ACM-Reference-Format}
\bibliography{paper}


\begin{thebibliography}{94}


\ifx \showCODEN    \undefined \def \showCODEN     #1{\unskip}     \fi
\ifx \showDOI      \undefined \def \showDOI       #1{#1}\fi
\ifx \showISBNx    \undefined \def \showISBNx     #1{\unskip}     \fi
\ifx \showISBNxiii \undefined \def \showISBNxiii  #1{\unskip}     \fi
\ifx \showISSN     \undefined \def \showISSN      #1{\unskip}     \fi
\ifx \showLCCN     \undefined \def \showLCCN      #1{\unskip}     \fi
\ifx \shownote     \undefined \def \shownote      #1{#1}          \fi
\ifx \showarticletitle \undefined \def \showarticletitle #1{#1}   \fi
\ifx \showURL      \undefined \def \showURL       {\relax}        \fi
\providecommand\bibfield[2]{#2}
\providecommand\bibinfo[2]{#2}
\providecommand\natexlab[1]{#1}
\providecommand\showeprint[2][]{arXiv:#2}

\bibitem[Akiba et~al\mbox{.}(2019)]%
        {optuna}
\bibfield{author}{\bibinfo{person}{Takuya Akiba}, \bibinfo{person}{Shotaro Sano}, \bibinfo{person}{Toshihiko Yanase}, \bibinfo{person}{Takeru Ohta}, {and} \bibinfo{person}{Masanori Koyama}.} \bibinfo{year}{2019}\natexlab{}.
\newblock \showarticletitle{Optuna: A Next-Generation Hyperparameter Optimization Framework}. In \bibinfo{booktitle}{\emph{Proceedings of the 25th ACM SIGKDD International Conference on Knowledge Discovery and Data Mining}} \emph{(\bibinfo{series}{KDD 19})}. \bibinfo{publisher}{Association for Computing Machinery}, \bibinfo{address}{Anchorage, AK, USA}, \bibinfo{pages}{2623–2631}.
\newblock
\showISBNx{9781450362016}
\urldef\tempurl%
\url{https://doi.org/10.1145/3292500.3330701}
\showDOI{\tempurl}


\bibitem[Ali et~al\mbox{.}(2020)]%
        {batch}
\bibfield{author}{\bibinfo{person}{Ahsan Ali}, \bibinfo{person}{Riccardo Pinciroli}, \bibinfo{person}{Feng Yan}, {and} \bibinfo{person}{Evgenia Smirni}.} \bibinfo{year}{2020}\natexlab{}.
\newblock \showarticletitle{BATCH: Machine Learning Inference Serving on Serverless Platforms with Adaptive Batching}. In \bibinfo{booktitle}{\emph{Proceedings of the 2020 International Conference for High Performance Computing, Networking, Storage and Analysis}} \emph{(\bibinfo{series}{SC 20})}. \bibinfo{publisher}{{IEEE}}, \bibinfo{address}{Atlanta, GA, USA}, \bibinfo{pages}{1--15}.
\newblock
\urldef\tempurl%
\url{https://doi.org/10.1109/SC41405.2020.00073}
\showDOI{\tempurl}


\bibitem[AuYoung et~al\mbox{.}(2006)]%
        {wilkes06}
\bibfield{author}{\bibinfo{person}{Alvin AuYoung}, \bibinfo{person}{Laura Rit}, \bibinfo{person}{Sohn Wiener}, {and} \bibinfo{person}{John Wilkes}.} \bibinfo{year}{2006}\natexlab{}.
\newblock \showarticletitle{Service contracts and aggregate utility functions}. In \bibinfo{booktitle}{\emph{Proceedings of the 15th {IEEE} International Symposium on High Performance Distributed Computing}} \emph{(\bibinfo{series}{HPDC 15})}. \bibinfo{publisher}{{IEEE}}, \bibinfo{address}{Paris, France}, \bibinfo{pages}{119--131}.
\newblock
\urldef\tempurl%
\url{https://doi.org/10.1109/HPDC.2006.1652143}
\showDOI{\tempurl}


\bibitem[Baset et~al\mbox{.}(2012)]%
        {towards2021baset}
\bibfield{author}{\bibinfo{person}{Salman~A. Baset}, \bibinfo{person}{Long Wang}, {and} \bibinfo{person}{Chunqiang Tang}.} \bibinfo{year}{2012}\natexlab{}.
\newblock \showarticletitle{Towards an Understanding of Oversubscription in Cloud}. In \bibinfo{booktitle}{\emph{2nd USENIX Workshop on Hot Topics in Management of Internet, Cloud, and Enterprise Networks and Services (Hot-ICE 12)}}. \bibinfo{publisher}{USENIX Association}, \bibinfo{address}{San Jose, CA}.
\newblock
\urldef\tempurl%
\url{https://www.usenix.org/conference/hot-ice12/workshop-program/presentation/baset}
\showURL{%
\tempurl}


\bibitem[Bhardwaj et~al\mbox{.}(2023)]%
        {cilantro}
\bibfield{author}{\bibinfo{person}{Romil Bhardwaj}, \bibinfo{person}{Kirthevasan Kandasamy}, \bibinfo{person}{Asim Biswal}, \bibinfo{person}{Wenshuo Guo}, \bibinfo{person}{Benjamin Hindman}, \bibinfo{person}{Joseph Gonzalez}, \bibinfo{person}{Michael Jordan}, {and} \bibinfo{person}{Ion Stoica}.} \bibinfo{year}{2023}\natexlab{}.
\newblock \showarticletitle{Cilantro: {Performance-Aware} Resource Allocation for General Objectives via Online Feedback}. In \bibinfo{booktitle}{\emph{Proceedings of 17th USENIX Symposium on Operating Systems Design and Implementation}} \emph{(\bibinfo{series}{OSDI 23})}. \bibinfo{publisher}{USENIX Association}, \bibinfo{address}{Boston, MA}, \bibinfo{pages}{623--643}.
\newblock
\showISBNx{978-1-939133-34-2}
\urldef\tempurl%
\url{https://www.usenix.org/conference/osdi23/presentation/bhardwaj}
\showURL{%
\tempurl}


\bibitem[Bhattacharjee et~al\mbox{.}(2019)]%
        {barista}
\bibfield{author}{\bibinfo{person}{Anirban Bhattacharjee}, \bibinfo{person}{Ajay~Dev Chhokra}, \bibinfo{person}{Zhuangwei Kang}, \bibinfo{person}{Hongyang Sun}, \bibinfo{person}{Aniruddha Gokhale}, {and} \bibinfo{person}{Gabor Karsai}.} \bibinfo{year}{2019}\natexlab{}.
\newblock \showarticletitle{BARISTA: Efficient and Scalable Serverless Serving System for Deep Learning Prediction Services}. In \bibinfo{booktitle}{\emph{Proceedings of the 2019 IEEE International Conference on Cloud Engineering}} \emph{(\bibinfo{series}{IC2E 2019})}. \bibinfo{publisher}{{IEEE}}, \bibinfo{address}{Prague, Czech Republic}, \bibinfo{pages}{23--33}.
\newblock
\urldef\tempurl%
\url{https://doi.org/10.1109/IC2E.2019.00-10}
\showDOI{\tempurl}


\bibitem[Boutin et~al\mbox{.}(2014)]%
        {apollo}
\bibfield{author}{\bibinfo{person}{Eric Boutin}, \bibinfo{person}{Jaliya Ekanayake}, \bibinfo{person}{Wei Lin}, \bibinfo{person}{Bing Shi}, \bibinfo{person}{Jingren Zhou}, \bibinfo{person}{Zhengping Qian}, \bibinfo{person}{Ming Wu}, {and} \bibinfo{person}{Lidong Zhou}.} \bibinfo{year}{2014}\natexlab{}.
\newblock \showarticletitle{Apollo: Scalable and Coordinated Scheduling for {Cloud-Scale} Computing}. In \bibinfo{booktitle}{\emph{11th USENIX Symposium on Operating Systems Design and Implementation (OSDI 14)}}. \bibinfo{publisher}{USENIX Association}, \bibinfo{address}{Broomfield, CO}, \bibinfo{pages}{285--300}.
\newblock
\showISBNx{978-1-931971-16-4}
\urldef\tempurl%
\url{https://www.usenix.org/conference/osdi14/technical-sessions/presentation/boutin}
\showURL{%
\tempurl}


\bibitem[Challu et~al\mbox{.}(2023)]%
        {nhits}
\bibfield{author}{\bibinfo{person}{Cristian Challu}, \bibinfo{person}{Kin~G. Olivares}, \bibinfo{person}{Boris~N. Oreshkin}, \bibinfo{person}{Federico Garza~Ramirez}, \bibinfo{person}{Max Mergenthaler~Canseco}, {and} \bibinfo{person}{Artur Dubrawski}.} \bibinfo{year}{2023}\natexlab{}.
\newblock \showarticletitle{NHITS: Neural Hierarchical Interpolation for Time Series Forecasting}.
\newblock \bibinfo{journal}{\emph{Proceedings of the 37th AAAI Conference on Artificial Intelligence (AAAI 23)}} \bibinfo{volume}{37}, \bibinfo{number}{6} (\bibinfo{date}{Jun.} \bibinfo{year}{2023}), \bibinfo{pages}{6989--6997}.
\newblock
\urldef\tempurl%
\url{https://doi.org/10.1609/aaai.v37i6.25854}
\showDOI{\tempurl}


\bibitem[Chaudhary et~al\mbox{.}(2020a)]%
        {gandiva_fair}
\bibfield{author}{\bibinfo{person}{Shubham Chaudhary}, \bibinfo{person}{Ramachandran Ramjee}, \bibinfo{person}{Muthian Sivathanu}, \bibinfo{person}{Nipun Kwatra}, {and} \bibinfo{person}{Srinidhi Viswanatha}.} \bibinfo{year}{2020}\natexlab{a}.
\newblock \showarticletitle{Balancing Efficiency and Fairness in Heterogeneous GPU Clusters for Deep Learning}. In \bibinfo{booktitle}{\emph{Proceedings of the Fifteenth European Conference on Computer Systems}} (Heraklion, Greece) \emph{(\bibinfo{series}{EuroSys '20})}. \bibinfo{publisher}{Association for Computing Machinery}, \bibinfo{address}{New York, NY, USA}, Article \bibinfo{articleno}{1}, \bibinfo{numpages}{16}~pages.
\newblock
\showISBNx{9781450368827}


\bibitem[Chaudhary et~al\mbox{.}(2020b)]%
        {slaq}
\bibfield{author}{\bibinfo{person}{Shubham Chaudhary}, \bibinfo{person}{Ramachandran Ramjee}, \bibinfo{person}{Muthian Sivathanu}, \bibinfo{person}{Nipun Kwatra}, {and} \bibinfo{person}{Srinidhi Viswanatha}.} \bibinfo{year}{2020}\natexlab{b}.
\newblock \showarticletitle{Balancing Efficiency and Fairness in Heterogeneous GPU Clusters for Deep Learning}. In \bibinfo{booktitle}{\emph{Proceedings of the Fifteenth European Conference on Computer Systems}} (Heraklion, Greece) \emph{(\bibinfo{series}{EuroSys '20})}. \bibinfo{publisher}{Association for Computing Machinery}, \bibinfo{address}{New York, NY, USA}, Article \bibinfo{articleno}{1}, \bibinfo{numpages}{16}~pages.
\newblock
\showISBNx{9781450368827}
\urldef\tempurl%
\url{https://doi.org/10.1145/3342195.3387555}
\showDOI{\tempurl}


\bibitem[Cho et~al\mbox{.}(2022)]%
        {ml_inference_for_clouds}
\bibfield{author}{\bibinfo{person}{Junguk Cho}, \bibinfo{person}{Diman Zad~Tootaghaj}, \bibinfo{person}{Lianjie Cao}, {and} \bibinfo{person}{Puneet Sharma}.} \bibinfo{year}{2022}\natexlab{}.
\newblock \showarticletitle{SLA-Driven ML Inference Framework for Clouds with Heterogeneous Accelerators}. In \bibinfo{booktitle}{\emph{Proceedings of 2022 Machine Learning and Systems (MLSys 22)}}, \bibfield{editor}{\bibinfo{person}{D.~Marculescu}, \bibinfo{person}{Y.~Chi}, {and} \bibinfo{person}{C.~Wu}} (Eds.), Vol.~\bibinfo{volume}{4}. \bibinfo{address}{Santa Clara, CA, USA}, \bibinfo{pages}{20--32}.
\newblock
\urldef\tempurl%
\url{https://proceedings.mlsys.org/paper/2022/file/0777d5c17d4066b82ab86dff8a46af6f-Paper.pdf}
\showURL{%
\tempurl}


\bibitem[Comer(2000)]%
        {comer2000internetworking}
\bibfield{author}{\bibinfo{person}{D. Comer}.} \bibinfo{year}{2000}\natexlab{}.
\newblock \bibinfo{booktitle}{\emph{Internetworking with TCP/IP}}.
\newblock \bibinfo{publisher}{Prentice Hall}, \bibinfo{address}{Upper Saddle River, NJ, USA}.
\newblock
\showISBNx{9780130183804}
\showLCCN{00027439}


\bibitem[Crankshaw et~al\mbox{.}(2017)]%
        {clipper}
\bibfield{author}{\bibinfo{person}{Daniel Crankshaw}, \bibinfo{person}{Xin Wang}, \bibinfo{person}{Guilio Zhou}, \bibinfo{person}{Michael~J. Franklin}, \bibinfo{person}{Joseph~E. Gonzalez}, {and} \bibinfo{person}{Ion Stoica}.} \bibinfo{year}{2017}\natexlab{}.
\newblock \showarticletitle{{Clipper}: A {Low-Latency} Online Prediction Serving System}. In \bibinfo{booktitle}{\emph{Proceedings of 14th USENIX Symposium on Networked Systems Design and Implementation (NSDI 17)}}. \bibinfo{publisher}{USENIX Association}, \bibinfo{address}{Boston, MA}, \bibinfo{pages}{613--627}.
\newblock
\showISBNx{978-1-931971-37-9}
\urldef\tempurl%
\url{https://www.usenix.org/conference/nsdi17/technical-sessions/presentation/crankshaw}
\showURL{%
\tempurl}


\bibitem[Dauphin et~al\mbox{.}(2014)]%
        {dauphin2014identifying}
\bibfield{author}{\bibinfo{person}{Yann~N. Dauphin}, \bibinfo{person}{Razvan Pascanu}, \bibinfo{person}{Caglar Gulcehre}, \bibinfo{person}{Kyunghyun Cho}, \bibinfo{person}{Surya Ganguli}, {and} \bibinfo{person}{Yoshua Bengio}.} \bibinfo{year}{2014}\natexlab{}.
\newblock \showarticletitle{Identifying and Attacking the Saddle Point Problem in High-Dimensional Non-Convex Optimization}. In \bibinfo{booktitle}{\emph{Proceedings of the 27th International Conference on Neural Information Processing Systems - Volume 2}} \emph{(\bibinfo{series}{NIPS 14})}. \bibinfo{publisher}{MIT Press}, \bibinfo{address}{Montreal, Canada}, \bibinfo{pages}{2933–2941}.
\newblock


\bibitem[Delimitrou and Kozyrakis(2013)]%
        {paragon}
\bibfield{author}{\bibinfo{person}{Christina Delimitrou} {and} \bibinfo{person}{Christos Kozyrakis}.} \bibinfo{year}{2013}\natexlab{}.
\newblock \showarticletitle{Paragon: QoS-aware scheduling for heterogeneous datacenters}.
\newblock \bibinfo{journal}{\emph{SIGPLAN Not.}} \bibinfo{volume}{48}, \bibinfo{number}{4} (\bibinfo{date}{mar} \bibinfo{year}{2013}), \bibinfo{pages}{77–88}.
\newblock
\showISSN{0362-1340}
\urldef\tempurl%
\url{https://doi.org/10.1145/2499368.2451125}
\showDOI{\tempurl}


\bibitem[Delimitrou and Kozyrakis(2014)]%
        {quasar}
\bibfield{author}{\bibinfo{person}{Christina Delimitrou} {and} \bibinfo{person}{Christos Kozyrakis}.} \bibinfo{year}{2014}\natexlab{}.
\newblock \showarticletitle{Quasar: Resource-Efficient and QoS-Aware Cluster Management}. In \bibinfo{booktitle}{\emph{Proceedings of the 19th International Conference on Architectural Support for Programming Languages and Operating Systems}} \emph{(\bibinfo{series}{ASPLOS 14})}. \bibinfo{publisher}{Association for Computing Machinery}, \bibinfo{address}{Salt Lake City, Utah, USA}, \bibinfo{pages}{127–144}.
\newblock
\showISBNx{9781450323055}
\urldef\tempurl%
\url{https://doi.org/10.1145/2541940.2541941}
\showDOI{\tempurl}


\bibitem[Elhabbash et~al\mbox{.}(2019)]%
        {sloml}
\bibfield{author}{\bibinfo{person}{Abdessalam Elhabbash}, \bibinfo{person}{Assylbek Jumagaliyev}, \bibinfo{person}{Gordon~S. Blair}, {and} \bibinfo{person}{Yehia Elkhatib}.} \bibinfo{year}{2019}\natexlab{}.
\newblock \showarticletitle{SLO-ML: A Language for Service Level Objective Modelling in Multi-Cloud Applications}. In \bibinfo{booktitle}{\emph{Proceedings of the 12th IEEE/ACM International Conference on Utility and Cloud Computing}} \emph{(\bibinfo{series}{UCC 19})}. \bibinfo{publisher}{Association for Computing Machinery}, \bibinfo{address}{Auckland, New Zealand}, \bibinfo{pages}{241–250}.
\newblock
\showISBNx{9781450368940}
\urldef\tempurl%
\url{https://doi.org/10.1145/3344341.3368805}
\showDOI{\tempurl}


\bibitem[Ferguson et~al\mbox{.}(2012)]%
        {jockey}
\bibfield{author}{\bibinfo{person}{Andrew~D. Ferguson}, \bibinfo{person}{Peter Bodik}, \bibinfo{person}{Srikanth Kandula}, \bibinfo{person}{Eric Boutin}, {and} \bibinfo{person}{Rodrigo Fonseca}.} \bibinfo{year}{2012}\natexlab{}.
\newblock \showarticletitle{Jockey: guaranteed job latency in data parallel clusters}. In \bibinfo{booktitle}{\emph{Proceedings of the 7th ACM European Conference on Computer Systems}} (Bern, Switzerland) \emph{(\bibinfo{series}{EuroSys '12})}. \bibinfo{publisher}{Association for Computing Machinery}, \bibinfo{address}{New York, NY, USA}, \bibinfo{pages}{99–112}.
\newblock
\showISBNx{9781450312233}
\urldef\tempurl%
\url{https://doi.org/10.1145/2168836.2168847}
\showDOI{\tempurl}


\bibitem[Foundation(2024)]%
        {kubernetes}
\bibfield{author}{\bibinfo{person}{Cloud Native~Computing Foundation}.} \bibinfo{year}{2024}\natexlab{}.
\newblock \bibinfo{title}{Kubernetes}.
\newblock
\newblock
\urldef\tempurl%
\url{https://kubernetes.io/}
\showURL{%
\tempurl}


\bibitem[Franx(2001)]%
        {mdc_queue}
\bibfield{author}{\bibinfo{person}{G.J. Franx}.} \bibinfo{year}{2001}\natexlab{}.
\newblock \showarticletitle{A simple solution for the M/D/c waiting time distribution}.
\newblock \bibinfo{journal}{\emph{Operations Research Letters}} \bibinfo{volume}{29}, \bibinfo{number}{5} (\bibinfo{year}{2001}), \bibinfo{pages}{221--229}.
\newblock
\showISSN{0167-6377}
\urldef\tempurl%
\url{https://doi.org/10.1016/S0167-6377(01)00108-0}
\showDOI{\tempurl}


\bibitem[Freedman et~al\mbox{.}(2004)]%
        {coral}
\bibfield{author}{\bibinfo{person}{Michael Freedman}, \bibinfo{person}{Eric Freudenthal}, {and} \bibinfo{person}{David Mazi{\`e}res}.} \bibinfo{year}{2004}\natexlab{}.
\newblock \showarticletitle{Democratizing Content Publication with Coral}. In \bibinfo{booktitle}{\emph{First Symposium on Networked Systems Design and Implementation (NSDI 04)}}. \bibinfo{publisher}{USENIX Association}, \bibinfo{address}{San Francisco, CA}.
\newblock
\urldef\tempurl%
\url{https://www.usenix.org/conference/nsdi-04/democratizing-content-publication-coral}
\showURL{%
\tempurl}


\bibitem[Friedrich et~al\mbox{.}(2010)]%
        {FRIEDRICH2010854}
\bibfield{author}{\bibinfo{person}{Tobias Friedrich}, \bibinfo{person}{Nils Hebbinghaus}, {and} \bibinfo{person}{Frank Neumann}.} \bibinfo{year}{2010}\natexlab{}.
\newblock \showarticletitle{Plateaus can be harder in multi-objective optimization}.
\newblock \bibinfo{journal}{\emph{Theoretical Computer Science}} \bibinfo{volume}{411}, \bibinfo{number}{6} (\bibinfo{year}{2010}), \bibinfo{pages}{854--864}.
\newblock
\showISSN{0304-3975}
\urldef\tempurl%
\url{https://doi.org/10.1016/j.tcs.2009.06.020}
\showDOI{\tempurl}


\bibitem[Fu et~al\mbox{.}(2021)]%
        {fu2023blackboxprediction}
\bibfield{author}{\bibinfo{person}{Silvery Fu}, \bibinfo{person}{Saurabh Gupta}, \bibinfo{person}{Radhika Mittal}, {and} \bibinfo{person}{Sylvia Ratnasamy}.} \bibinfo{year}{2021}\natexlab{}.
\newblock \showarticletitle{On the Use of {ML} for Blackbox System Performance Prediction}. In \bibinfo{booktitle}{\emph{Proceedings of 18th USENIX Symposium on Networked Systems Design and Implementation (NSDI 21)}}. \bibinfo{publisher}{USENIX Association}, \bibinfo{address}{Virtual Event}, \bibinfo{pages}{763--784}.
\newblock
\showISBNx{978-1-939133-21-2}
\urldef\tempurl%
\url{https://www.usenix.org/conference/nsdi21/presentation/fu}
\showURL{%
\tempurl}


\bibitem[Gandhi et~al\mbox{.}(2012)]%
        {autoscale}
\bibfield{author}{\bibinfo{person}{Anshul Gandhi}, \bibinfo{person}{Mor Harchol-Balter}, \bibinfo{person}{Ram Raghunathan}, {and} \bibinfo{person}{Michael~A. Kozuch}.} \bibinfo{year}{2012}\natexlab{}.
\newblock \showarticletitle{AutoScale: Dynamic, Robust Capacity Management for Multi-Tier Data Centers}.
\newblock \bibinfo{journal}{\emph{ACM Trans. Comput. Syst.}} \bibinfo{volume}{30}, \bibinfo{number}{4}, Article \bibinfo{articleno}{14} (\bibinfo{date}{nov} \bibinfo{year}{2012}), \bibinfo{numpages}{26}~pages.
\newblock
\showISSN{0734-2071}
\urldef\tempurl%
\url{https://doi.org/10.1145/2382553.2382556}
\showDOI{\tempurl}


\bibitem[Gneiting and Katzfuss(2014)]%
        {prob_forecast}
\bibfield{author}{\bibinfo{person}{Tilmann Gneiting} {and} \bibinfo{person}{Matthias Katzfuss}.} \bibinfo{year}{2014}\natexlab{}.
\newblock \showarticletitle{Probabilistic forecasting}.
\newblock \bibinfo{journal}{\emph{Annual Review of Statistics and Its Application}}  \bibinfo{volume}{1} (\bibinfo{year}{2014}), \bibinfo{pages}{125--151}.
\newblock


\bibitem[Grandl et~al\mbox{.}(2014)]%
        {tetris}
\bibfield{author}{\bibinfo{person}{Robert Grandl}, \bibinfo{person}{Ganesh Ananthanarayanan}, \bibinfo{person}{Srikanth Kandula}, \bibinfo{person}{Sriram Rao}, {and} \bibinfo{person}{Aditya Akella}.} \bibinfo{year}{2014}\natexlab{}.
\newblock \showarticletitle{Multi-Resource Packing for Cluster Schedulers}. In \bibinfo{booktitle}{\emph{Proceedings of the 2014 ACM Conference on SIGCOMM}} \emph{(\bibinfo{series}{SIGCOMM 14})}. \bibinfo{publisher}{Association for Computing Machinery}, \bibinfo{address}{Chicago, IL, USA}, \bibinfo{pages}{455–466}.
\newblock
\showISBNx{9781450328364}
\urldef\tempurl%
\url{https://doi.org/10.1145/2619239.2626334}
\showDOI{\tempurl}


\bibitem[Greenberg et~al\mbox{.}(2009)]%
        {vl2}
\bibfield{author}{\bibinfo{person}{Albert Greenberg}, \bibinfo{person}{James~R. Hamilton}, \bibinfo{person}{Navendu Jain}, \bibinfo{person}{Srikanth Kandula}, \bibinfo{person}{Changhoon Kim}, \bibinfo{person}{Parantap Lahiri}, \bibinfo{person}{David~A. Maltz}, \bibinfo{person}{Parveen Patel}, {and} \bibinfo{person}{Sudipta Sengupta}.} \bibinfo{year}{2009}\natexlab{}.
\newblock \showarticletitle{VL2: a scalable and flexible data center network}. In \bibinfo{booktitle}{\emph{Proceedings of the ACM SIGCOMM 2009 Conference on Data Communication}} (Barcelona, Spain) \emph{(\bibinfo{series}{SIGCOMM '09})}. \bibinfo{publisher}{Association for Computing Machinery}, \bibinfo{address}{New York, NY, USA}, \bibinfo{pages}{51–62}.
\newblock
\showISBNx{9781605585949}
\urldef\tempurl%
\url{https://doi.org/10.1145/1592568.1592576}
\showDOI{\tempurl}


\bibitem[Gujarati et~al\mbox{.}(2017)]%
        {swayam}
\bibfield{author}{\bibinfo{person}{Arpan Gujarati}, \bibinfo{person}{Sameh Elnikety}, \bibinfo{person}{Yuxiong He}, \bibinfo{person}{Kathryn~S. McKinley}, {and} \bibinfo{person}{Bj\"{o}rn~B. Brandenburg}.} \bibinfo{year}{2017}\natexlab{}.
\newblock \showarticletitle{Swayam: Distributed Autoscaling to Meet SLAs of Machine Learning Inference Services with Resource Efficiency}. In \bibinfo{booktitle}{\emph{Proceedings of the 18th ACM/IFIP/USENIX Middleware Conference}} \emph{(\bibinfo{series}{Middleware 17})}. \bibinfo{publisher}{Association for Computing Machinery}, \bibinfo{address}{Las Vegas, Nevada, USA}, \bibinfo{pages}{109–120}.
\newblock
\showISBNx{9781450347204}


\bibitem[Gujarati et~al\mbox{.}(2020)]%
        {clockwork}
\bibfield{author}{\bibinfo{person}{Arpan Gujarati}, \bibinfo{person}{Reza Karimi}, \bibinfo{person}{Safya Alzayat}, \bibinfo{person}{Wei Hao}, \bibinfo{person}{Antoine Kaufmann}, \bibinfo{person}{Ymir Vigfusson}, {and} \bibinfo{person}{Jonathan Mace}.} \bibinfo{year}{2020}\natexlab{}.
\newblock \showarticletitle{Serving {DNNs} like Clockwork: Performance Predictability from the Bottom Up}. In \bibinfo{booktitle}{\emph{Proceedings of the 14th USENIX Symposium on Operating Systems Design and Implementation (OSDI 20)}}. \bibinfo{publisher}{USENIX Association}, \bibinfo{address}{Virtual Event}, \bibinfo{pages}{443--462}.
\newblock
\showISBNx{978-1-939133-19-9}


\bibitem[Gunasekaran et~al\mbox{.}(2022)]%
        {cocktail}
\bibfield{author}{\bibinfo{person}{Jashwant~Raj Gunasekaran}, \bibinfo{person}{Cyan~Subhra Mishra}, \bibinfo{person}{Prashanth Thinakaran}, \bibinfo{person}{Bikash Sharma}, \bibinfo{person}{Mahmut~Taylan Kandemir}, {and} \bibinfo{person}{Chita~R. Das}.} \bibinfo{year}{2022}\natexlab{}.
\newblock \showarticletitle{Cocktail: A Multidimensional Optimization for Model Serving in Cloud}. In \bibinfo{booktitle}{\emph{Proceedings of the 19th USENIX Symposium on Networked Systems Design and Implementation (NSDI 22)}}. \bibinfo{publisher}{USENIX Association}, \bibinfo{address}{Renton, WA}, \bibinfo{pages}{1041--1057}.
\newblock
\showISBNx{978-1-939133-27-4}
\urldef\tempurl%
\url{https://www.usenix.org/conference/nsdi22/presentation/gunasekaran}
\showURL{%
\tempurl}


\bibitem[Gupta et~al\mbox{.}(2020)]%
        {deeprecsys}
\bibfield{author}{\bibinfo{person}{Udit Gupta}, \bibinfo{person}{Samuel Hsia}, \bibinfo{person}{Vikram Saraph}, \bibinfo{person}{Xiaodong Wang}, \bibinfo{person}{Brandon Reagen}, \bibinfo{person}{Gu-Yeon Wei}, \bibinfo{person}{Hsien-Hsin~S. Lee}, \bibinfo{person}{David Brooks}, {and} \bibinfo{person}{Carole-Jean Wu}.} \bibinfo{year}{2020}\natexlab{}.
\newblock \showarticletitle{DeepRecSys: A System for Optimizing End-to-End at-Scale Neural Recommendation Inference}. In \bibinfo{booktitle}{\emph{Proceedings of the ACM/IEEE 47th Annual International Symposium on Computer Architecture}} \emph{(\bibinfo{series}{ISCA 20})}. \bibinfo{publisher}{IEEE Press}, \bibinfo{address}{Virtual Event}, \bibinfo{pages}{982–995}.
\newblock
\showISBNx{9781728146614}
\urldef\tempurl%
\url{https://doi.org/10.1109/ISCA45697.2020.00084}
\showDOI{\tempurl}


\bibitem[Hat(2024)]%
        {openshift}
\bibfield{author}{\bibinfo{person}{Red Hat}.} \bibinfo{year}{2024}\natexlab{}.
\newblock \bibinfo{title}{Red Hat OpenShift}.
\newblock
\newblock
\urldef\tempurl%
\url{https://www.redhat.com/en/technologies/cloud-computing/openshift}
\showURL{%
\tempurl}


\bibitem[He et~al\mbox{.}(2016)]%
        {resnet}
\bibfield{author}{\bibinfo{person}{Kaiming He}, \bibinfo{person}{Xiangyu Zhang}, \bibinfo{person}{Shaoqing Ren}, {and} \bibinfo{person}{Jian Sun}.} \bibinfo{year}{2016}\natexlab{}.
\newblock \showarticletitle{Deep Residual Learning for Image Recognition}. In \bibinfo{booktitle}{\emph{Proceedings of the 2016 IEEE Conference on Computer Vision and Pattern Recognition (CVPR 16)}}. \bibinfo{publisher}{{IEEE}}, \bibinfo{address}{Las Vegas, NV, USA}, \bibinfo{pages}{770--778}.
\newblock
\urldef\tempurl%
\url{https://doi.org/10.1109/CVPR.2016.90}
\showDOI{\tempurl}


\bibitem[Herzen et~al\mbox{.}(2022)]%
        {darts}
\bibfield{author}{\bibinfo{person}{Julien Herzen}, \bibinfo{person}{Francesco Lässig}, \bibinfo{person}{Samuele~Giuliano Piazzetta}, \bibinfo{person}{Thomas Neuer}, \bibinfo{person}{Léo Tafti}, \bibinfo{person}{Guillaume Raille}, \bibinfo{person}{Tomas~Van Pottelbergh}, \bibinfo{person}{Marek Pasieka}, \bibinfo{person}{Andrzej Skrodzki}, \bibinfo{person}{Nicolas Huguenin}, \bibinfo{person}{Maxime Dumonal}, \bibinfo{person}{Jan Kościsz}, \bibinfo{person}{Dennis Bader}, \bibinfo{person}{Frédérick Gusset}, \bibinfo{person}{Mounir Benheddi}, \bibinfo{person}{Camila Williamson}, \bibinfo{person}{Michal Kosinski}, \bibinfo{person}{Matej Petrik}, {and} \bibinfo{person}{Gaël Grosch}.} \bibinfo{year}{2022}\natexlab{}.
\newblock \showarticletitle{Darts: User-Friendly Modern Machine Learning for Time Series}.
\newblock \bibinfo{journal}{\emph{Journal of Machine Learning Research}} \bibinfo{volume}{23}, \bibinfo{number}{124} (\bibinfo{year}{2022}), \bibinfo{pages}{1--6}.
\newblock
\urldef\tempurl%
\url{http://jmlr.org/papers/v23/21-1177.html}
\showURL{%
\tempurl}


\bibitem[Hu et~al\mbox{.}(2022)]%
        {uber_deepeta}
\bibfield{author}{\bibinfo{person}{Xinyu Hu}, \bibinfo{person}{Olcay Cirit}, \bibinfo{person}{Tanmay Binaykiya}, {and} \bibinfo{person}{Ramit Hora}.} \bibinfo{year}{2022}\natexlab{}.
\newblock \bibinfo{title}{DeepETA: How Uber Predicts Arrival Times Using Deep Learning, https://www.uber.com/blog/deepeta-how-uber-predicts-arrival-times/}.
\newblock
\newblock


\bibitem[IBM(2024a)]%
        {ibmcloud_vpc}
\bibfield{author}{\bibinfo{person}{IBM}.} \bibinfo{year}{2024}\natexlab{a}.
\newblock \bibinfo{title}{{IBM Cloud VPC Solutions}}.
\newblock
\newblock
\urldef\tempurl%
\url{https://www.ibm.com/cloud/vpc}
\showURL{%
\tempurl}


\bibitem[IBM(2024b)]%
        {ibmcloud_sla}
\bibfield{author}{\bibinfo{person}{IBM}.} \bibinfo{year}{2024}\natexlab{b}.
\newblock \bibinfo{title}{Service Level Agreements (SLAs) for IBM Cloud}.
\newblock
\newblock
\urldef\tempurl%
\url{https://cloud.ibm.com/docs/overview?topic=overview-slas}
\showURL{%
\tempurl}


\bibitem[Jyothi et~al\mbox{.}(2016)]%
        {morpheus}
\bibfield{author}{\bibinfo{person}{Sangeetha~Abdu Jyothi}, \bibinfo{person}{Carlo Curino}, \bibinfo{person}{Ishai Menache}, \bibinfo{person}{Shravan~Matthur Narayanamurthy}, \bibinfo{person}{Alexey Tumanov}, \bibinfo{person}{Jonathan Yaniv}, \bibinfo{person}{Ruslan Mavlyutov}, \bibinfo{person}{Inigo Goiri}, \bibinfo{person}{Subru Krishnan}, \bibinfo{person}{Janardhan Kulkarni}, {and} \bibinfo{person}{Sriram Rao}.} \bibinfo{year}{2016}\natexlab{}.
\newblock \showarticletitle{Morpheus: Towards Automated {SLOs} for Enterprise Clusters}. In \bibinfo{booktitle}{\emph{12th USENIX Symposium on Operating Systems Design and Implementation (OSDI 16)}}. \bibinfo{publisher}{USENIX Association}, \bibinfo{address}{Savannah, GA}, \bibinfo{pages}{117--134}.
\newblock
\showISBNx{978-1-931971-33-1}
\urldef\tempurl%
\url{https://www.usenix.org/conference/osdi16/technical-sessions/presentation/jyothi}
\showURL{%
\tempurl}


\bibitem[Kalim et~al\mbox{.}(2018)]%
        {henge}
\bibfield{author}{\bibinfo{person}{Faria Kalim}, \bibinfo{person}{Le Xu}, \bibinfo{person}{Sharanya Bathey}, \bibinfo{person}{Richa Meherwal}, {and} \bibinfo{person}{Indranil Gupta}.} \bibinfo{year}{2018}\natexlab{}.
\newblock \showarticletitle{Henge: Intent-Driven Multi-Tenant Stream Processing}. In \bibinfo{booktitle}{\emph{Proceedings of the 2018 ACM Symposium on Cloud Computing}} \emph{(\bibinfo{series}{SoCC 18})}. \bibinfo{publisher}{Association for Computing Machinery}, \bibinfo{address}{Carlsbad, CA, USA}, \bibinfo{pages}{249–262}.
\newblock
\showISBNx{9781450360111}
\urldef\tempurl%
\url{https://doi.org/10.1145/3267809.3267832}
\showDOI{\tempurl}


\bibitem[Kandasamy et~al\mbox{.}(2020)]%
        {kandasamy2020online}
\bibfield{author}{\bibinfo{person}{Kirthevasan Kandasamy}, \bibinfo{person}{Gur-Eyal Sela}, \bibinfo{person}{Joseph~E Gonzalez}, \bibinfo{person}{Michael~I Jordan}, {and} \bibinfo{person}{Ion Stoica}.} \bibinfo{year}{2020}\natexlab{}.
\newblock \bibinfo{title}{Online Learning Demands in Max-min Fairness}.
\newblock
\newblock
\showeprint[arxiv]{2012.08648}~[stat.ML]


\bibitem[Kendall(1938)]%
        {kendall_tau}
\bibfield{author}{\bibinfo{person}{M.~G. Kendall}.} \bibinfo{year}{1938}\natexlab{}.
\newblock \showarticletitle{A New Measure of Rank Correlation}.
\newblock \bibinfo{journal}{\emph{Biometrika}} \bibinfo{volume}{30}, \bibinfo{number}{1/2} (\bibinfo{year}{1938}), \bibinfo{pages}{81--93}.
\newblock
\showISSN{00063444}
\urldef\tempurl%
\url{http://www.jstor.org/stable/2332226}
\showURL{%
\tempurl}


\bibitem[Kraft(1988)]%
        {slsqp}
\bibfield{author}{\bibinfo{person}{Dieter Kraft}.} \bibinfo{year}{1988}\natexlab{}.
\newblock \bibinfo{booktitle}{\emph{A software package for sequential quadratic programming, Ein Software-Paket zur sequentiellen quadratischen Optimierung, Forschungsbericht. Deutsche Forschungs- und Versuchsanstalt für Luft- und Raumfahrt, DFVLR}}.
\newblock \bibinfo{type}{{T}echnical {R}eport}. \bibinfo{institution}{Institut für Dynamik der Flugsysteme, Deutsche Forschungs- und Versuchsanstalt für Luft- und Raumfahrt \(DFVLR\), Oberpfaffenhofen}, \bibinfo{address}{Köln}.
\newblock
\urldef\tempurl%
\url{https://www.tib.eu/de/suchen/id/TIBKAT%3A016896521}
\showURL{%
\tempurl}


\bibitem[Kumar et~al\mbox{.}(2016)]%
        {cedar}
\bibfield{author}{\bibinfo{person}{Gautam Kumar}, \bibinfo{person}{Ganesh Ananthanarayanan}, \bibinfo{person}{Sylvia Ratnasamy}, {and} \bibinfo{person}{Ion Stoica}.} \bibinfo{year}{2016}\natexlab{}.
\newblock \showarticletitle{Hold 'em or fold 'em? aggregation queries under performance variations}. In \bibinfo{booktitle}{\emph{Proceedings of the Eleventh European Conference on Computer Systems}} (London, United Kingdom) \emph{(\bibinfo{series}{EuroSys '16})}. \bibinfo{publisher}{Association for Computing Machinery}, \bibinfo{address}{New York, NY, USA}, Article \bibinfo{articleno}{7}, \bibinfo{numpages}{14}~pages.
\newblock
\showISBNx{9781450342407}
\urldef\tempurl%
\url{https://doi.org/10.1145/2901318.2901351}
\showDOI{\tempurl}


\bibitem[Lam et~al\mbox{.}(2015)]%
        {numba}
\bibfield{author}{\bibinfo{person}{Siu~Kwan Lam}, \bibinfo{person}{Antoine Pitrou}, {and} \bibinfo{person}{Stanley Seibert}.} \bibinfo{year}{2015}\natexlab{}.
\newblock \showarticletitle{Numba: a LLVM-based Python JIT compiler}. In \bibinfo{booktitle}{\emph{Proceedings of the Second Workshop on the LLVM Compiler Infrastructure in HPC}} (Austin, Texas) \emph{(\bibinfo{series}{LLVM '15})}. \bibinfo{publisher}{Association for Computing Machinery}, \bibinfo{address}{New York, NY, USA}, Article \bibinfo{articleno}{7}, \bibinfo{numpages}{6}~pages.
\newblock
\showISBNx{9781450340052}
\urldef\tempurl%
\url{https://doi.org/10.1145/2833157.2833162}
\showDOI{\tempurl}


\bibitem[Li et~al\mbox{.}(2019)]%
        {asplos2019li}
\bibfield{author}{\bibinfo{person}{Chen Li}, \bibinfo{person}{Rachata Ausavarungnirun}, \bibinfo{person}{Christopher~J. Rossbach}, \bibinfo{person}{Youtao Zhang}, \bibinfo{person}{Onur Mutlu}, \bibinfo{person}{Yang Guo}, {and} \bibinfo{person}{Jun Yang}.} \bibinfo{year}{2019}\natexlab{}.
\newblock \showarticletitle{A Framework for Memory Oversubscription Management in Graphics Processing Units}. In \bibinfo{booktitle}{\emph{Proceedings of the Twenty-Fourth International Conference on Architectural Support for Programming Languages and Operating Systems}} (Providence, RI, USA) \emph{(\bibinfo{series}{ASPLOS '19})}. \bibinfo{publisher}{Association for Computing Machinery}, \bibinfo{address}{New York, NY, USA}, \bibinfo{pages}{49–63}.
\newblock
\showISBNx{9781450362405}
\urldef\tempurl%
\url{https://doi.org/10.1145/3297858.3304044}
\showDOI{\tempurl}


\bibitem[Liaw et~al\mbox{.}(2018)]%
        {ray_tune}
\bibfield{author}{\bibinfo{person}{Richard Liaw}, \bibinfo{person}{Eric Liang}, \bibinfo{person}{Robert Nishihara}, \bibinfo{person}{Philipp Moritz}, \bibinfo{person}{Joseph~E. Gonzalez}, {and} \bibinfo{person}{Ion Stoica}.} \bibinfo{year}{2018}\natexlab{}.
\newblock \bibinfo{title}{Tune: A Research Platform for Distributed Model Selection and Training}.
\newblock
\newblock
\showeprint[arxiv]{1807.05118}~[cs.LG]


\bibitem[Mahajan et~al\mbox{.}(2020)]%
        {themis}
\bibfield{author}{\bibinfo{person}{Kshiteej Mahajan}, \bibinfo{person}{Arjun Balasubramanian}, \bibinfo{person}{Arjun Singhvi}, \bibinfo{person}{Shivaram Venkataraman}, \bibinfo{person}{Aditya Akella}, \bibinfo{person}{Amar Phanishayee}, {and} \bibinfo{person}{Shuchi Chawla}.} \bibinfo{year}{2020}\natexlab{}.
\newblock \showarticletitle{Themis: Fair and Efficient {GPU} Cluster Scheduling}. In \bibinfo{booktitle}{\emph{17th USENIX Symposium on Networked Systems Design and Implementation (NSDI 20)}}. \bibinfo{publisher}{USENIX Association}, \bibinfo{address}{Santa Clara, CA}, \bibinfo{pages}{289--304}.
\newblock
\showISBNx{978-1-939133-13-7}
\urldef\tempurl%
\url{https://www.usenix.org/conference/nsdi20/presentation/mahajan}
\showURL{%
\tempurl}


\bibitem[Makridakis and Hibon(1997)]%
        {arma}
\bibfield{author}{\bibinfo{person}{Spyros Makridakis} {and} \bibinfo{person}{Michèle Hibon}.} \bibinfo{year}{1997}\natexlab{}.
\newblock \showarticletitle{ARMA Models and the Box–Jenkins Methodology}.
\newblock \bibinfo{journal}{\emph{Journal of Forecasting}} \bibinfo{volume}{16}, \bibinfo{number}{3} (\bibinfo{year}{1997}), \bibinfo{pages}{147--163}.
\newblock
\urldef\tempurl%
\url{https://doi.org/10.1002/(SICI)1099-131X(199705)16:3<147::AID-FOR652>3.0.CO;2-X}
\showDOI{\tempurl}


\bibitem[Mao et~al\mbox{.}(2016)]%
        {mao2016resource}
\bibfield{author}{\bibinfo{person}{Hongzi Mao}, \bibinfo{person}{Mohammad Alizadeh}, \bibinfo{person}{Ishai Menache}, {and} \bibinfo{person}{Srikanth Kandula}.} \bibinfo{year}{2016}\natexlab{}.
\newblock \showarticletitle{Resource Management with Deep Reinforcement Learning}. In \bibinfo{booktitle}{\emph{Proceedings of the 15th ACM Workshop on Hot Topics in Networks}} \emph{(\bibinfo{series}{HotNets 16})}. \bibinfo{publisher}{Association for Computing Machinery}, \bibinfo{address}{Atlanta, GA, USA}, \bibinfo{pages}{50–56}.
\newblock
\showISBNx{9781450346610}
\urldef\tempurl%
\url{https://doi.org/10.1145/3005745.3005750}
\showDOI{\tempurl}


\bibitem[McCarthy(2023)]%
        {xbox_content_moderation}
\bibfield{author}{\bibinfo{person}{Dave McCarthy}.} \bibinfo{year}{2023}\natexlab{}.
\newblock \bibinfo{title}{Xbox Releases Second Transparency Report Demonstrating the Integral Role of Proactive Content Moderation, https://news.xbox.com/en-us/2023/05/22/xbox-releases-second-transparency-report/}.
\newblock
\newblock


\bibitem[Mickulicz et~al\mbox{.}(2013)]%
        {scale_or_not}
\bibfield{author}{\bibinfo{person}{Nathan~D. Mickulicz}, \bibinfo{person}{Priya Narasimhan}, {and} \bibinfo{person}{Rajeev Gandhi}.} \bibinfo{year}{2013}\natexlab{}.
\newblock \showarticletitle{To Auto Scale or Not to Auto Scale}. In \bibinfo{booktitle}{\emph{Proceedings of the 10th International Conference on Autonomic Computing (ICAC 13)}}. \bibinfo{publisher}{USENIX Association}, \bibinfo{address}{San Jose, CA}, \bibinfo{pages}{145--151}.
\newblock
\showISBNx{978-1-931971-02-7}
\urldef\tempurl%
\url{https://www.usenix.org/conference/icac13/technical-sessions/presentation/mickulicz}
\showURL{%
\tempurl}


\bibitem[Moritz et~al\mbox{.}(2018)]%
        {ray}
\bibfield{author}{\bibinfo{person}{Philipp Moritz}, \bibinfo{person}{Robert Nishihara}, \bibinfo{person}{Stephanie Wang}, \bibinfo{person}{Alexey Tumanov}, \bibinfo{person}{Richard Liaw}, \bibinfo{person}{Eric Liang}, \bibinfo{person}{Melih Elibol}, \bibinfo{person}{Zongheng Yang}, \bibinfo{person}{William Paul}, \bibinfo{person}{Michael~I. Jordan}, {and} \bibinfo{person}{Ion Stoica}.} \bibinfo{year}{2018}\natexlab{}.
\newblock \showarticletitle{Ray: A Distributed Framework for Emerging {AI} Applications}. In \bibinfo{booktitle}{\emph{Proceedings of the 13th USENIX Symposium on Operating Systems Design and Implementation (OSDI 18)}}. \bibinfo{publisher}{USENIX Association}, \bibinfo{address}{Carlsbad, CA, USA}, \bibinfo{pages}{561--577}.
\newblock
\showISBNx{978-1-939133-08-3}
\urldef\tempurl%
\url{https://www.usenix.org/conference/osdi18/presentation/moritz}
\showURL{%
\tempurl}


\bibitem[Narayanan et~al\mbox{.}(2020)]%
        {gavel}
\bibfield{author}{\bibinfo{person}{Deepak Narayanan}, \bibinfo{person}{Keshav Santhanam}, \bibinfo{person}{Fiodar Kazhamiaka}, \bibinfo{person}{Amar Phanishayee}, {and} \bibinfo{person}{Matei Zaharia}.} \bibinfo{year}{2020}\natexlab{}.
\newblock \showarticletitle{Heterogeneity-Aware Cluster Scheduling Policies for Deep Learning Workloads}. In \bibinfo{booktitle}{\emph{14th {USENIX} Symposium on Operating Systems Design and Implementation ({OSDI} 20)}}. \bibinfo{publisher}{{USENIX} Association}, \bibinfo{pages}{481--498}.
\newblock
\showISBNx{978-1-939133-19-9}


\bibitem[Netapp(2023)]%
        {netapp_latency_req}
\bibfield{author}{\bibinfo{person}{Netapp}.} \bibinfo{year}{2023}\natexlab{}.
\newblock \bibinfo{title}{A little more talk, a little more action}.
\newblock
\newblock
\urldef\tempurl%
\url{https://www.netapp.com/media/57081-NA-644-0721-More-talk-more-action.pdf}
\showURL{%
\tempurl}


\bibitem[Ning et~al\mbox{.}(2022)]%
        {arima_comparison2}
\bibfield{author}{\bibinfo{person}{Yanrui Ning}, \bibinfo{person}{Hossein Kazemi}, {and} \bibinfo{person}{Pejman Tahmasebi}.} \bibinfo{year}{2022}\natexlab{}.
\newblock \showarticletitle{A comparative machine learning study for time series oil production forecasting: ARIMA, LSTM, and Prophet}.
\newblock \bibinfo{journal}{\emph{Computers \& Geosciences}}  \bibinfo{volume}{164} (\bibinfo{year}{2022}), \bibinfo{pages}{105126}.
\newblock
\showISSN{0098-3004}
\urldef\tempurl%
\url{https://doi.org/10.1016/j.cageo.2022.105126}
\showDOI{\tempurl}


\bibitem[Olston et~al\mbox{.}(2017)]%
        {tf_serving}
\bibfield{author}{\bibinfo{person}{Christopher Olston}, \bibinfo{person}{Fangwei Li}, \bibinfo{person}{Jeremiah Harmsen}, \bibinfo{person}{Jordan Soyke}, \bibinfo{person}{Kiril Gorovoy}, \bibinfo{person}{Li Lao}, \bibinfo{person}{Noah Fiedel}, \bibinfo{person}{Sukriti Ramesh}, {and} \bibinfo{person}{Vinu Rajashekhar}.} \bibinfo{year}{2017}\natexlab{}.
\newblock \showarticletitle{{TensorFlow-Serving}: Flexible, High-Performance ML Serving}. In \bibinfo{booktitle}{\emph{Workshop on ML Systems at NIPS 2017}}. \bibinfo{address}{Long Beach, CA, USA}.
\newblock


\bibitem[Ousterhout et~al\mbox{.}(2013)]%
        {sparrow}
\bibfield{author}{\bibinfo{person}{Kay Ousterhout}, \bibinfo{person}{Patrick Wendell}, \bibinfo{person}{Matei Zaharia}, {and} \bibinfo{person}{Ion Stoica}.} \bibinfo{year}{2013}\natexlab{}.
\newblock \showarticletitle{Sparrow: Distributed, Low Latency Scheduling}. In \bibinfo{booktitle}{\emph{Proceedings of the Twenty-Fourth ACM Symposium on Operating Systems Principles}} \emph{(\bibinfo{series}{SOSP 13})}. \bibinfo{publisher}{Association for Computing Machinery}, \bibinfo{address}{Farminton, PA, USA}, \bibinfo{pages}{69–84}.
\newblock
\showISBNx{9781450323888}


\bibitem[Pascanu et~al\mbox{.}(2014)]%
        {pascanu2014saddle}
\bibfield{author}{\bibinfo{person}{Razvan Pascanu}, \bibinfo{person}{Yann~N. Dauphin}, \bibinfo{person}{Surya Ganguli}, {and} \bibinfo{person}{Yoshua Bengio}.} \bibinfo{year}{2014}\natexlab{}.
\newblock \bibinfo{title}{On the saddle point problem for non-convex optimization}.
\newblock
\newblock
\showeprint[arxiv]{1405.4604}~[cs.LG]


\bibitem[Paszke et~al\mbox{.}(2019)]%
        {pytorch}
\bibfield{author}{\bibinfo{person}{Adam Paszke}, \bibinfo{person}{Sam Gross}, \bibinfo{person}{Francisco Massa}, \bibinfo{person}{Adam Lerer}, \bibinfo{person}{James Bradbury}, \bibinfo{person}{Gregory Chanan}, \bibinfo{person}{Trevor Killeen}, \bibinfo{person}{Zeming Lin}, \bibinfo{person}{Natalia Gimelshein}, \bibinfo{person}{Luca Antiga}, \bibinfo{person}{Alban Desmaison}, \bibinfo{person}{Andreas K\"{o}pf}, \bibinfo{person}{Edward Yang}, \bibinfo{person}{Zach DeVito}, \bibinfo{person}{Martin Raison}, \bibinfo{person}{Alykhan Tejani}, \bibinfo{person}{Sasank Chilamkurthy}, \bibinfo{person}{Benoit Steiner}, \bibinfo{person}{Lu Fang}, \bibinfo{person}{Junjie Bai}, {and} \bibinfo{person}{Soumith Chintala}.} \bibinfo{year}{2019}\natexlab{}.
\newblock \showarticletitle{PyTorch: An Imperative Style, High-Performance Deep Learning Library}. In \bibinfo{booktitle}{\emph{Proceedings of the 33rd International Conference on Neural Information Processing Systems}} \emph{(\bibinfo{series}{NeurIPS 19})}. \bibinfo{publisher}{Curran Associates Inc.}, \bibinfo{address}{Vancouver, Canada}, Article \bibinfo{articleno}{721}, \bibinfo{numpages}{12}~pages.
\newblock
\urldef\tempurl%
\url{http://papers.neurips.cc/paper/9015-pytorch-an-imperative-style-high-performance-deep-learning-library.pdf}
\showURL{%
\tempurl}


\bibitem[Patel et~al\mbox{.}(2024)]%
        {polca}
\bibfield{author}{\bibinfo{person}{Pratyush Patel}, \bibinfo{person}{Esha Choukse}, \bibinfo{person}{Chaojie Zhang}, \bibinfo{person}{\'{I}\~{n}igo Goiri}, \bibinfo{person}{Brijesh Warrier}, \bibinfo{person}{Nithish Mahalingam}, {and} \bibinfo{person}{Ricardo Bianchini}.} \bibinfo{year}{2024}\natexlab{}.
\newblock \showarticletitle{Characterizing Power Management Opportunities for LLMs in the Cloud}. In \bibinfo{booktitle}{\emph{Proceedings of the 29th ACM International Conference on Architectural Support for Programming Languages and Operating Systems, Volume 3}} (La Jolla, CA, USA) \emph{(\bibinfo{series}{ASPLOS '24})}. \bibinfo{publisher}{Association for Computing Machinery}, \bibinfo{address}{New York, NY, USA}, \bibinfo{pages}{207–222}.
\newblock
\showISBNx{9798400703867}
\urldef\tempurl%
\url{https://doi.org/10.1145/3620666.3651329}
\showDOI{\tempurl}


\bibitem[Powell(1994)]%
        {cobyla}
\bibfield{author}{\bibinfo{person}{M.~J.~D. Powell}.} \bibinfo{year}{1994}\natexlab{}.
\newblock \bibinfo{booktitle}{\emph{A Direct Search Optimization Method That Models the Objective and Constraint Functions by Linear Interpolation}}.
\newblock \bibinfo{publisher}{Springer Netherlands}, \bibinfo{address}{Dordrecht}, \bibinfo{pages}{51--67}.
\newblock
\showISBNx{978-94-015-8330-5}
\urldef\tempurl%
\url{https://doi.org/10.1007/978-94-015-8330-5_4}
\showDOI{\tempurl}


\bibitem[Pratt(2022)]%
        {visa_fraud_detection}
\bibfield{author}{\bibinfo{person}{Mary~K. Pratt}.} \bibinfo{year}{2022}\natexlab{}.
\newblock \bibinfo{title}{How Visa fights fraud}.
\newblock
\newblock
\urldef\tempurl%
\url{https://www.csoonline.com/article/573009/how-visa-fights-fraud.html}
\showURL{%
\tempurl}


\bibitem[Qiao et~al\mbox{.}(2021)]%
        {pollux}
\bibfield{author}{\bibinfo{person}{Aurick Qiao}, \bibinfo{person}{Sang~Keun Choe}, \bibinfo{person}{Suhas~Jayaram Subramanya}, \bibinfo{person}{Willie Neiswanger}, \bibinfo{person}{Qirong Ho}, \bibinfo{person}{Hao Zhang}, \bibinfo{person}{Gregory~R. Ganger}, {and} \bibinfo{person}{Eric~P. Xing}.} \bibinfo{year}{2021}\natexlab{}.
\newblock \showarticletitle{Pollux: Co-adaptive Cluster Scheduling for Goodput-Optimized Deep Learning}. In \bibinfo{booktitle}{\emph{15th {USENIX} Symposium on Operating Systems Design and Implementation ({OSDI} 21)}}. \bibinfo{publisher}{{USENIX} Association}, \bibinfo{pages}{1--18}.
\newblock
\showISBNx{978-1-939133-22-9}
\urldef\tempurl%
\url{https://www.usenix.org/conference/osdi21/presentation/qiao}
\showURL{%
\tempurl}


\bibitem[Rasley et~al\mbox{.}(2020)]%
        {deepspeed}
\bibfield{author}{\bibinfo{person}{Jeff Rasley}, \bibinfo{person}{Samyam Rajbhandari}, \bibinfo{person}{Olatunji Ruwase}, {and} \bibinfo{person}{Yuxiong He}.} \bibinfo{year}{2020}\natexlab{}.
\newblock \showarticletitle{DeepSpeed: System Optimizations Enable Training Deep Learning Models with Over 100 Billion Parameters}. In \bibinfo{booktitle}{\emph{Proceedings of the 26th ACM SIGKDD International Conference on Knowledge Discovery and Data Mining}} \emph{(\bibinfo{series}{KDD 20})}. \bibinfo{publisher}{Association for Computing Machinery}, \bibinfo{address}{Virtual Event, CA, USA}, \bibinfo{pages}{3505–3506}.
\newblock
\showISBNx{9781450379984}
\urldef\tempurl%
\url{https://doi.org/10.1145/3394486.3406703}
\showDOI{\tempurl}


\bibitem[Reddi et~al\mbox{.}(2020)]%
        {mlperf_inference}
\bibfield{author}{\bibinfo{person}{Vijay~Janapa Reddi}, \bibinfo{person}{Christine Cheng}, \bibinfo{person}{David Kanter}, \bibinfo{person}{Peter Mattson}, \bibinfo{person}{Guenther Schmuelling}, \bibinfo{person}{Carole-Jean Wu}, \bibinfo{person}{Brian Anderson}, \bibinfo{person}{Maximilien Breughe}, \bibinfo{person}{Mark Charlebois}, \bibinfo{person}{William Chou}, {et~al\mbox{.}}} \bibinfo{year}{2020}\natexlab{}.
\newblock \showarticletitle{MLPerf Inference Benchmark}. In \bibinfo{booktitle}{\emph{Proceedings of the 2020 ACM/IEEE 47th Annual International Symposium on Computer Architecture (ISCA 20)}}. \bibinfo{publisher}{{IEEE}}, \bibinfo{address}{Virtual Event}, \bibinfo{pages}{446--459}.
\newblock


\bibitem[Romero et~al\mbox{.}(2021)]%
        {infaas}
\bibfield{author}{\bibinfo{person}{Francisco Romero}, \bibinfo{person}{Qian Li}, \bibinfo{person}{Neeraja~J. Yadwadkar}, {and} \bibinfo{person}{Christos Kozyrakis}.} \bibinfo{year}{2021}\natexlab{}.
\newblock \showarticletitle{{INFaaS}: Automated Model-less Inference Serving}. In \bibinfo{booktitle}{\emph{Proceedings of the 2021 USENIX Annual Technical Conference (USENIX ATC 21)}}. \bibinfo{publisher}{USENIX Association}, \bibinfo{address}{Virtual Event}, \bibinfo{pages}{397--411}.
\newblock
\showISBNx{978-1-939133-23-6}


\bibitem[Russell and Norvig(2009)]%
        {ai_textbook}
\bibfield{author}{\bibinfo{person}{Stuart Russell} {and} \bibinfo{person}{Peter Norvig}.} \bibinfo{year}{2009}\natexlab{}.
\newblock \bibinfo{booktitle}{\emph{Artificial Intelligence: A Modern Approach} (\bibinfo{edition}{3rd} ed.)}.
\newblock \bibinfo{publisher}{Prentice Hall Press}, \bibinfo{address}{USA}.
\newblock
\showISBNx{0136042597}


\bibitem[Rzadca et~al\mbox{.}(2020)]%
        {autopilot}
\bibfield{author}{\bibinfo{person}{Krzysztof Rzadca}, \bibinfo{person}{Pawel Findeisen}, \bibinfo{person}{Jacek Swiderski}, \bibinfo{person}{Przemyslaw Zych}, \bibinfo{person}{Przemyslaw Broniek}, \bibinfo{person}{Jarek Kusmierek}, \bibinfo{person}{Pawel Nowak}, \bibinfo{person}{Beata Strack}, \bibinfo{person}{Piotr Witusowski}, \bibinfo{person}{Steven Hand}, {and} \bibinfo{person}{John Wilkes}.} \bibinfo{year}{2020}\natexlab{}.
\newblock \showarticletitle{Autopilot: Workload Autoscaling at Google}. In \bibinfo{booktitle}{\emph{Proceedings of the Fifteenth European Conference on Computer Systems}} \emph{(\bibinfo{series}{EuroSys 20})}. \bibinfo{publisher}{Association for Computing Machinery}, \bibinfo{address}{Heraklion, Greece}, Article \bibinfo{articleno}{16}, \bibinfo{numpages}{16}~pages.
\newblock
\showISBNx{9781450368827}
\urldef\tempurl%
\url{https://doi.org/10.1145/3342195.3387524}
\showDOI{\tempurl}


\bibitem[Schuler et~al\mbox{.}(2021)]%
        {serverless_autoscaling}
\bibfield{author}{\bibinfo{person}{L. Schuler}, \bibinfo{person}{S. Jamil}, {and} \bibinfo{person}{N. Kuhl}.} \bibinfo{year}{2021}\natexlab{}.
\newblock \showarticletitle{{AI}-based Resource Allocation: Reinforcement Learning for Adaptive Auto-scaling in Serverless Environments}. In \bibinfo{booktitle}{\emph{Proceedings of the 2021 IEEE/ACM 21st International Symposium on Cluster, Cloud and Internet Computing (CCGrid 21)}}. \bibinfo{publisher}{IEEE Computer Society}, \bibinfo{address}{Los Alamitos, CA, USA}, \bibinfo{pages}{804--811}.
\newblock
\urldef\tempurl%
\url{https://doi.org/10.1109/CCGrid51090.2021.00098}
\showDOI{\tempurl}


\bibitem[Services(2024)]%
        {aws_sla}
\bibfield{author}{\bibinfo{person}{Amazon~Web Services}.} \bibinfo{year}{2024}\natexlab{}.
\newblock \bibinfo{title}{{AWS} Service Level Agreements (SLAs)}.
\newblock
\newblock
\urldef\tempurl%
\url{https://aws.amazon.com/legal/service-level-agreements}
\showURL{%
\tempurl}


\bibitem[Shahrad et~al\mbox{.}(2020)]%
        {azure_function_2019}
\bibfield{author}{\bibinfo{person}{Mohammad Shahrad}, \bibinfo{person}{Rodrigo Fonseca}, \bibinfo{person}{Inigo Goiri}, \bibinfo{person}{Gohar Chaudhry}, \bibinfo{person}{Paul Batum}, \bibinfo{person}{Jason Cooke}, \bibinfo{person}{Eduardo Laureano}, \bibinfo{person}{Colby Tresness}, \bibinfo{person}{Mark Russinovich}, {and} \bibinfo{person}{Ricardo Bianchini}.} \bibinfo{year}{2020}\natexlab{}.
\newblock \showarticletitle{Serverless in the Wild: Characterizing and Optimizing the Serverless Workload at a Large Cloud Provider}. In \bibinfo{booktitle}{\emph{Proceedings of the 2020 USENIX Annual Technical Conference (USENIX ATC 20)}}. \bibinfo{publisher}{USENIX Association}, \bibinfo{address}{Virtual Event}, \bibinfo{pages}{205--218}.
\newblock
\showISBNx{978-1-939133-14-4}
\urldef\tempurl%
\url{https://www.usenix.org/conference/atc20/presentation/shahrad}
\showURL{%
\tempurl}


\bibitem[Shen et~al\mbox{.}(2011)]%
        {cloudscale}
\bibfield{author}{\bibinfo{person}{Zhiming Shen}, \bibinfo{person}{Sethuraman Subbiah}, \bibinfo{person}{Xiaohui Gu}, {and} \bibinfo{person}{John Wilkes}.} \bibinfo{year}{2011}\natexlab{}.
\newblock \showarticletitle{CloudScale: Elastic Resource Scaling for Multi-Tenant Cloud Systems}. In \bibinfo{booktitle}{\emph{Proceedings of the 2nd ACM Symposium on Cloud Computing}} \emph{(\bibinfo{series}{SoCC 11})}. \bibinfo{publisher}{Association for Computing Machinery}, \bibinfo{address}{Cascais, Portugal}, Article \bibinfo{articleno}{5}, \bibinfo{numpages}{14}~pages.
\newblock
\showISBNx{9781450309769}
\urldef\tempurl%
\url{https://doi.org/10.1145/2038916.2038921}
\showDOI{\tempurl}


\bibitem[Shortle et~al\mbox{.}(2018)]%
        {shortle2018fundamentals}
\bibfield{author}{\bibinfo{person}{J.F. Shortle}, \bibinfo{person}{J.M. Thompson}, \bibinfo{person}{D. Gross}, {and} \bibinfo{person}{C.M. Harris}.} \bibinfo{year}{2018}\natexlab{}.
\newblock \bibinfo{booktitle}{\emph{Fundamentals of Queueing Theory}}.
\newblock \bibinfo{publisher}{John Wiley \& Sons, Ltd}, \bibinfo{address}{Hoboken, NJ, USA}.
\newblock
\showISBNx{9781118943564}
\showLCCN{2017041116}


\bibitem[Siami-Namini et~al\mbox{.}(2018)]%
        {arima_comparison1}
\bibfield{author}{\bibinfo{person}{Sima Siami-Namini}, \bibinfo{person}{Neda Tavakoli}, {and} \bibinfo{person}{Akbar Siami~Namin}.} \bibinfo{year}{2018}\natexlab{}.
\newblock \showarticletitle{A Comparison of ARIMA and LSTM in Forecasting Time Series}. In \bibinfo{booktitle}{\emph{2018 17th IEEE International Conference on Machine Learning and Applications (ICMLA)}}. \bibinfo{pages}{1394--1401}.
\newblock
\urldef\tempurl%
\url{https://doi.org/10.1109/ICMLA.2018.00227}
\showDOI{\tempurl}


\bibitem[Storn and Price(1997)]%
        {de}
\bibfield{author}{\bibinfo{person}{Rainer Storn} {and} \bibinfo{person}{Kenneth Price}.} \bibinfo{year}{1997}\natexlab{}.
\newblock \showarticletitle{Differential evolution-a simple and efficient heuristic for global optimization over continuous spaces}.
\newblock \bibinfo{journal}{\emph{Journal of global optimization}} \bibinfo{volume}{11}, \bibinfo{number}{4} (\bibinfo{year}{1997}), \bibinfo{pages}{341}.
\newblock


\bibitem[Team(2024a)]%
        {k8s_hpa}
\bibfield{author}{\bibinfo{person}{Kubernetes Team}.} \bibinfo{year}{2024}\natexlab{a}.
\newblock \bibinfo{title}{Kubernetes Horizontal Pod Autoscaler}.
\newblock
\newblock
\urldef\tempurl%
\url{https://kubernetes.io/docs/tasks/run-application/horizontal-pod-autoscale/}
\showURL{%
\tempurl}


\bibitem[Team(2024b)]%
        {k8s_resource_quota}
\bibfield{author}{\bibinfo{person}{Kubernetes Team}.} \bibinfo{year}{2024}\natexlab{b}.
\newblock \bibinfo{title}{Resource Quotas}.
\newblock
\newblock
\urldef\tempurl%
\url{https://kubernetes.io/docs/concepts/policy/resource-quotas/}
\showURL{%
\tempurl}


\bibitem[Team(2024c)]%
        {ray_serve_autoscaler}
\bibfield{author}{\bibinfo{person}{Ray Team}.} \bibinfo{year}{2024}\natexlab{c}.
\newblock \bibinfo{title}{Ray Serve Autoscaling}.
\newblock
\newblock
\urldef\tempurl%
\url{https://docs.ray.io/en/releases-2.0.0/serve/scaling-and-resource-allocation.html}
\showURL{%
\tempurl}


\bibitem[Team(2024d)]%
        {ray_serve}
\bibfield{author}{\bibinfo{person}{Ray Team}.} \bibinfo{year}{2024}\natexlab{d}.
\newblock \bibinfo{title}{Ray Serve: Scalable and Programmable Serving}.
\newblock
\newblock
\urldef\tempurl%
\url{https://docs.ray.io/en/releases-2.0.0/serve}
\showURL{%
\tempurl}


\bibitem[Tijms(2006)]%
        {mdc_approximation}
\bibfield{author}{\bibinfo{person}{Henk Tijms}.} \bibinfo{year}{2006}\natexlab{}.
\newblock \showarticletitle{New and old results for the M/D/c queue}.
\newblock \bibinfo{journal}{\emph{AEU - International Journal of Electronics and Communications}} \bibinfo{volume}{60}, \bibinfo{number}{2} (\bibinfo{year}{2006}), \bibinfo{pages}{125--130}.
\newblock
\showISSN{1434-8411}
\urldef\tempurl%
\url{https://doi.org/10.1016/j.aeue.2005.11.008}
\showDOI{\tempurl}


\bibitem[Twitter(2018)]%
        {twitter_trace}
\bibfield{author}{\bibinfo{person}{Twitter}.} \bibinfo{year}{2018}\natexlab{}.
\newblock \bibinfo{title}{Twitter stream traces}.
\newblock
\newblock
\urldef\tempurl%
\url{https://archive.org/details/archiveteam-twitter-stream-2018-04}
\showURL{%
\tempurl}


\bibitem[Verma et~al\mbox{.}(2015)]%
        {borg}
\bibfield{author}{\bibinfo{person}{Abhishek Verma}, \bibinfo{person}{Luis Pedrosa}, \bibinfo{person}{Madhukar Korupolu}, \bibinfo{person}{David Oppenheimer}, \bibinfo{person}{Eric Tune}, {and} \bibinfo{person}{John Wilkes}.} \bibinfo{year}{2015}\natexlab{}.
\newblock \showarticletitle{Large-Scale Cluster Management at Google with Borg}. In \bibinfo{booktitle}{\emph{Proceedings of the Tenth European Conference on Computer Systems}} \emph{(\bibinfo{series}{EuroSys 15})}. \bibinfo{publisher}{Association for Computing Machinery}, \bibinfo{address}{Bordeaux, France}, Article \bibinfo{articleno}{18}, \bibinfo{numpages}{17}~pages.
\newblock
\showISBNx{9781450332385}
\urldef\tempurl%
\url{https://doi.org/10.1145/2741948.2741964}
\showDOI{\tempurl}


\bibitem[Virtanen et~al\mbox{.}(2020)]%
        {scipy}
\bibfield{author}{\bibinfo{person}{Pauli Virtanen}, \bibinfo{person}{Ralf Gommers}, \bibinfo{person}{Travis~E. Oliphant}, \bibinfo{person}{Matt Haberland}, \bibinfo{person}{Tyler Reddy}, \bibinfo{person}{David Cournapeau}, \bibinfo{person}{Evgeni Burovski}, \bibinfo{person}{Pearu Peterson}, \bibinfo{person}{Warren Weckesser}, \bibinfo{person}{Jonathan Bright}, \bibinfo{person}{St{\'e}fan~J. {van der Walt}}, \bibinfo{person}{Matthew Brett}, \bibinfo{person}{Joshua Wilson}, \bibinfo{person}{K.~Jarrod Millman}, \bibinfo{person}{Nikolay Mayorov}, \bibinfo{person}{Andrew R.~J. Nelson}, \bibinfo{person}{Eric Jones}, \bibinfo{person}{Robert Kern}, \bibinfo{person}{Eric Larson}, \bibinfo{person}{C~J Carey}, \bibinfo{person}{{\.I}lhan Polat}, \bibinfo{person}{Yu Feng}, \bibinfo{person}{Eric~W. Moore}, \bibinfo{person}{Jake {VanderPlas}}, \bibinfo{person}{Denis Laxalde}, \bibinfo{person}{Josef Perktold}, \bibinfo{person}{Robert Cimrman}, \bibinfo{person}{Ian Henriksen}, \bibinfo{person}{E.~A. Quintero},
  \bibinfo{person}{Charles~R. Harris}, \bibinfo{person}{Anne~M. Archibald}, \bibinfo{person}{Ant{\^o}nio~H. Ribeiro}, \bibinfo{person}{Fabian Pedregosa}, \bibinfo{person}{Paul {van Mulbregt}}, {and} \bibinfo{person}{{SciPy 1.0 Contributors}}.} \bibinfo{year}{2020}\natexlab{}.
\newblock \showarticletitle{{{SciPy} 1.0: Fundamental Algorithms for Scientific Computing in Python}}.
\newblock \bibinfo{journal}{\emph{Nature Methods}}  \bibinfo{volume}{17} (\bibinfo{year}{2020}), \bibinfo{pages}{261--272}.
\newblock
\urldef\tempurl%
\url{https://doi.org/10.1038/s41592-019-0686-2}
\showDOI{\tempurl}


\bibitem[Wang et~al\mbox{.}(2012)]%
        {cake}
\bibfield{author}{\bibinfo{person}{Andrew Wang}, \bibinfo{person}{Shivaram Venkataraman}, \bibinfo{person}{Sara Alspaugh}, \bibinfo{person}{Randy Katz}, {and} \bibinfo{person}{Ion Stoica}.} \bibinfo{year}{2012}\natexlab{}.
\newblock \showarticletitle{Cake: enabling high-level SLOs on shared storage systems}. In \bibinfo{booktitle}{\emph{Proceedings of the Third ACM Symposium on Cloud Computing}} (San Jose, California) \emph{(\bibinfo{series}{SoCC '12})}. \bibinfo{publisher}{Association for Computing Machinery}, \bibinfo{address}{New York, NY, USA}, Article \bibinfo{articleno}{14}, \bibinfo{numpages}{14}~pages.
\newblock
\showISBNx{9781450317610}
\urldef\tempurl%
\url{https://doi.org/10.1145/2391229.2391243}
\showDOI{\tempurl}


\bibitem[Wang et~al\mbox{.}(2013)]%
        {wang2013slo}
\bibfield{author}{\bibinfo{person}{Jie Wang}, \bibinfo{person}{Qingzhong Li}, {and} \bibinfo{person}{Yuliang Shi}.} \bibinfo{year}{2013}\natexlab{}.
\newblock \showarticletitle{SLO-Driven Task Scheduling in MapReduce Environments}. In \bibinfo{booktitle}{\emph{Proceedings of the 2013 10th Web Information System and Application Conference (WISA 13)}}. \bibinfo{publisher}{{IEEE}}, \bibinfo{address}{Yangzhou, China}, \bibinfo{pages}{308--313}.
\newblock
\urldef\tempurl%
\url{https://doi.org/10.1109/WISA.2013.64}
\showDOI{\tempurl}


\bibitem[Wang et~al\mbox{.}(2021)]%
        {ownership}
\bibfield{author}{\bibinfo{person}{Stephanie Wang}, \bibinfo{person}{Eric Liang}, \bibinfo{person}{Edward Oakes}, \bibinfo{person}{Ben Hindman}, \bibinfo{person}{Frank~Sifei Luan}, \bibinfo{person}{Audrey Cheng}, {and} \bibinfo{person}{Ion Stoica}.} \bibinfo{year}{2021}\natexlab{}.
\newblock \showarticletitle{Ownership: A Distributed Futures System for {Fine-Grained} Tasks}. In \bibinfo{booktitle}{\emph{Proceedings of the 18th USENIX Symposium on Networked Systems Design and Implementation (NSDI 21)}}. \bibinfo{publisher}{USENIX Association}, \bibinfo{address}{Virtual Event}, \bibinfo{pages}{671--686}.
\newblock
\showISBNx{978-1-939133-21-2}
\urldef\tempurl%
\url{https://www.usenix.org/conference/nsdi21/presentation/cheng}
\showURL{%
\tempurl}


\bibitem[Weng et~al\mbox{.}(2022)]%
        {mlaas_in_the_workload}
\bibfield{author}{\bibinfo{person}{Qizhen Weng}, \bibinfo{person}{Wencong Xiao}, \bibinfo{person}{Yinghao Yu}, \bibinfo{person}{Wei Wang}, \bibinfo{person}{Cheng Wang}, \bibinfo{person}{Jian He}, \bibinfo{person}{Yong Li}, \bibinfo{person}{Liping Zhang}, \bibinfo{person}{Wei Lin}, {and} \bibinfo{person}{Yu Ding}.} \bibinfo{year}{2022}\natexlab{}.
\newblock \showarticletitle{{MLaaS} in the Wild: Workload Analysis and Scheduling in {Large-Scale} Heterogeneous {GPU} Clusters}. In \bibinfo{booktitle}{\emph{Proceedings of the 19th USENIX Symposium on Networked Systems Design and Implementation (NSDI 22)}}. \bibinfo{publisher}{USENIX Association}, \bibinfo{address}{Renton, WA}, \bibinfo{pages}{945--960}.
\newblock
\showISBNx{978-1-939133-27-4}
\urldef\tempurl%
\url{https://www.usenix.org/conference/nsdi22/presentation/weng}
\showURL{%
\tempurl}


\bibitem[Wilkes(2009)]%
        {wilkes2009utility}
\bibfield{author}{\bibinfo{person}{John Wilkes}.} \bibinfo{year}{2009}\natexlab{}.
\newblock \showarticletitle{Utility Functions, Prices, and Negotiation}. In \bibinfo{booktitle}{\emph{Market‐Oriented Grid and Utility Computing}}. \bibinfo{publisher}{John Wiley \& Sons, Ltd}, \bibinfo{address}{Hoboken, NJ, USA}, \bibinfo{pages}{67--88}.
\newblock
\showISBNx{9780470455432}
\urldef\tempurl%
\url{https://doi.org/10.1002/9780470455432.ch4}
\showDOI{\tempurl}


\bibitem[Xiao et~al\mbox{.}(2018)]%
        {gandiva}
\bibfield{author}{\bibinfo{person}{Wencong Xiao}, \bibinfo{person}{Romil Bhardwaj}, \bibinfo{person}{Ramachandran Ramjee}, \bibinfo{person}{Muthian Sivathanu}, \bibinfo{person}{Nipun Kwatra}, \bibinfo{person}{Zhenhua Han}, \bibinfo{person}{Pratyush Patel}, \bibinfo{person}{Xuan Peng}, \bibinfo{person}{Hanyu Zhao}, \bibinfo{person}{Quanlu Zhang}, \bibinfo{person}{Fan Yang}, {and} \bibinfo{person}{Lidong Zhou}.} \bibinfo{year}{2018}\natexlab{}.
\newblock \showarticletitle{Gandiva: Introspective Cluster Scheduling for Deep Learning}. In \bibinfo{booktitle}{\emph{13th {USENIX} Symposium on Operating Systems Design and Implementation ({OSDI} 18)}}. \bibinfo{publisher}{{USENIX} Association}, \bibinfo{address}{Carlsbad, CA}, \bibinfo{pages}{595--610}.
\newblock
\showISBNx{978-1-939133-08-3}
\urldef\tempurl%
\url{https://www.usenix.org/conference/osdi18/presentation/xiao}
\showURL{%
\tempurl}


\bibitem[Xue et~al\mbox{.}(2022)]%
        {Xue_2022}
\bibfield{author}{\bibinfo{person}{Siqiao Xue}, \bibinfo{person}{Chao Qu}, \bibinfo{person}{Xiaoming Shi}, \bibinfo{person}{Cong Liao}, \bibinfo{person}{Shiyi Zhu}, \bibinfo{person}{Xiaoyu Tan}, \bibinfo{person}{Lintao Ma}, \bibinfo{person}{Shiyu Wang}, \bibinfo{person}{Shijun Wang}, \bibinfo{person}{Yun Hu}, \bibinfo{person}{Lei Lei}, \bibinfo{person}{Yangfei Zheng}, \bibinfo{person}{Jianguo Li}, {and} \bibinfo{person}{James Zhang}.} \bibinfo{year}{2022}\natexlab{}.
\newblock \showarticletitle{A Meta Reinforcement Learning Approach for Predictive Autoscaling in the Cloud}. In \bibinfo{booktitle}{\emph{Proceedings of the 28th ACM SIGKDD Conference on Knowledge Discovery and Data Mining}} \emph{(\bibinfo{series}{KDD 22})}. \bibinfo{publisher}{Association for Computing Machinery}, \bibinfo{address}{Washington DC, USA}, \bibinfo{pages}{4290–4299}.
\newblock
\showISBNx{9781450393850}
\urldef\tempurl%
\url{https://doi.org/10.1145/3534678.3539063}
\showDOI{\tempurl}


\bibitem[Zhang et~al\mbox{.}(2019)]%
        {mark}
\bibfield{author}{\bibinfo{person}{Chengliang Zhang}, \bibinfo{person}{Minchen Yu}, \bibinfo{person}{Wei Wang}, {and} \bibinfo{person}{Feng Yan}.} \bibinfo{year}{2019}\natexlab{}.
\newblock \showarticletitle{{MArk}: Exploiting Cloud Services for {Cost-Effective}, {SLO-Aware} Machine Learning Inference Serving}. In \bibinfo{booktitle}{\emph{Proceedings of the 2019 USENIX Annual Technical Conference (USENIX ATC 19)}}. \bibinfo{publisher}{USENIX Association}, \bibinfo{address}{Renton, WA}, \bibinfo{pages}{1049--1062}.
\newblock
\showISBNx{978-1-939133-03-8}
\urldef\tempurl%
\url{https://www.usenix.org/conference/atc19/presentation/zhang-chengliang}
\showURL{%
\tempurl}


\bibitem[Zhang et~al\mbox{.}(2017)]%
        {zhang2017live}
\bibfield{author}{\bibinfo{person}{Haoyu Zhang}, \bibinfo{person}{Ganesh Ananthanarayanan}, \bibinfo{person}{Peter Bodik}, \bibinfo{person}{Matthai Philipose}, \bibinfo{person}{Paramvir Bahl}, {and} \bibinfo{person}{Michael~J. Freedman}.} \bibinfo{year}{2017}\natexlab{}.
\newblock \showarticletitle{Live Video Analytics at Scale with Approximation and {Delay-Tolerance}}. In \bibinfo{booktitle}{\emph{Proceedings of the 14th USENIX Symposium on Networked Systems Design and Implementation (NSDI 17)}}. \bibinfo{publisher}{USENIX Association}, \bibinfo{address}{Boston, MA}, \bibinfo{pages}{377--392}.
\newblock
\showISBNx{978-1-931971-37-9}
\urldef\tempurl%
\url{https://www.usenix.org/conference/nsdi17/technical-sessions/presentation/zhang}
\showURL{%
\tempurl}


\bibitem[Zhao and Uta(2022)]%
        {tiny_autoscaler}
\bibfield{author}{\bibinfo{person}{Yuxuan Zhao} {and} \bibinfo{person}{Alexandru Uta}.} \bibinfo{year}{2022}\natexlab{}.
\newblock \showarticletitle{Tiny Autoscalers for Tiny Workloads: Dynamic CPU Allocation for Serverless Functions}. In \bibinfo{booktitle}{\emph{Proceedings of the 2022 22nd IEEE International Symposium on Cluster, Cloud and Internet Computing (CCGrid 22)}}. \bibinfo{publisher}{{IEEE}}, \bibinfo{address}{Taormina, Italy}, \bibinfo{pages}{170--179}.
\newblock
\urldef\tempurl%
\url{https://doi.org/10.1109/CCGrid54584.2022.00026}
\showDOI{\tempurl}


\bibitem[Zheng et~al\mbox{.}(2023)]%
        {shockwave}
\bibfield{author}{\bibinfo{person}{Pengfei Zheng}, \bibinfo{person}{Rui Pan}, \bibinfo{person}{Tarannum Khan}, \bibinfo{person}{Shivaram Venkataraman}, {and} \bibinfo{person}{Aditya Akella}.} \bibinfo{year}{2023}\natexlab{}.
\newblock \showarticletitle{Shockwave: Fair and Efficient Cluster Scheduling for Dynamic Adaptation in Machine Learning}. In \bibinfo{booktitle}{\emph{20th USENIX Symposium on Networked Systems Design and Implementation (NSDI 23)}}. \bibinfo{publisher}{USENIX Association}, \bibinfo{address}{Boston, MA}, \bibinfo{pages}{703--723}.
\newblock
\showISBNx{978-1-939133-33-5}
\urldef\tempurl%
\url{https://www.usenix.org/conference/nsdi23/presentation/zheng}
\showURL{%
\tempurl}


\end{thebibliography}


\end{document}